\documentclass[twocolumn]{aastex63}
\usepackage{amsmath}

\newcommand{\lnu}{$L_{\nu}^d$(5000)}
\newcommand{\fluxa}{$f_{\rm F444W}$}
\newcommand{\fluxb}{$f_{\rm F150W}$}

\newcommand{\jwratio}{$f_{\rm F444W}/f_{\rm F150W}$}
\newcommand{\jwratioc}{$f_{\rm F444W}/f_{\rm F115W}$}

\newcommand{\afluxa}{450~$\mu$m}
\newcommand{\afluxb}{850~$\mu$m}

\newcommand{\fafluxbb}{$f_{850\,\mu{\rm m}}^d$}

\shortauthors{Barger \& Cowie}

\begin{document}

\title{Mapping the Decline with Redshift of Dusty Star-forming Galaxies Using JWST and SCUBA-2}

\correspondingauthor{Amy J. Barger}
\email{barger@astro.wisc.edu}

\author[0000-0002-3306-1606]{A.~J.~Barger}
\affiliation{Department of Astronomy, University of Wisconsin-Madison,
475 N. Charter Street, Madison, WI 53706, USA}
\affiliation{Department of Physics and Astronomy, University of Hawaii,
2505 Correa Road, Honolulu, HI 96822, USA}
\affiliation{Institute for Astronomy, University of Hawaii, 2680 Woodlawn Drive,
Honolulu, HI 96822, USA}

\author[0000-0002-6319-1575]{L.~L.~Cowie}
\affiliation{Institute for Astronomy, University of Hawaii,
2680 Woodlawn Drive, Honolulu, HI 96822, USA}

%-----------------------------------------------------------------------------
%   Abstract
%-----------------------------------------------------------------------------
\begin{abstract}
We use JWST NIRCam observations
of the massive lensing cluster field A2744 to develop a red
galaxy selection of \fluxa\ $>1~\mu$Jy and \jwratio\ $>3.5$ that picks
out all 9 $>4.5\sigma$ ALMA 1.1 or 1.2~mm sources and 17 of the 19 $>5\sigma$
SCUBA-2 \afluxb\ sources in the covered areas. 
We show that by using the red galaxies as priors, we 
can probe deeper in the SCUBA-2 \afluxb\ image.
This gives a sample of 44 $>3\sigma$ SCUBA-2 \afluxb\ sources with accurate positions,
photometric redshifts, and magnifications.
To investigate why our red galaxy selection picks out the \afluxb\ sources,
we next analyze an extended sample of 167 sources with \fluxa\ $>0.05~\mu$Jy and \jwratio\ $>3.5$.
We find that the fainter \fluxa\ sources in this sample are too faint to be
detected in the SCUBA-2 \afluxb\ image.
We also show that there is a strong drop between $z<4$ and $z>4$ (a factor of around 5) 
in the ratio of the far-infrared luminosity 
estimated from the \afluxb\ flux to the $\nu$\lnu\
at rest-frame 5000~\AA. We argue that this result may be due to the high-redshift sources having
less dust content than the lower redshift sources.
\end{abstract}


%-----------------------------------------------------------------------------
%   Introduction
%-----------------------------------------------------------------------------
\section{Introduction}
\label{secintro}
Distant, dusty, extremely luminous galaxies 
\citep[e.g.,][]{smail97,barger98,hughes98,eales99}
are some of the most powerfully star-forming galaxies in the universe and
are significant contributors to the total star formation history from
$z\sim2$ to at least $z\sim5$ 
\citep[e.g.,][]{barger00,barger14,chapman05,wardlow11,casey13,swinbank14,cowie17,zavala21}. 
These dusty star-forming galaxies  (DSFGs) (also known as
submillimeter galaxies, or SMGs) are most easily found through wide-field
submillimeter/millimeter imaging on single-dish telescopes, such as 
the James Clerk Maxwell Telescope (JCMT) with the
\hbox{SCUBA-2} camera \citep{holland13}, or, in the near future, the Large Millimeter Telescope
with the TolTEC camera \citep{wilson20}. 

The natural limit of single-dish submillimeter/millimeter surveys is the depth where 
confusion---the blending 
of sources, or where the noise is dominated by unresolved 
contributions from fainter sources---becomes important. For example,
\citet{cowie17} give a confusion limit of 1.65~mJy for
\afluxb\ observations using \hbox{SCUBA-2} on the 15~m JCMT.

The lack of positional accuracy is also a major problem when trying to ascertain
the properties of DSFGs. 
Such identifications are critical for estimating photometric redshifts, modeling
spectral energy distributions (SEDs), and determining morphologies.
Historically, deep radio interferometric 
images were used to identify counterparts to SMGs
\citep[e.g.,][]{barger00,smail00,ivison02,chapman03},
while, more recently, submillimeter/millimeter interferometry with 
the Submillimeter Array (SMA; \citealt{ho04}), 
NOEMA (and, previously, the IRAM Plateau de Bure interferometer),
and, most powerfully, the Atacama Large Millimeter/submillimeter Array (ALMA) have become essential tools
for obtaining accurate positions, as well as for resolving some single-dish sources into multiple
submillimeter/millimeter sources (e.g., \citealt{wang11}).
However, interferometric observations have small fields of view, which make direct
searches \citep[e.g.,][]{dunlop17,gonzalez17cont,franco18,umehata18,casey21,fujimoto23},
or even follow-up observations of sources detected in single-dish surveys
\citep[e.g.,][]{daddi09,barger12,walter12,hodge13,chen14,cowie18,stach19,jones21,cooper22,cairns23}, quite costly.

It has been recognized for some time that galaxies with extremely red infrared colors,
such as the $K$-4.5~$\mu$m selected KIEROs of \citet{wang12}
(the KIEROs acronym stands for $K_s$ and IRAC selected Extremely Red Objects)
or the $H$-4.5~$\mu$m
(also $H$-3.6~$\mu$m) selected HIEROs of \citet{caputi12}, \citet{wang16}, and
\citet{alcalde19} (the HIEROs acronym stands for $H$ and IRAC selected Extremely Red Objects)
are effective in picking out submillimeter/millimeter galaxies \citep{wang12,wang19}.
However, the advent of JWST, with its extremely deep, very high spatial resolution
near-infrared (NIR) observations, is set to revolutionize this field. 

Using the CEERS JWST NIRCam data (JWST-ERS-1345),
\citet{barrufet23} described the selection and properties of dark galaxies with 4.44~$\mu$m 
to 1.6~$\mu$m flux ratios $>8.3$ (based on the \citealt{caputi12} selection from Spitzer and HST). 
They showed that these are very 
dusty galaxies extending over a wide range of redshifts ($z=2$--8). Although
they suggested that their dark galaxies may be higher redshift, lower star formation rate 
(SFR) extensions of submillimeter/millimeter selected DSFGs,
they did not match to the
submillimeter/millimeter data in the field to relate their dark galaxies to DSFGs directly. 

In the present paper, we demonstrate using observations of the massive lensing 
cluster field A2744 how ideally suited JWST NIRCam data are to finding DSFGs.
The structure of the paper is as follows.
In Section~\ref{secdata}, we introduce the published datasets that we use in our analysis.
In Section~\ref{secalma}, we give our NIRCam color selection criteria
that identify all of the known ALMA sources in the field.
In Section~\ref{secscuba2cat}, we use these criteria to find NIRCam
counterparts to nearly all of the SCUBA-2 sources in the
NIRCam-observed region, which allows us to obtain accurate positions
for the SCUBA-2 sources.
In Section~\ref{secDSFG}, we invert this procedure and use our
NIRCam color selected sample as priors to obtain deeper submillimeter measurements 
in the SCUBA-2 images. 
In Section~\ref{secdisc}, we use the photometric redshifts and magnifications
of our NIRCam color selected sample to compare the 
far-infrared (FIR) luminosities to the rest-frame optical luminosities.
In Section~\ref{secsummary}, we summarize our results.

We assume 
a cosmology of $H_0=70.5$~km~s$^{-1}$~Mpc$^{-1}$,
$\Omega_{\rm M}=0.27$, and $\Omega_\Lambda=0.73$ 
\citep{larson11} throughout.

%---------------------------------------------------------------------
\section{Data}
\label{secdata}
%---------------------------------------------------------------------
A2744 is one of the six Hubble Frontier Field clusters \citep[HFF;][]{lotz17}.
A2744 has both deep SCUBA-2 observations from \citet{cowie22}
and deep ALMA mosaics from the ALMA Lensing Cluster Survey 
(ALCS; 1.2~mm; \citealt{kohno19,fujimoto23})
and the ALMA Frontier Fields Survey 
(AFFS; 1.1~mm; \citealt{gonzalez17cont}).
The AFFS has a $5\sigma$ threshold of 0.28~mJy. 
The very central regions of A2744 are relatively
rich in luminous DSFGs. The mosaicked ALMA images,
together with deeper follow-up ALMA observations 
(ALMA program \#2017.1.01219.S; PI: F.~Bauer), have yielded
9 $>4.5\sigma$ ALMA sources. These are listed in Table~9
of \citet{cowie22}, along with their known spectroscopic
redshifts. Note that we have updated the spectroscopic redshift
of the second A2744 source in the \citet{cowie22} table from 
$z=2.482$ to $z=2.585$ based on the ALMA CO
observations of F.~Bauer (priv. comm.); see \citet{kokorev23}
for a detailed analysis of this source.

%-----------------------------------------------------------------------------
%FIGURE 1:  jwst_overlay.pdf
%-----------------------------------------------------------------------------
\begin{figure}
\includegraphics[width=8cm,angle=0]{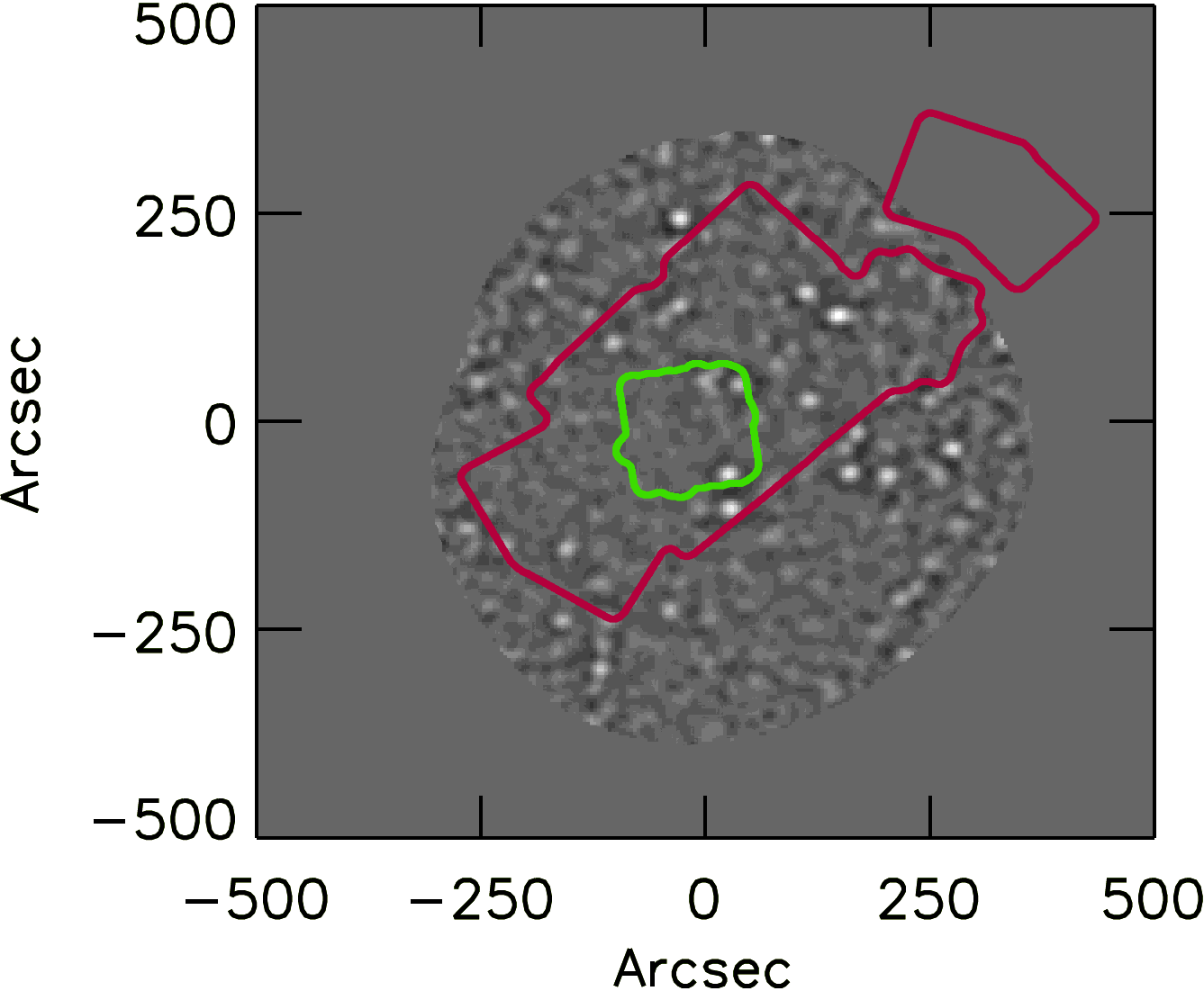}
\caption{SCUBA-2 \afluxb\ image of the massive lensing cluster field A2744 from \citet{cowie22}.
The footprints of the JWST NIRCam F444W image released by \citet{paris23}
(red) and the combined ALMA mosaics from AFFS and ALCS (green) are overlaid.
The imaging and catalog from \citet{weaver23}
do not cover a small portion of the lower-left corner of the NIRCAM footprint shown.
\label{footprint}
}
\end{figure}
%-----------------------------------------------------------------------------

A2744 was the target of multiple JWST NIRCam programs
(JWST-ERS-1324, JWST-GO-2561, JWST-DDT-2756). The combined images 
and a catalog were released by \citet{paris23} for the GLASS team, and images
and a catalog were released by \citet{weaver23} for the UNCOVER team.
In Figure~\ref{footprint}, we show the areas covered 
by the combined ALMA mosaics (green)
and the JWST NIRCam F444W data (red) of \citet{paris23},
overlaid on the SCUBA-2 \afluxb\ matched filter image of \citet{cowie22}.
In this work, we adopt the \citet{paris23} isophotal flux catalog.
We visually inspected the images at the catalog
positions in theses areas to flag artifacts and
identify objects that are likely parts of a single object. In these latter cases,
we combined the fluxes from the parts to provide a single flux for the object.
We note that there is substantial patterning in some regions of the images, 
so visual inspection is important.

The \citet{weaver23} catalog gives photometric redshifts obtained from the
EAZY code \citep{brammer08} and magnifications 
based on the source positions and redshifts \citep{furtak23}.
In this work, we adopt these photometric redshifts and 
magnifications ($\mu$), but we note that the \citet{weaver23} catalog does not
fully cover the JWST NIRCam image from \citet{paris23}.

%---------------------------------------------------------------------
% FIGURE 2:  alma#_new (#:1-9)
%---------------------------------------------------------------------
\begin{figure*}
\includegraphics[width=2.2in,angle=0]{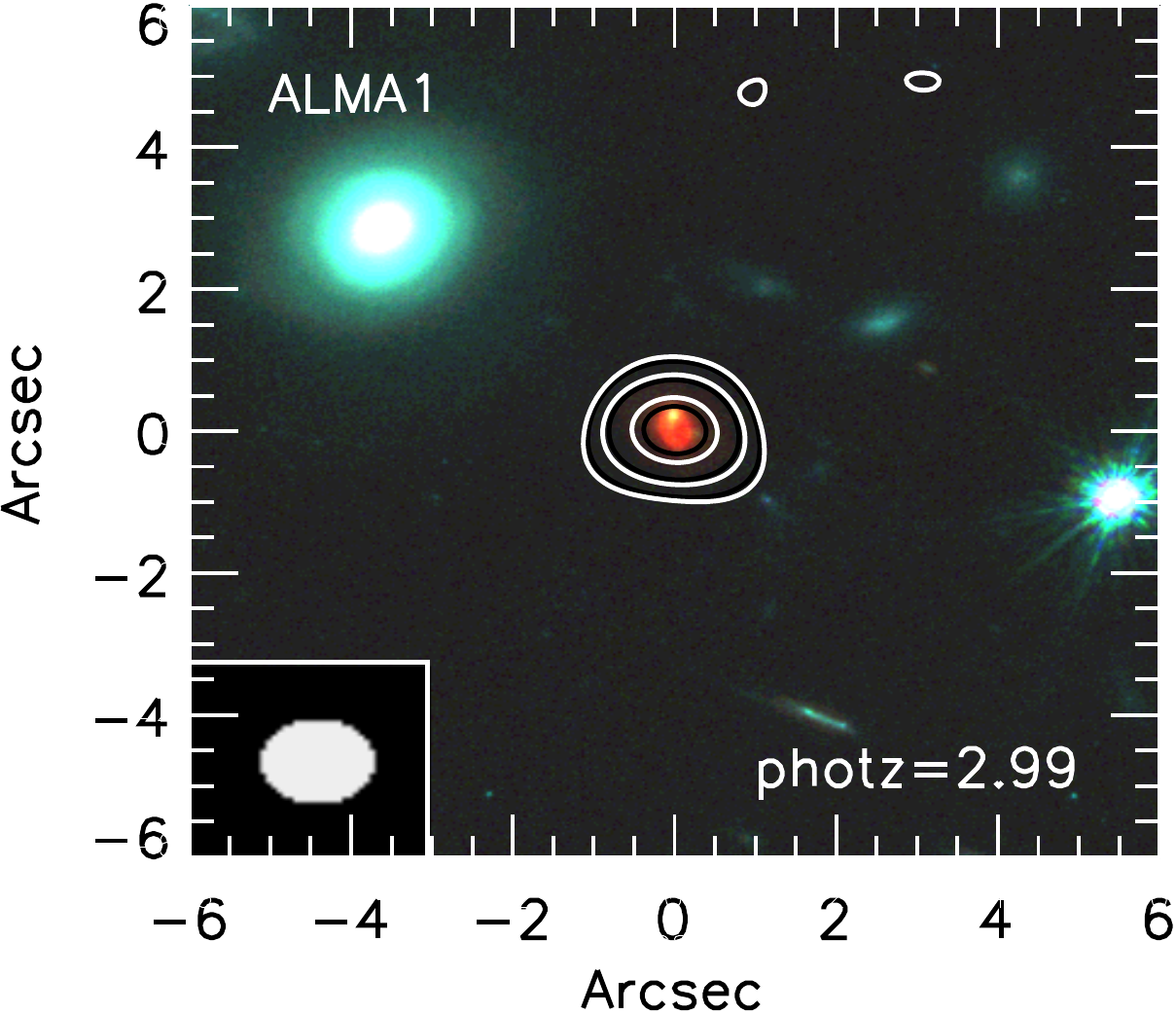}
\includegraphics[width=2.2in,angle=0]{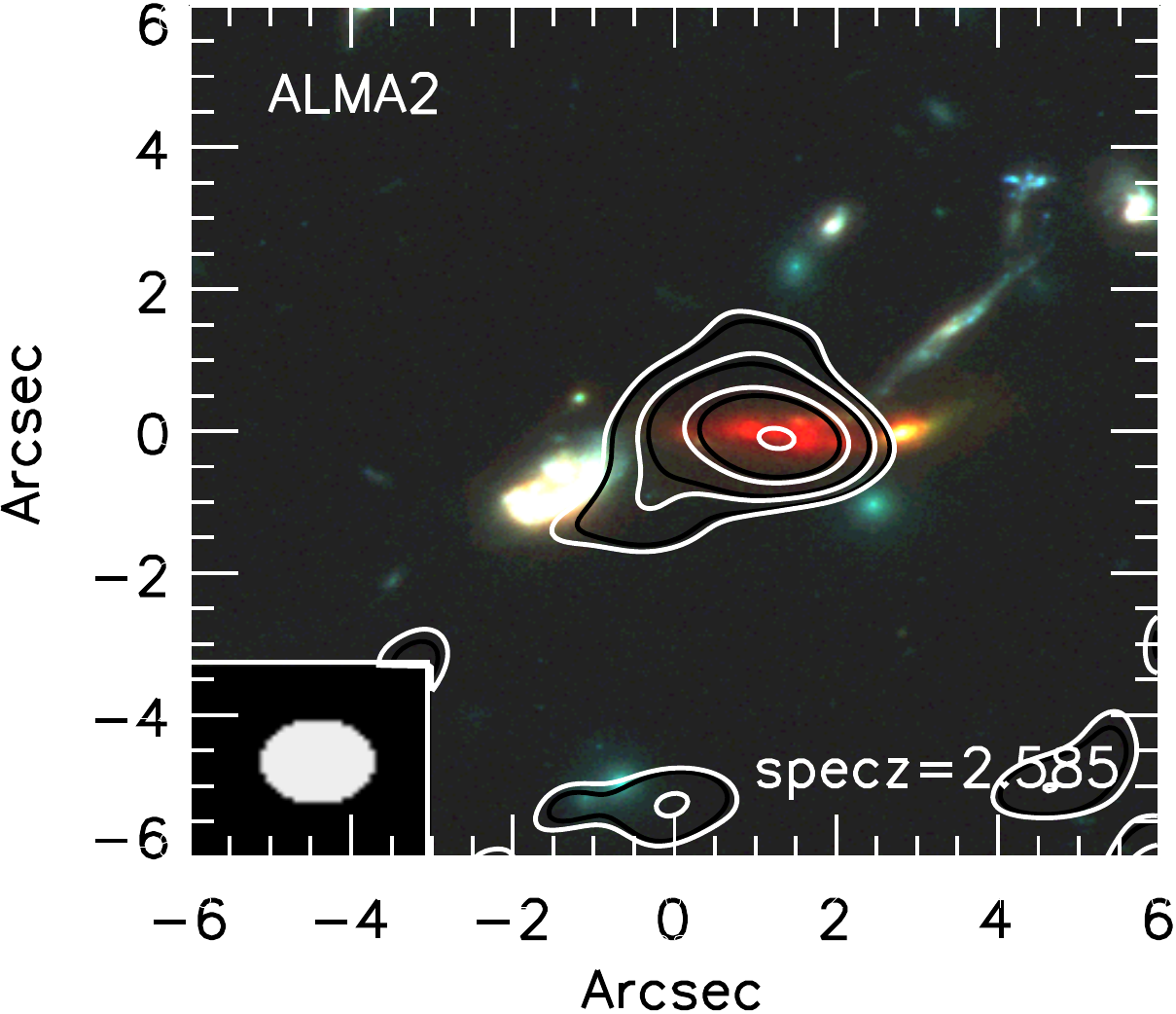}
\includegraphics[width=2.2in,angle=0]{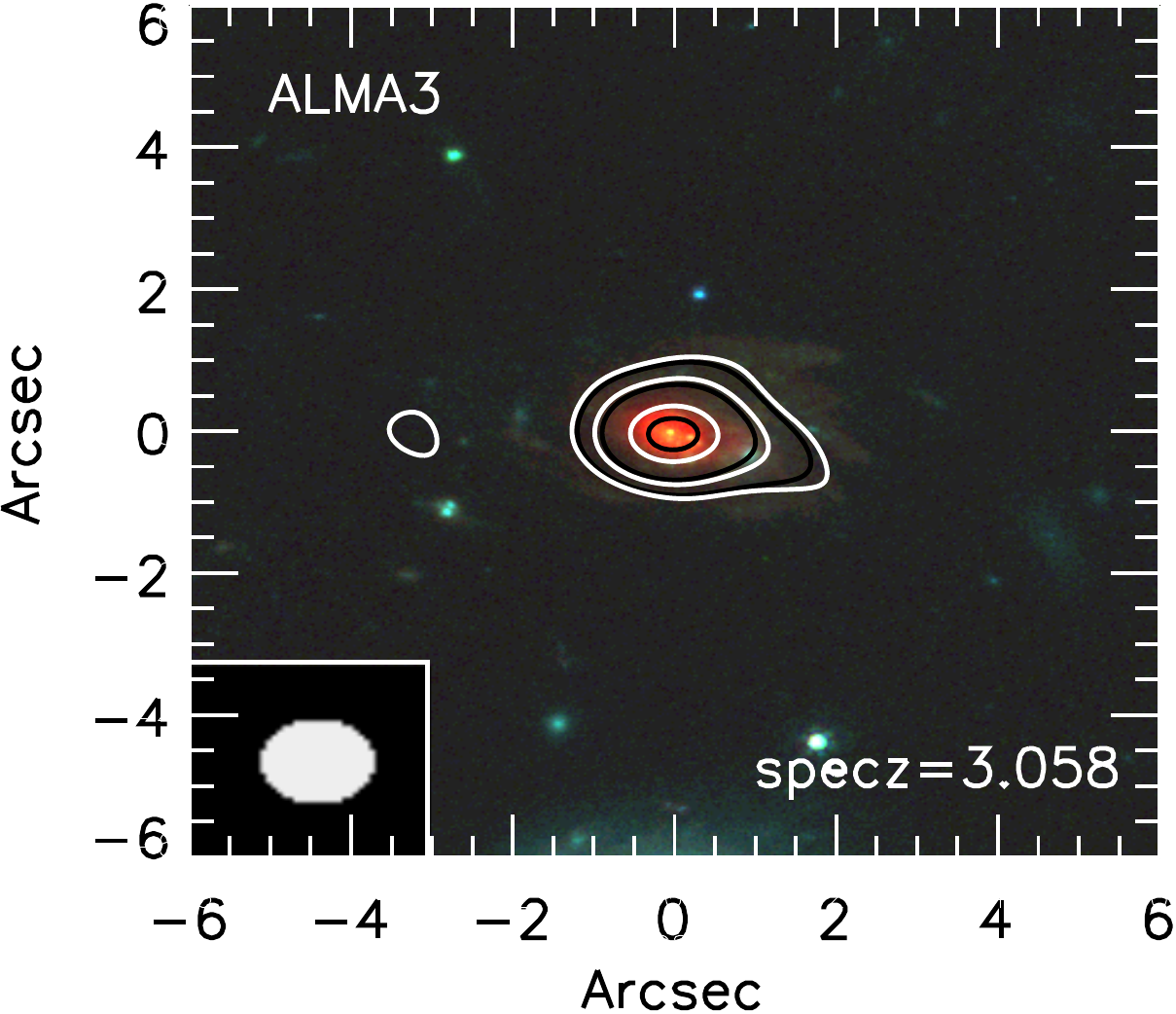}
\includegraphics[width=2.2in,angle=0]{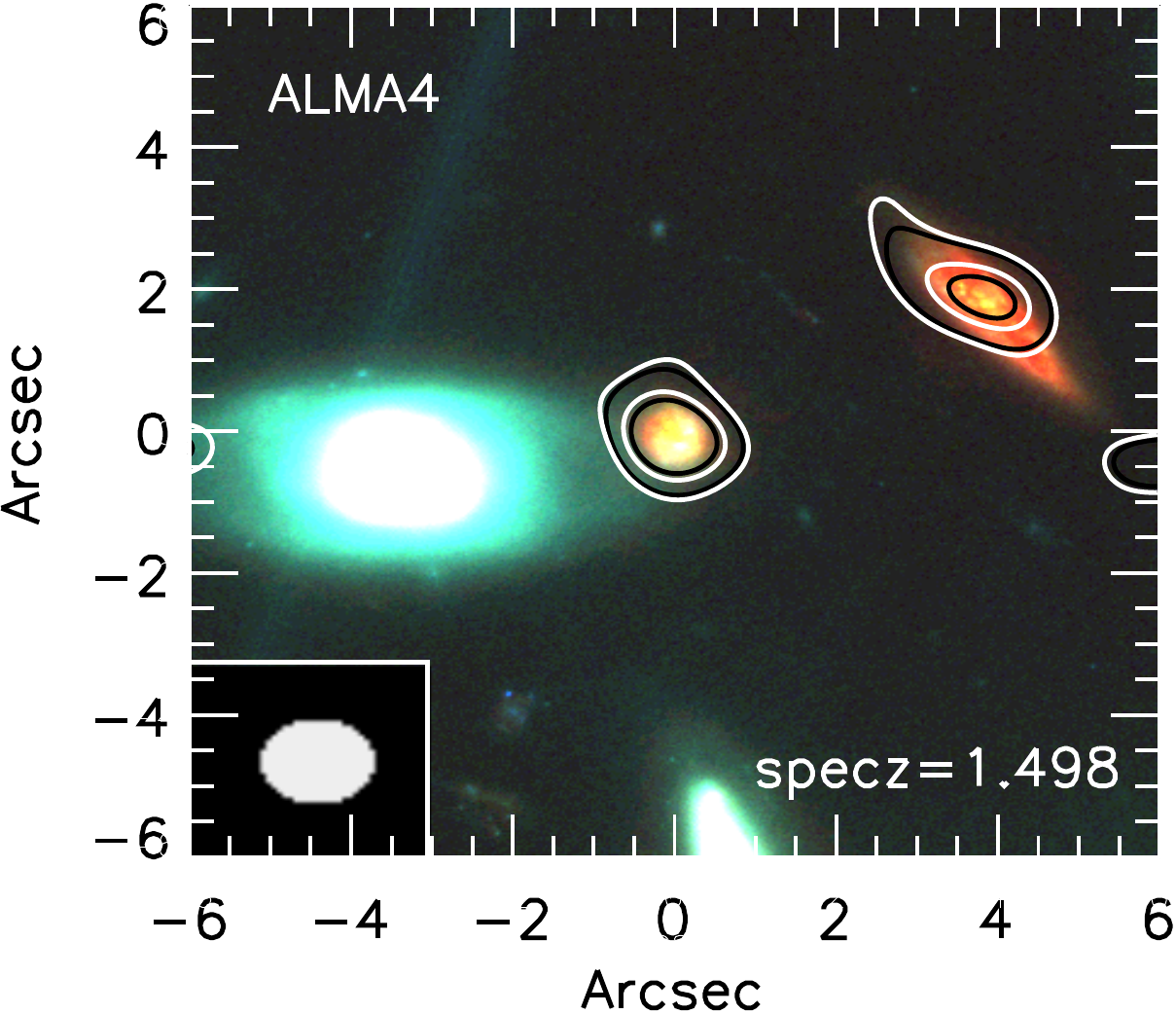}
\includegraphics[width=2.2in,angle=0]{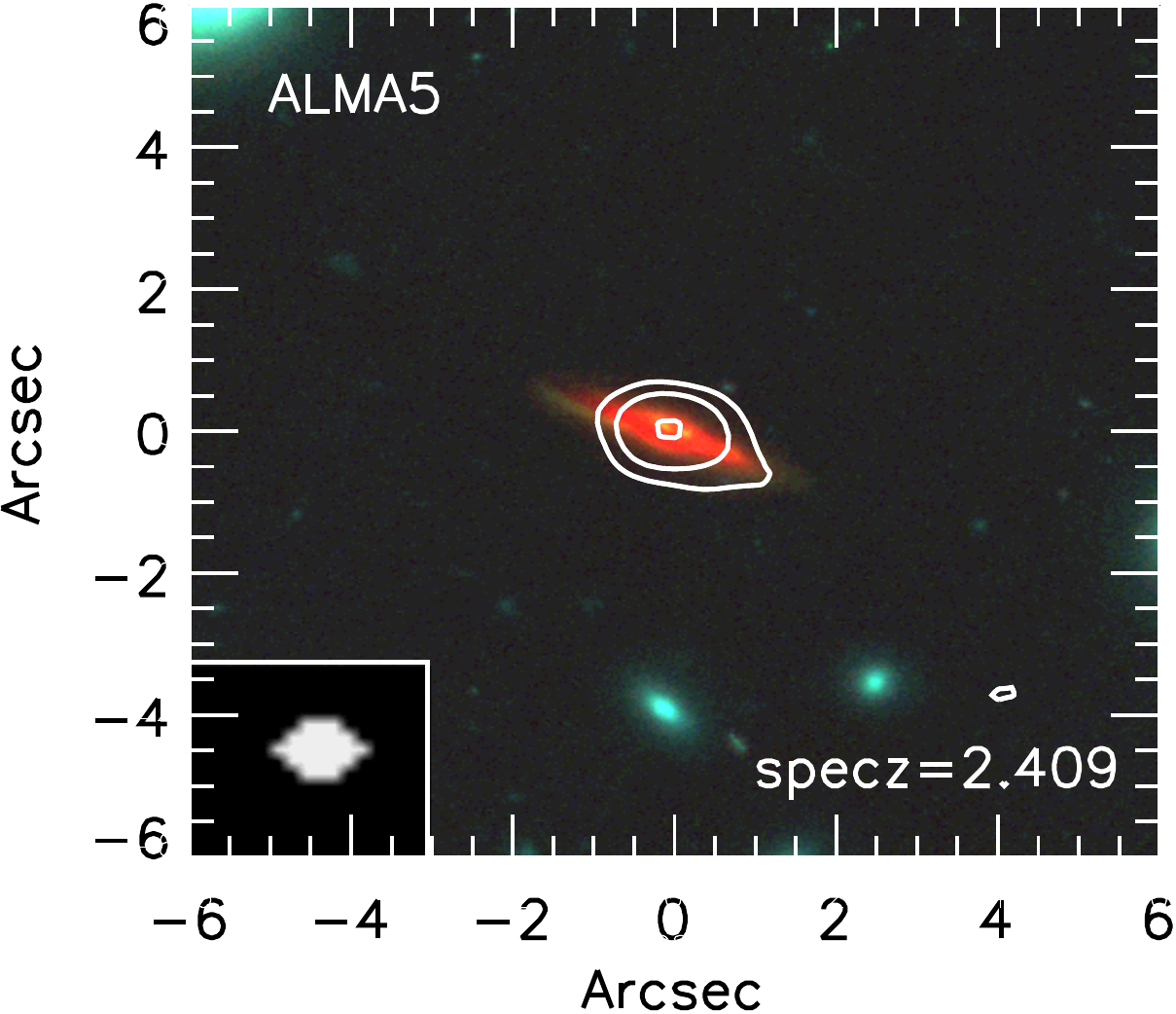}
\includegraphics[width=2.2in,angle=0]{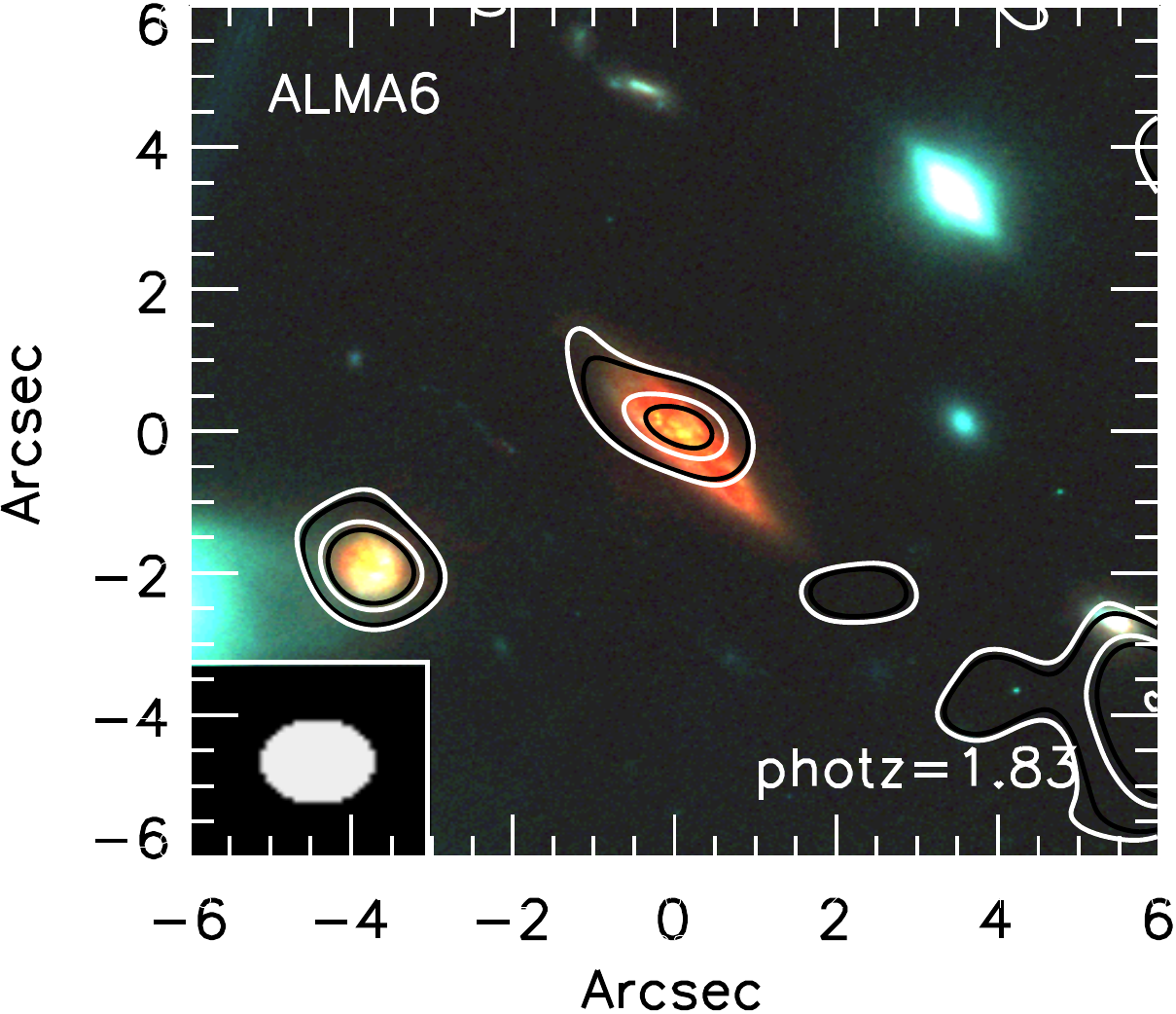}
\includegraphics[width=2.2in,angle=0]{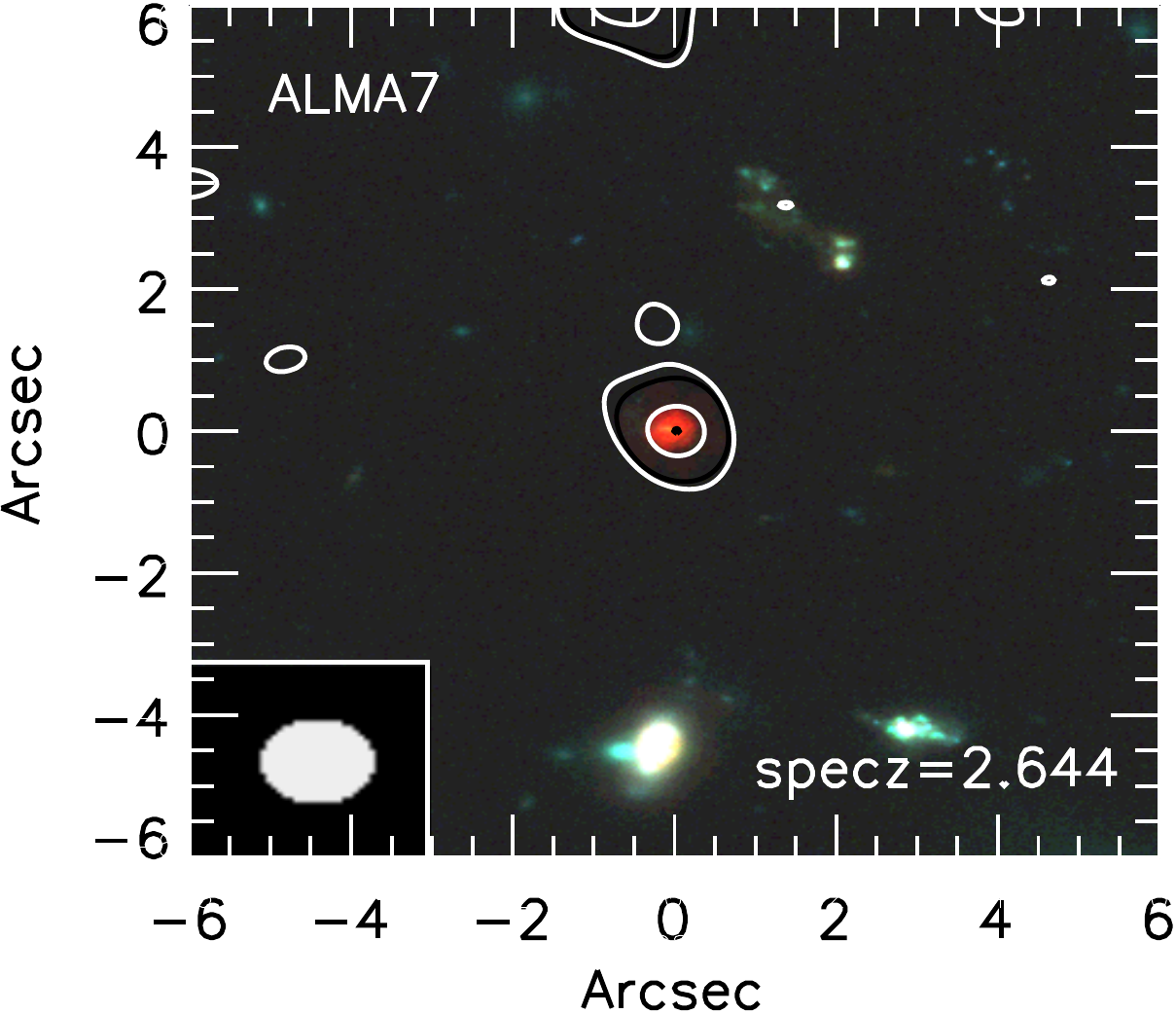}
\hspace{0.4cm}
\includegraphics[width=2.2in,angle=0]{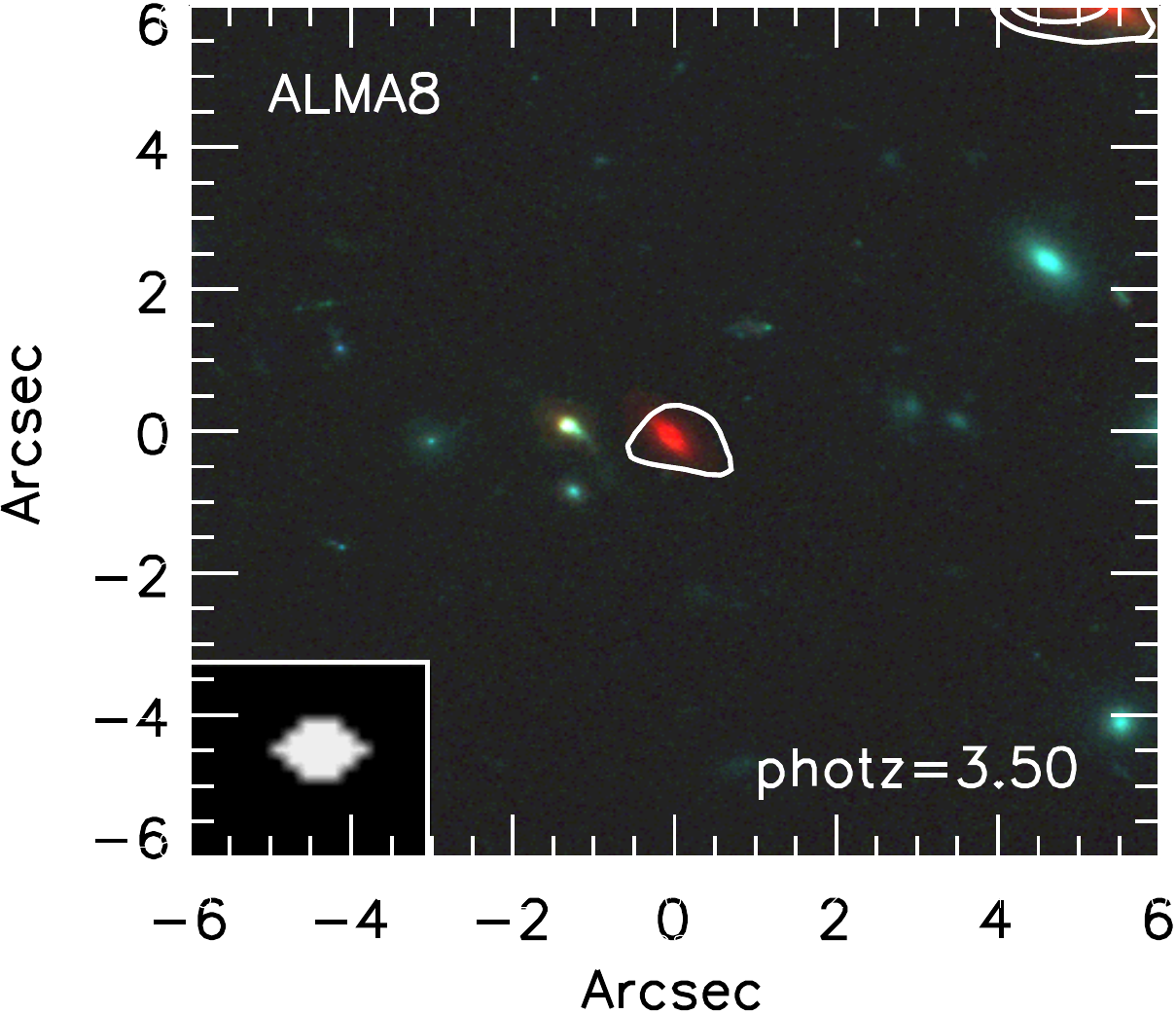}
\hspace{0.4cm}
\includegraphics[width=2.2in,angle=0]{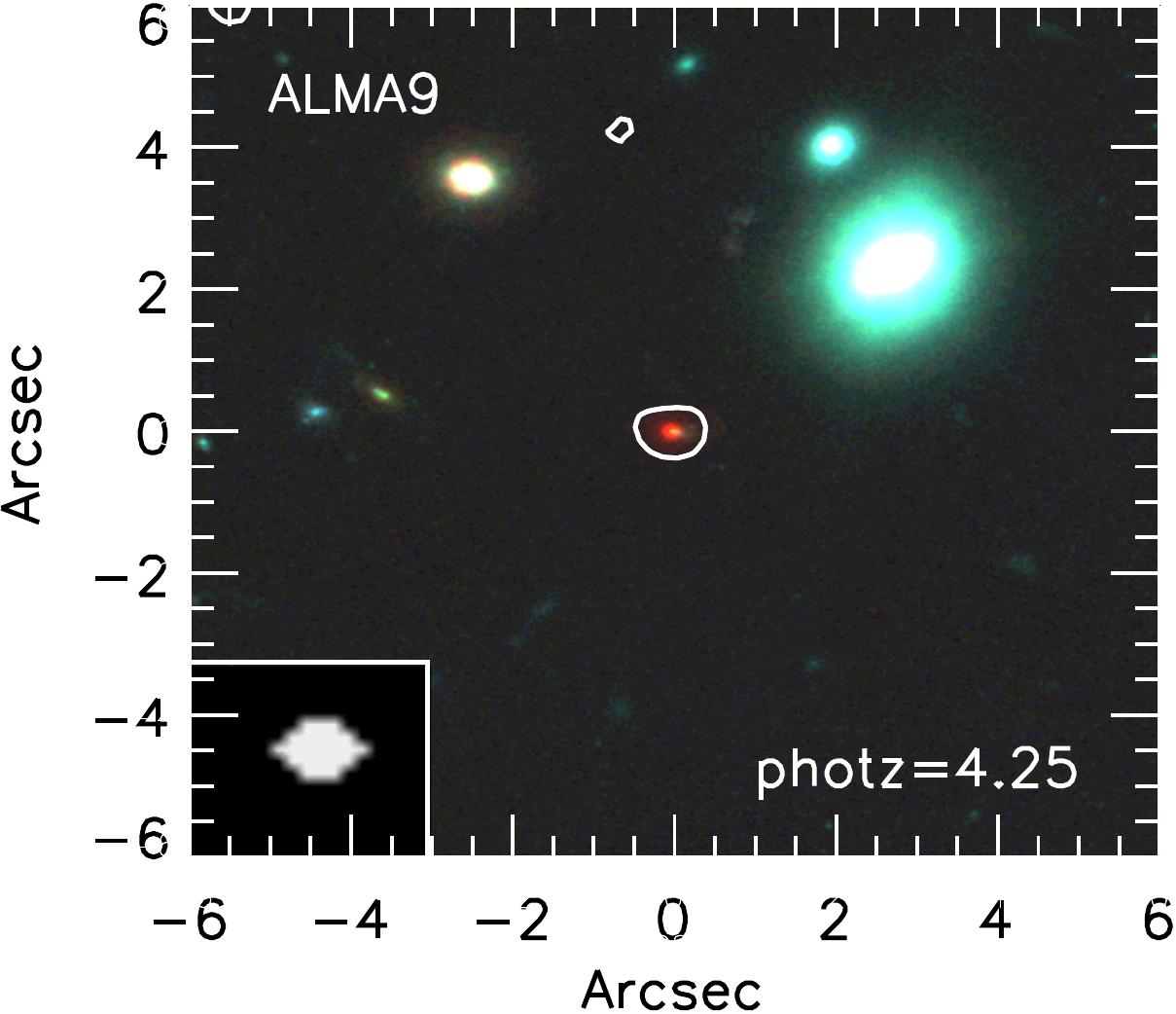}
\caption{Three-color JWST NIRCam images (blue = F115W, green = F150W, 
and red = F444W) for the 9 $>4.5\sigma$ ALMA sources in the A2744 field.
The thumbnails are $12''$ on a side, or $\sim100$~kpc at $z=2$.
The ALMA continuum emission is  shown with white contours. 
The redshifts (discussed in Section~\ref{secdata}) are marked as either spectroscopic
(specz) or photometric (photz).
\label{alma_images}}
\end{figure*}

%-----------------------------------------------------------------------------
%TABLE 1
%-----------------------------------------------------------------------------
\begin{deluxetable*}{ccccccccc}
\renewcommand\baselinestretch{1.0}
\tablewidth{0pt}
\tablecaption{SCUBA-2 and JWST NIRCam Fluxes of the 9 $>4.5\sigma$ ALMA Millimeter Sources  \label{tabALMA}}
\scriptsize
\tablehead{ALMA & \multicolumn{2}{c}{ALMA} & \multicolumn{2}{c}{SCUBA-2} & \multicolumn{3}{c}{JWST NIRCam} & Redshift \cr
No. &  R.A. &  Decl. &   \afluxb\   &  \afluxa\    &  \fluxa\  &  \fluxb\  &  Ratio  &  \\  &  \multicolumn{2}{c}{(J2000.0)} 
& \multicolumn{2}{c}{(mJy)}   &  \multicolumn{2}{c}{($\mu$Jy)}   & &   \\ (1)  &  (2)  &  (3)  &  (4)  &  (5)  &  (6)  &  (7)  &  (8)   & (9)}
\startdata
       1 &        3.5825000 &       -30.385473 & 1.82(0.27) & 3.07(2.93) & 7.21(0.019) & 0.73(0.011) & 9.84&2.99(2.92,3.09) \cr
       2 &        3.5764582 &       -30.413166 & 6.37(0.27) & 13.6(3.01) & 1.41(0.003) & 0.18(0.002) & 7.52&2.585 \cr
       3 &        3.5850000 &       -30.381777 & 3.27(0.28) & 9.30(2.97) & 20.2(0.047) & 4.40(0.023) & 4.60&3.058 \cr
       4 &        3.5732501 &       -30.383472 & 3.37(0.27) & 21.8(3.01) & 29.2(0.026) & 7.34(0.014) & 3.97&1.498 \cr
       5 &        3.5796666 &       -30.378389 & 1.67(0.29) & 5.38(3.04) & 19.5(0.039) & 1.87(0.015) & 10.4&2.409 \cr
       6 &        3.5720000 &       -30.382944 & 4.18(0.27) & 20.7(3.03) & 44.0(0.042) & 5.38(0.022) & 8.17&1.83(1.55,1.89) \cr
       7 &        3.5920832 &       -30.380472 & -0.3(0.29) & 4.88(3.02) & 8.20(0.013) & 0.22(0.008) & 36.4&2.644 \cr
       8 &        3.5812500 &       -30.380196 & 1.78(0.28) & -6.5(3.01) & 3.32(0.019) & 0.13(0.010) & 25.0&3.50(3.23,3.53) \cr
       9 &        3.5824583 &       -30.377167 & 0.49(0.30) & 8.30(3.07) & 2.14(0.019) & 0.35(0.008) & 5.96&4.25(4.11,4.41) \cr
\enddata
\tablecomments{The columns are (1) ALMA source number,
(2) and (3) ALMA R.A.  and decl., (4) and (5) SCUBA-2 \afluxb\ and \afluxa\ fluxes (uncertainties in parentheses)
measured at each ALMA position, (6), (7), and (8) F444W and F150W fluxes (uncertainties in parentheses)
and their ratio (these fluxes are from the \citealt{paris23} catalog), and (9) redshift (see Section~\ref{secdata};
spectroscopic has three digits after the decimal point,
while photometric has two digits after the decimal point, with the 16th and 84th percentiles of the posterior
given in parentheses).
}
\end{deluxetable*}
%-----------------------------------------------------------------------------

%-----------------------------------------------------------------------------
%FIGURE 3:  alma_color_plot and alma_color_color
%-----------------------------------------------------------------------------
\begin{figure*}
\includegraphics[width=3.4in]{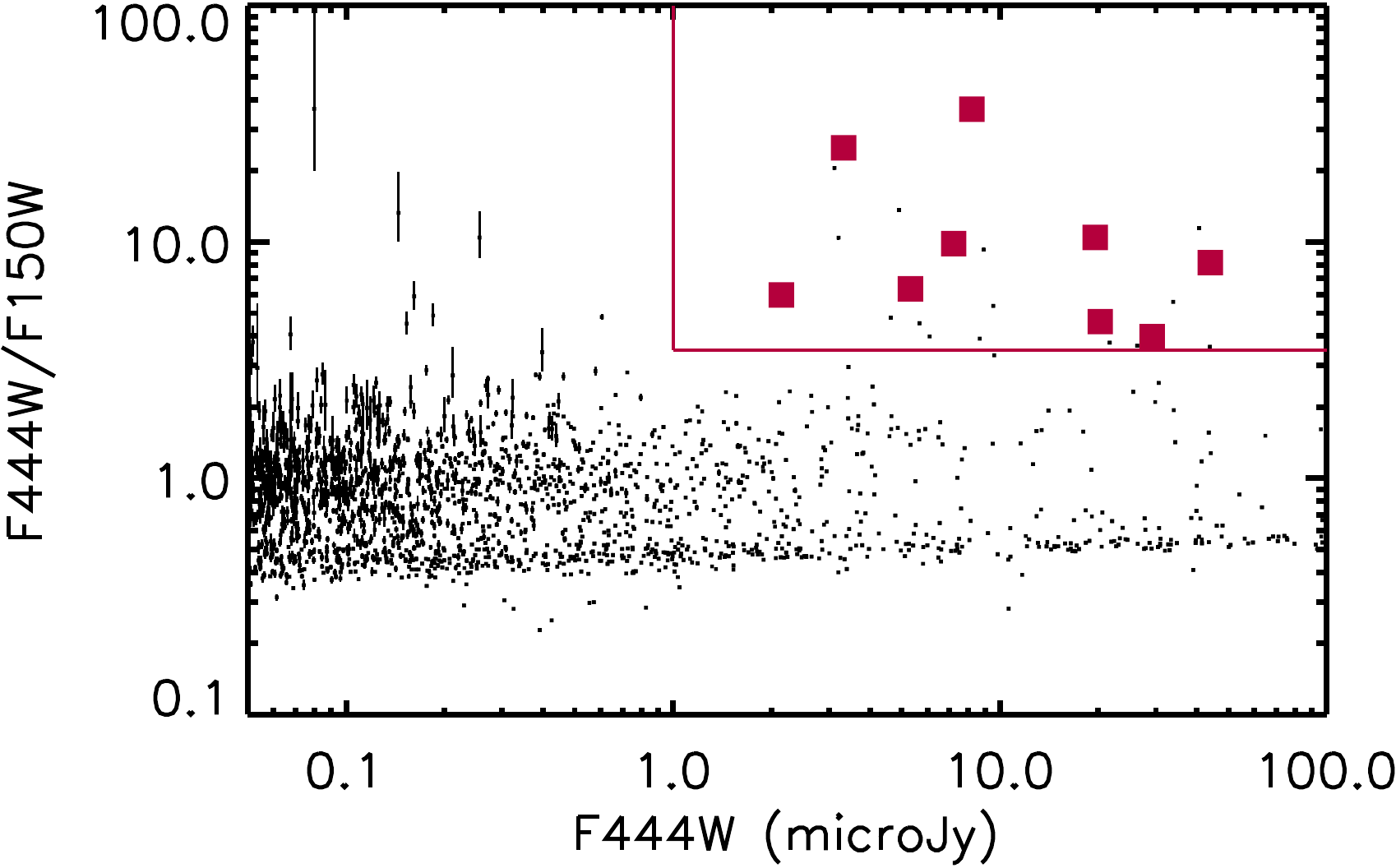}
\includegraphics[width=3.4in]{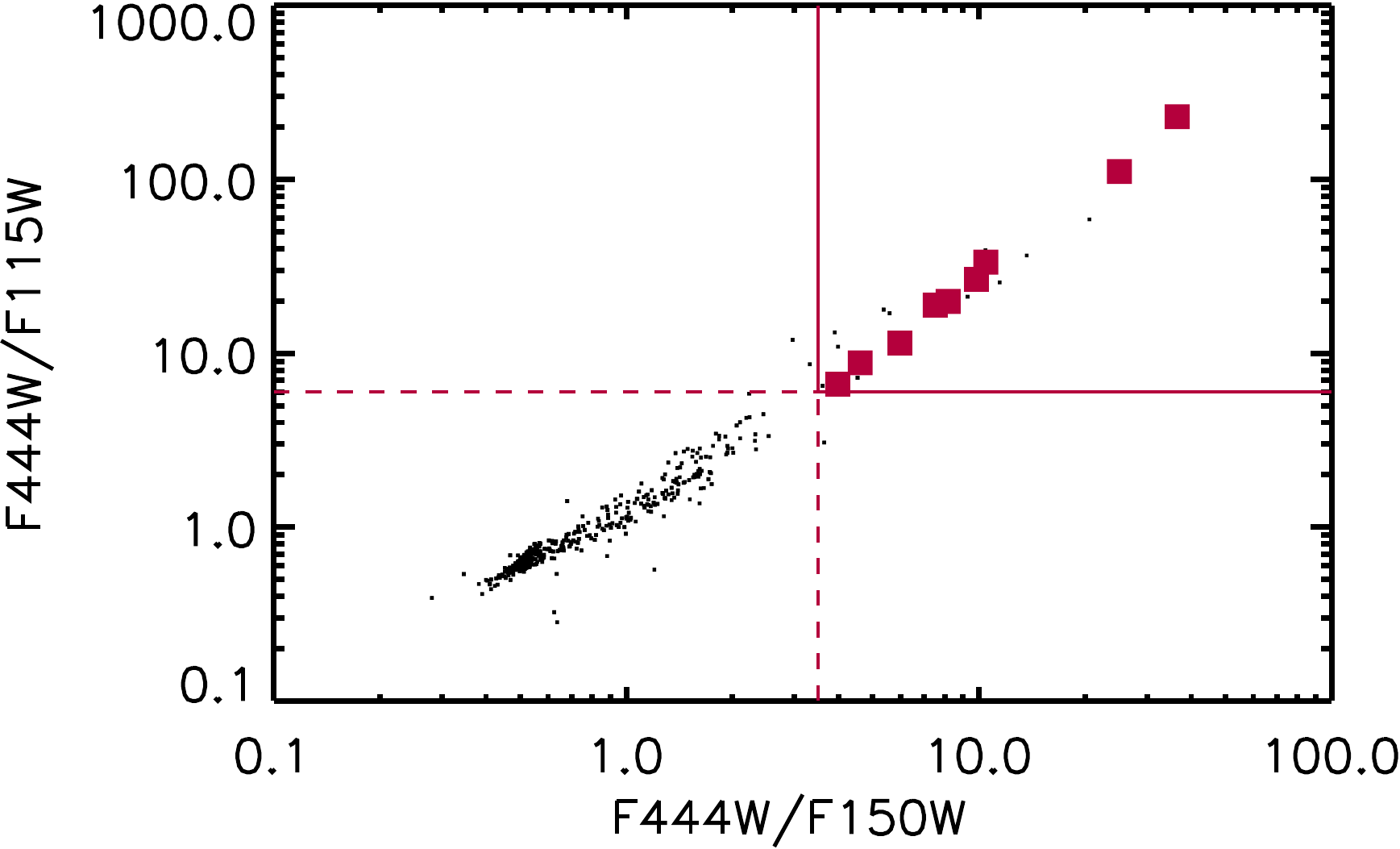}
\caption{
(Left) \jwratio\ vs. \fluxa\ 
for the JWST NIRCam sources that lie
in the area covered by the ALMA data (black dots). 
The red lines delineate our red galaxy selection region 
(\fluxa\ $>1~\mu$Jy and \jwratio\ $>3.5$).
All 9 $>4.5\sigma$ ALMA sources satisfy these criteria (red squares).
(Right) \jwratio\ vs. \jwratioc\ for sources
with \fluxa\ $>1~\mu$Jy that lie in the area covered by the ALMA data (black dots). 
The 9 $>4.5\sigma$ ALMA sources are again shown with red squares. The red vertical line
shows our \jwratio\ $>3.5$ selection, while the red horizontal line 
shows an alternate \jwratioc\ $>6$ selection.
The solid portions of the lines delineate a selection region that uses both criteria, but the improvement
is marginal.
\label{alma_color_plot}
}
\end{figure*}
%-----------------------------------------------------------------------------

%---------------------------------------------------------------------
% FIGURE 4:  new_scuba# (#: 1,3,4,5,6,7,9,11,12,15,16,18)
%---------------------------------------------------------------------
\begin{figure*}
\includegraphics[width=2.2in,angle=0]{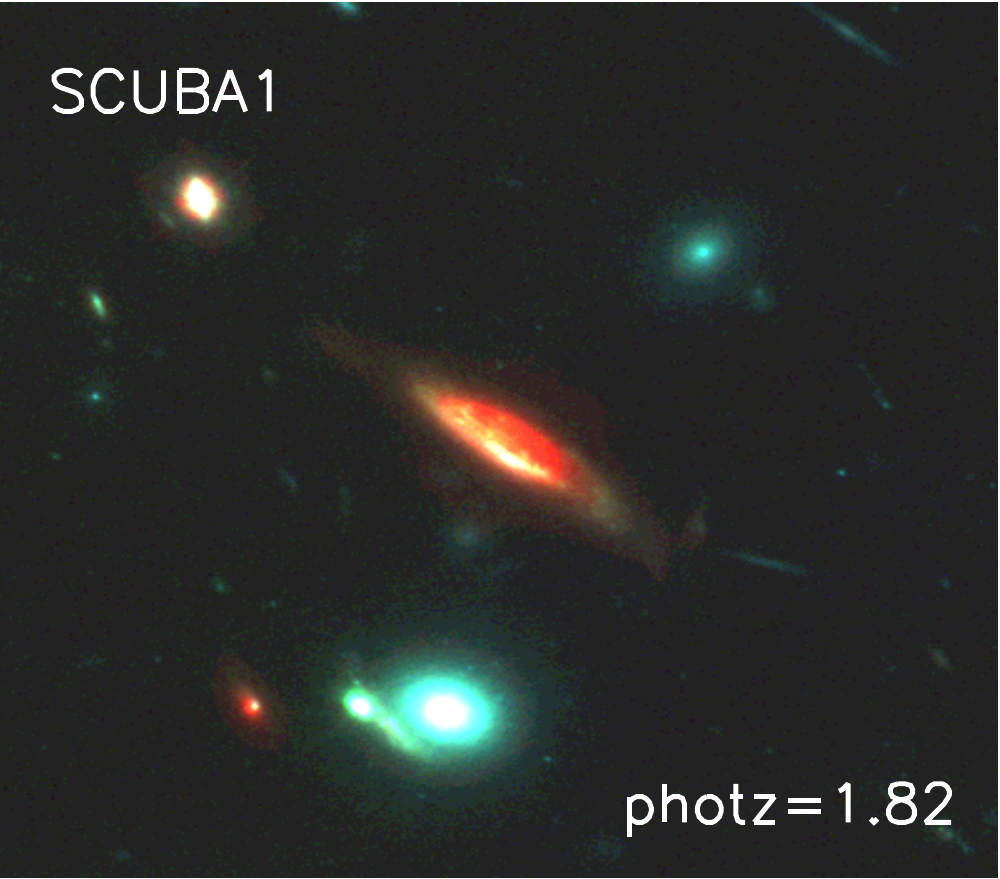}
\includegraphics[width=2.2in,angle=0]{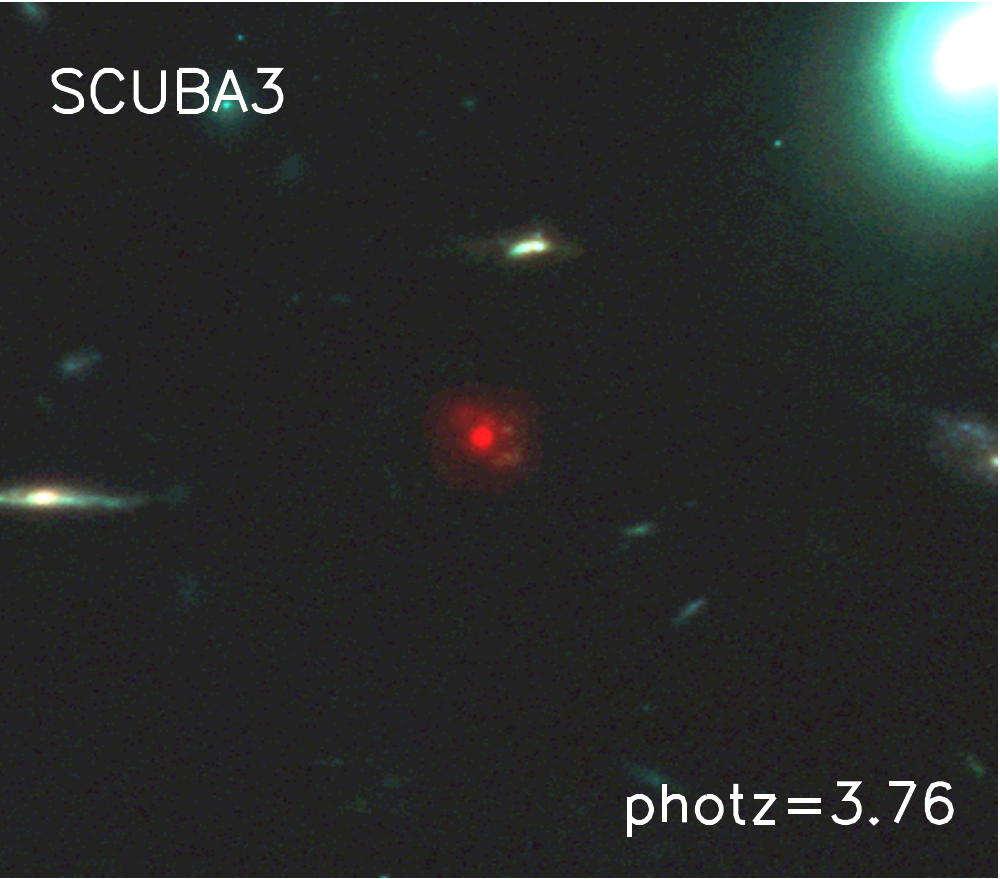}
\includegraphics[width=2.2in,angle=0]{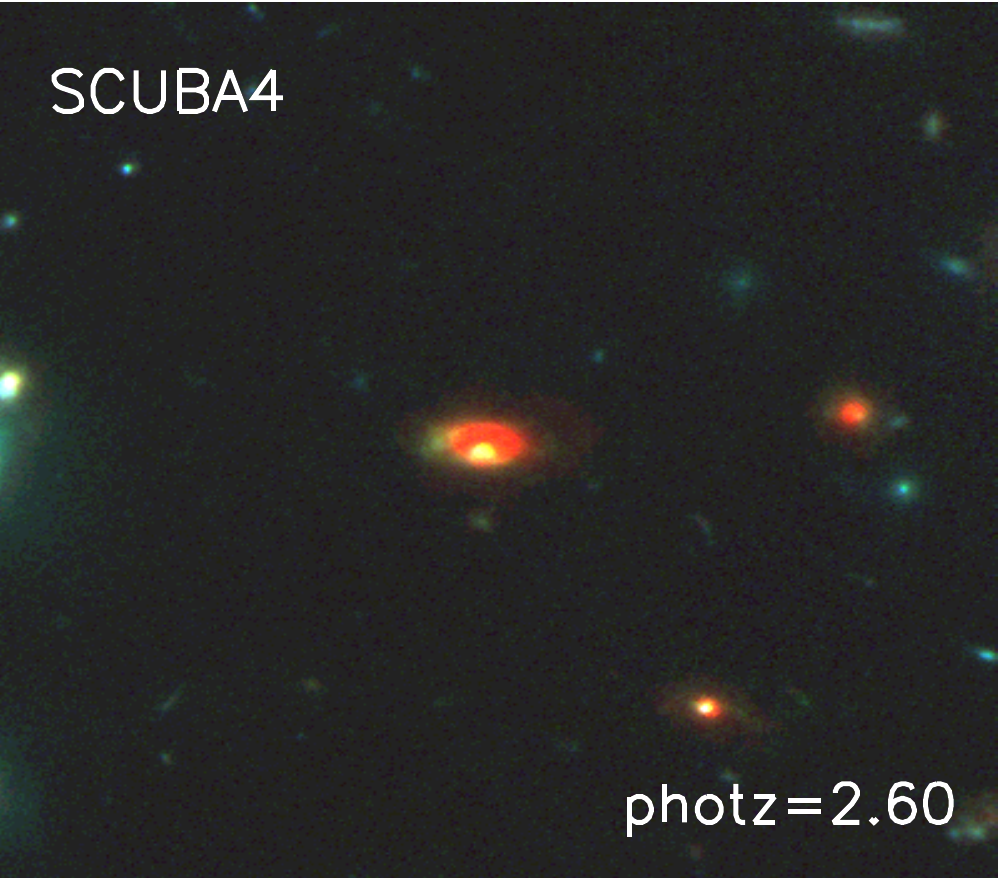}
\includegraphics[width=2.2in,angle=0]{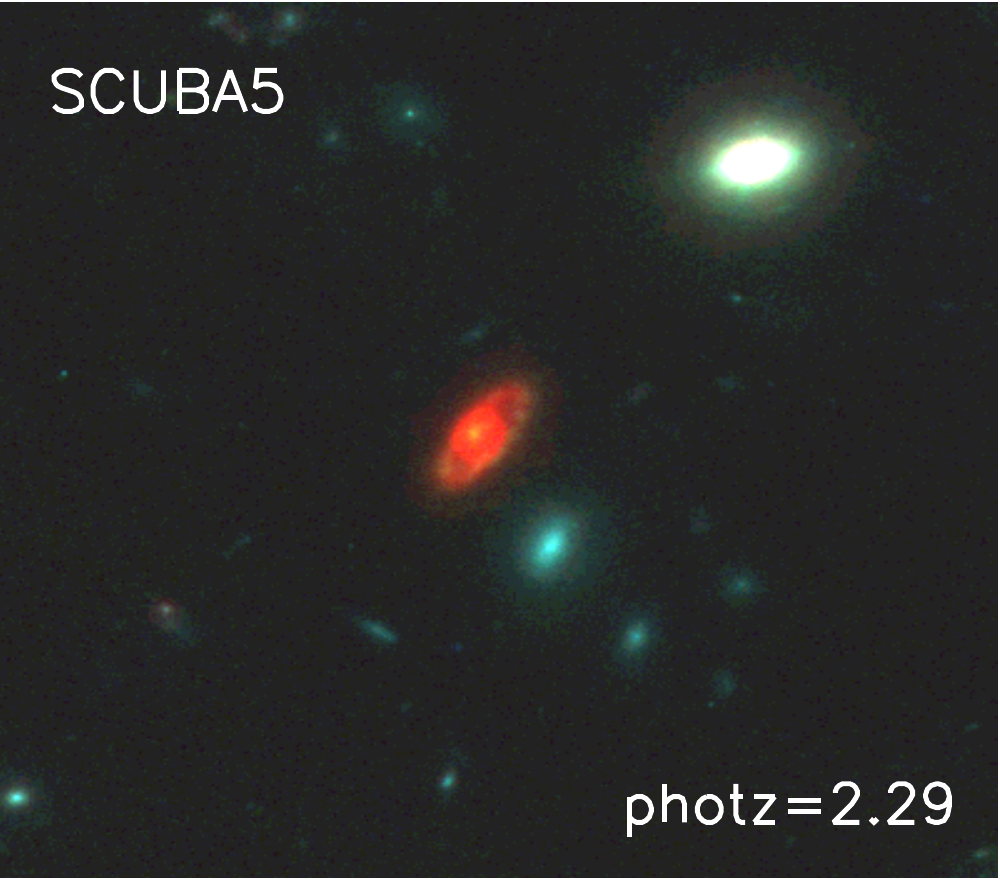}
\includegraphics[width=2.2in,angle=0]{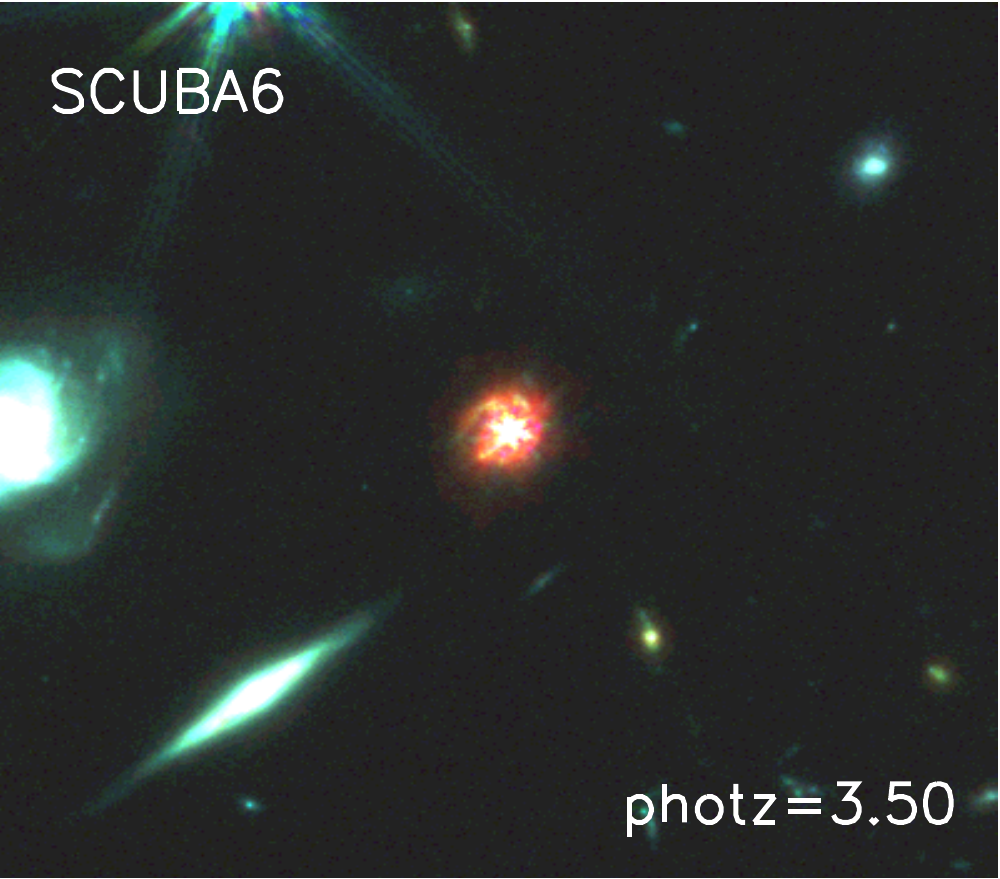}
\includegraphics[width=2.2in,angle=0]{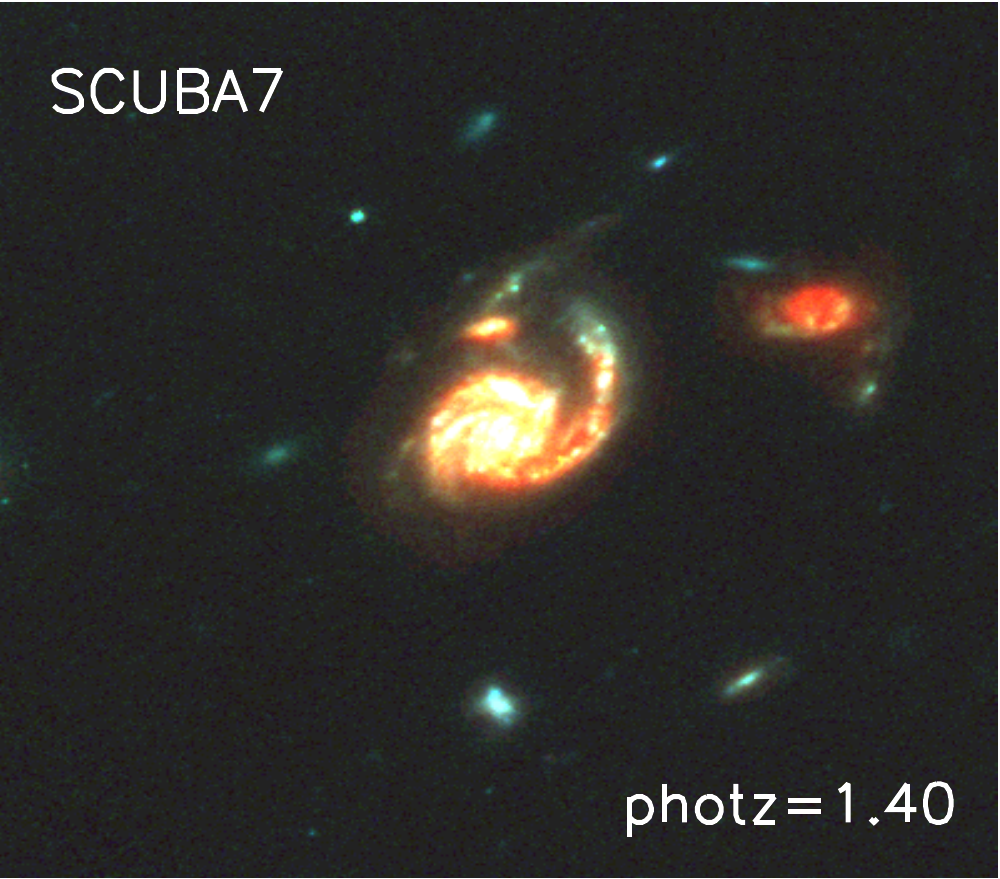}
\includegraphics[width=2.2in,angle=0]{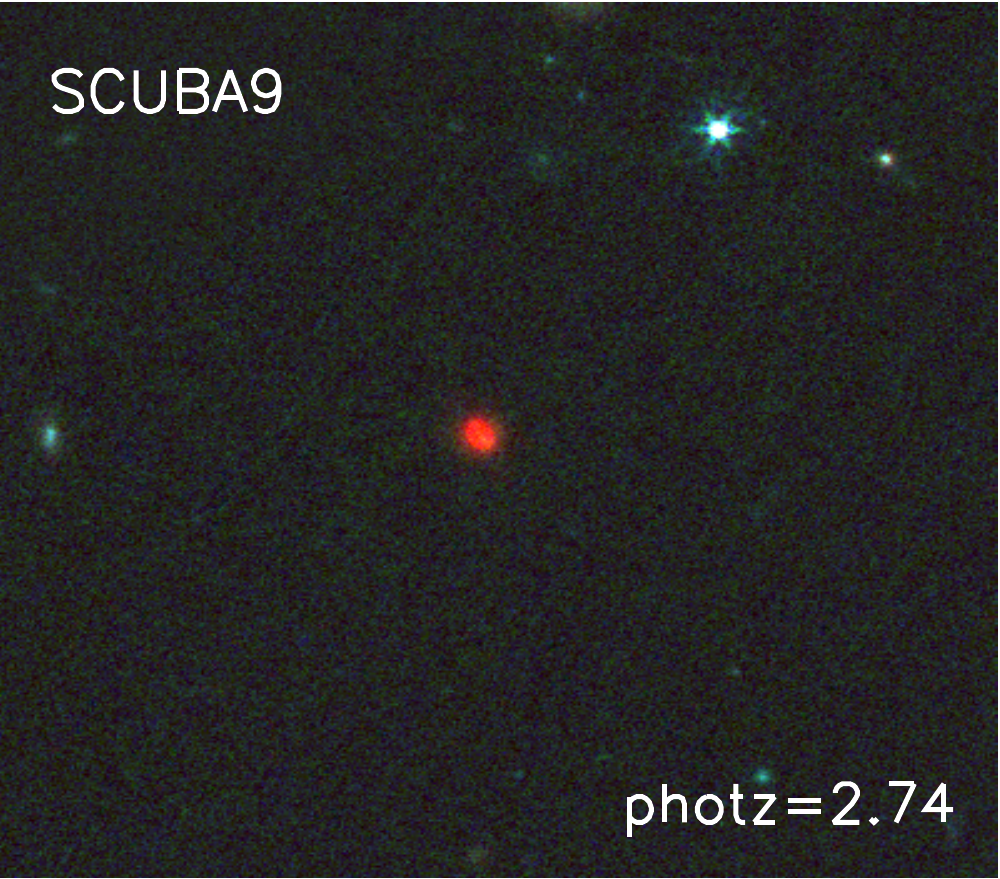}
\includegraphics[width=2.2in,angle=0]{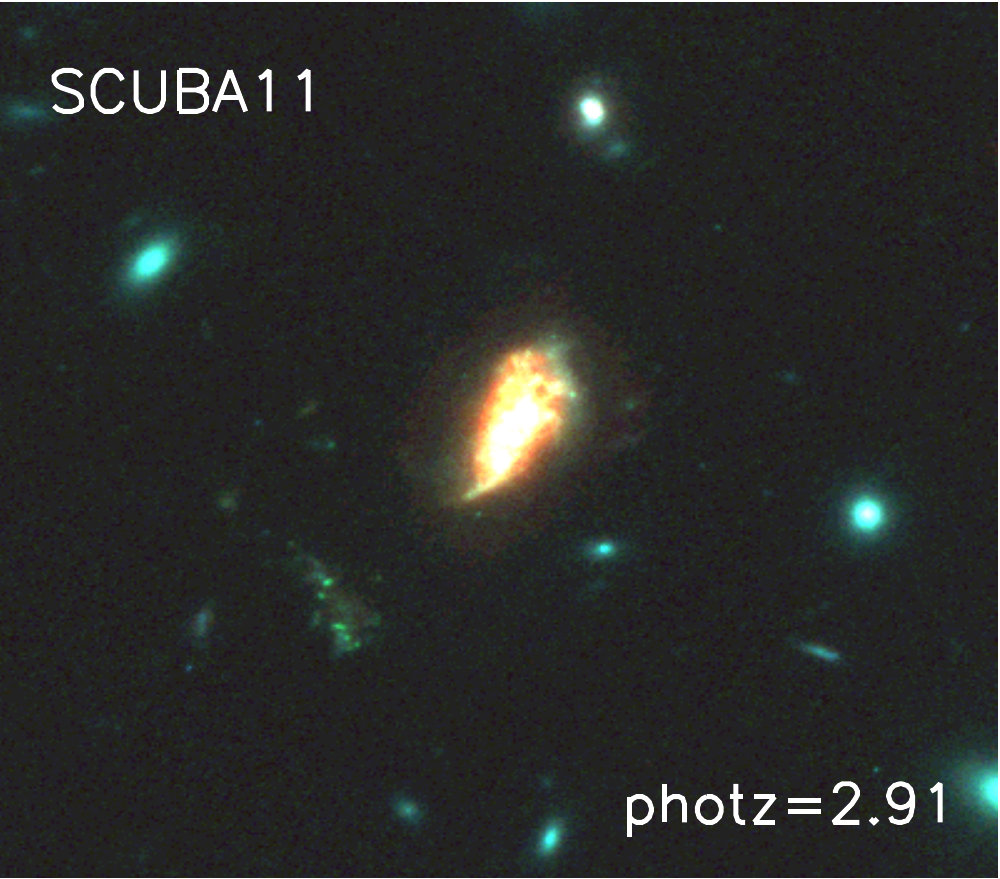}
\includegraphics[width=2.2in,angle=0]{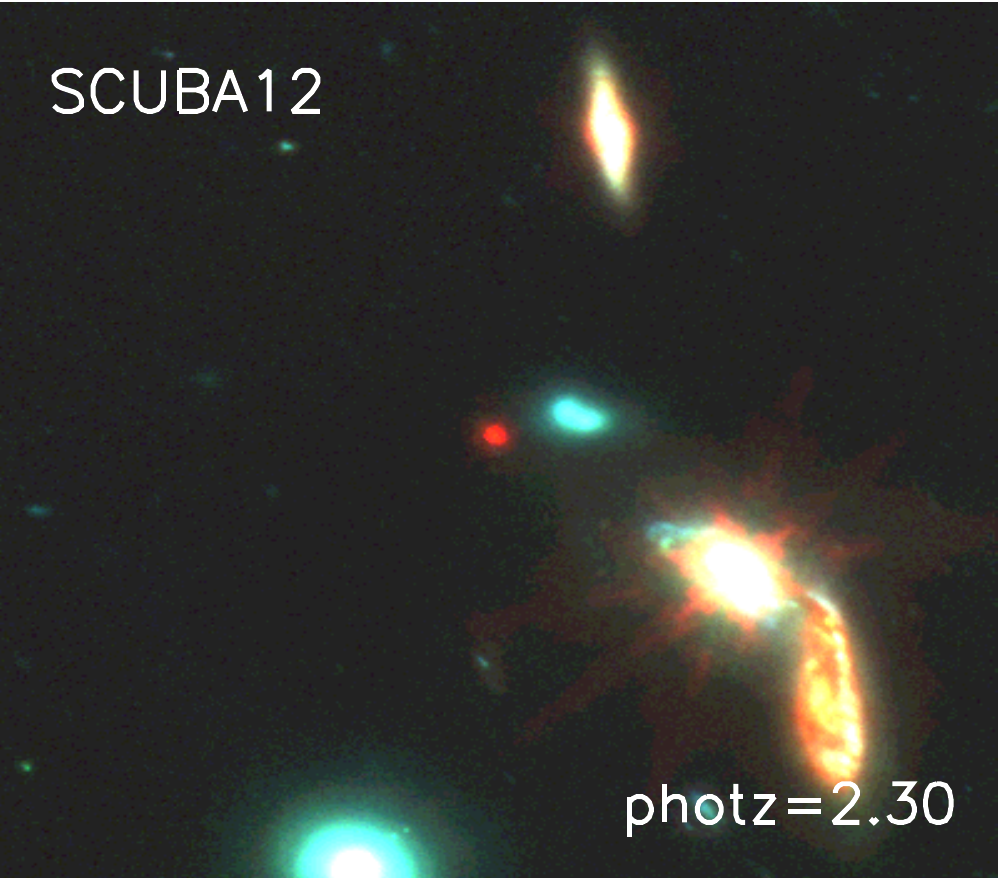}
\includegraphics[width=2.2in,angle=0]{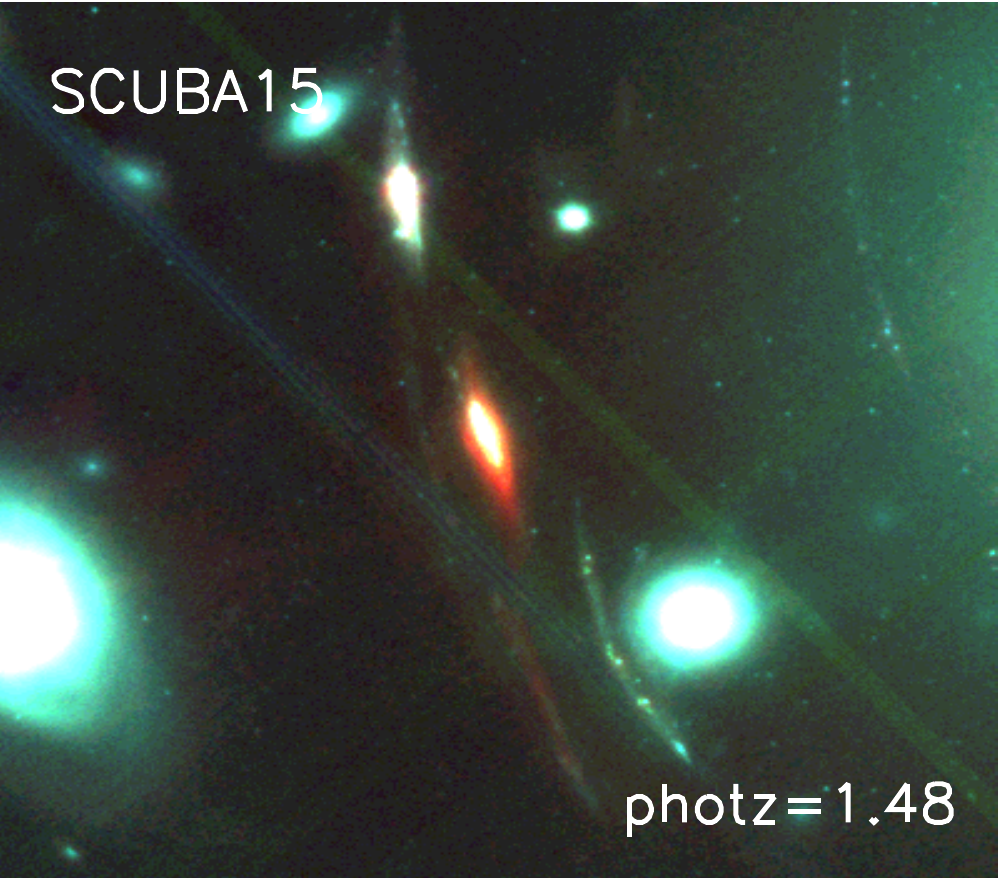}
\hspace{0.4cm}
\includegraphics[width=2.2in,angle=0]{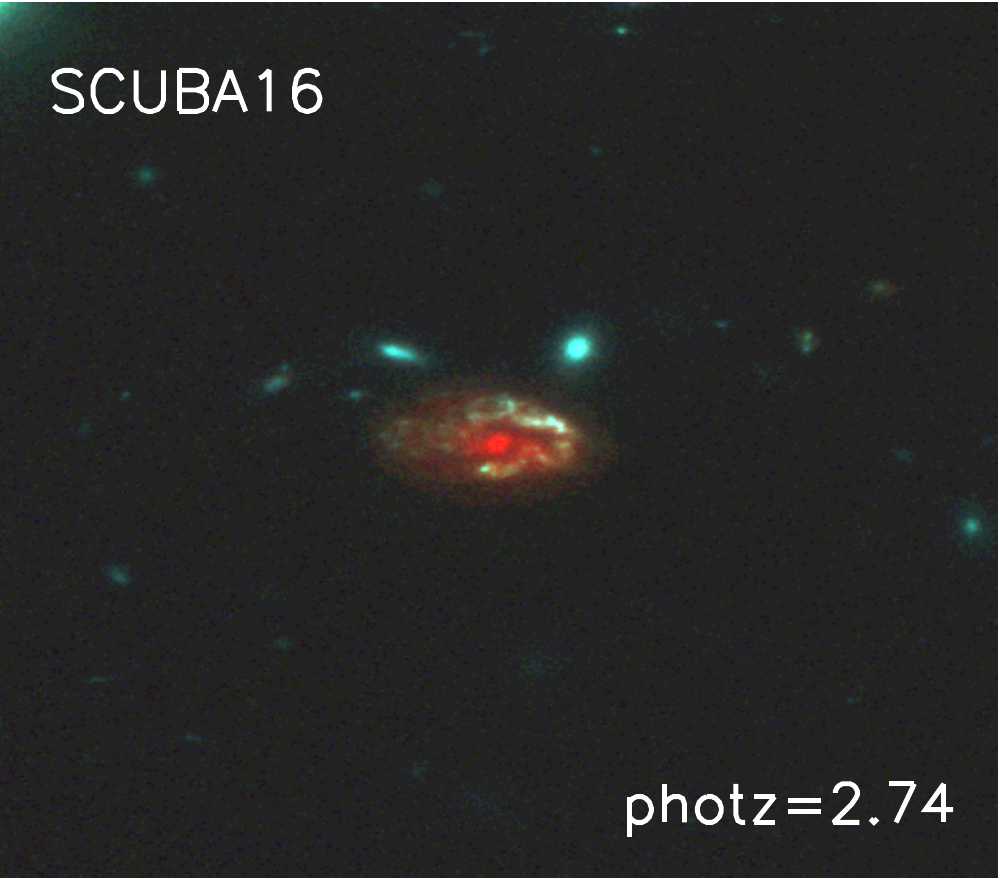}
\hspace{0.4cm}
\includegraphics[width=2.2in,angle=0]{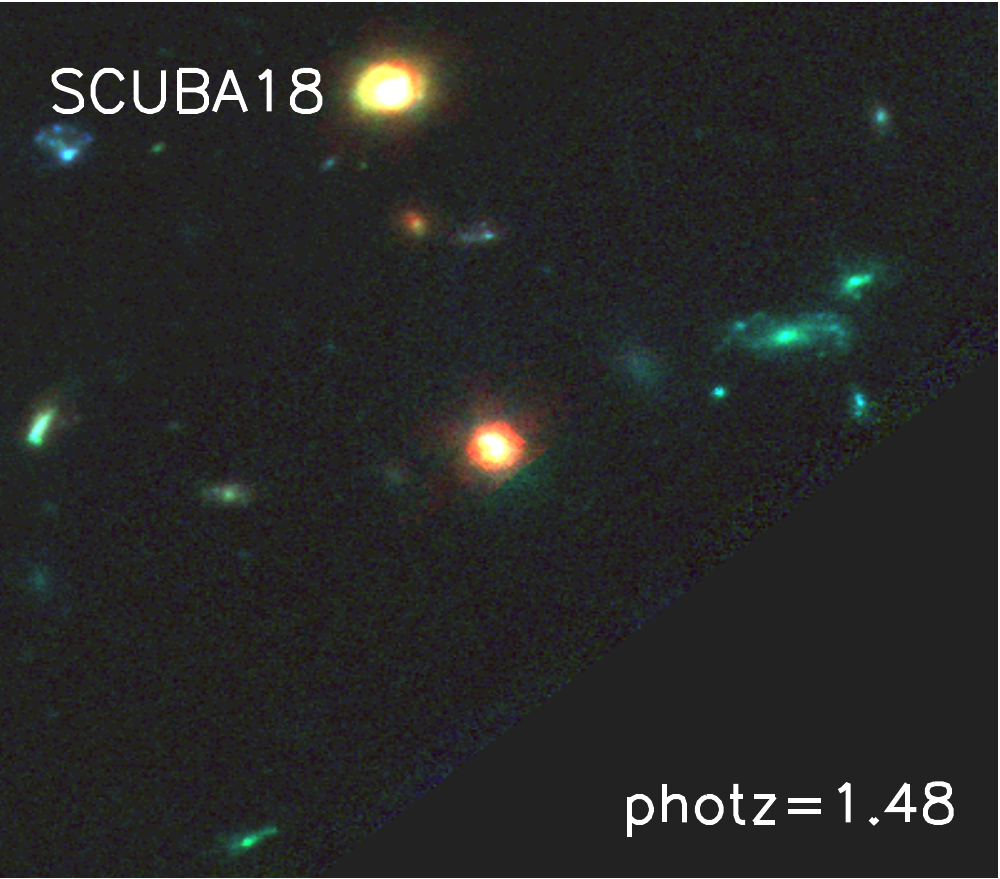}
\caption{Three-color JWST NIRCam images (blue = F115W, green = F150W, 
and red = F444W) for the $>5\sigma$ SCUBA-2 sources with accurate NIRCam positions. The
five SCUBA-2 sources with already known accurate positions from ALMA (see Figure~\ref{alma_images}
and Table~\ref{scuba2_acc}) are not shown. 
The thumbnails are $12''$ on a side, or roughly 100~kpc at $z=2$.
The thumbnails are labeled with the photometric redshifts from \citet{weaver23}.
\label{scuba2_images}
}
\end{figure*}
%---------------------------------------------------------------------

%-----------------------------------------------------------------------------
%TABLE 2
%-----------------------------------------------------------------------------
\begin{deluxetable*}{ccccrccccc}
\renewcommand\baselinestretch{1.0}
\tablewidth{0pt}
\tablecaption{A2744 $>5\sigma$ SCUBA-2 \afluxb\ Sources in the JWST NIRCam plus SCUBA-2 Footprint \label{scuba2_acc}}
\scriptsize
\tablehead{SCUBA-2 & \multicolumn{2}{c}{JWST NIRCam} & \multicolumn{2}{c}{SCUBA-2} & \multicolumn{3}{c}{JWST NIRCam} & ALMA & Redshift \\
No. &  R.A.  &  Decl. &   \afluxb\   &  \afluxa\   &  \fluxa\  &  \fluxb\  &  Ratio   &  No.  &   \\  
&  \multicolumn{2}{c}{(J2000.0)}  &  \multicolumn{2}{c}{(mJy)}  &  \multicolumn{2}{c}{(mJy)}  &  & &  \\ 
(1)  &  (2)  &  (3)  &  (4)  &  (5)  &  (6)  &  (7)  &  (8)  &  (9) & (10)}
\startdata
       1 &        3.5362918 &       -30.360361 & 7.63(0.38) & 32.1(4.01) & 43.5(0.02) & 7.06(0.02) & 6.16 & \nodata & 1.82(1.69,1.92) \cr
       2 &        3.5761251 &       -30.413166 & 6.36(0.27) & 13.9(3.01) & 3.11(0.00) & 0.15(0.00) & 20.5 & ALMA2 & 2.585 \cr
       3 &        3.5755000 &       -30.424389 & 6.18(0.29) & 8.72(3.35) & 5.00(0.02) & 0.14(0.01) & 34.0 & \nodata & 3.76(2.64,3.91) \cr
       4 &        3.5490835 &       -30.352222 & 5.29(0.41) & 15.7(3.93) & 16.9(0.02) & 2.23(0.01) & 7.61 & \nodata & 2.60(2.43,2.68) \cr
       5 &        3.5473332 &       -30.388306 & 5.12(0.30) & 16.9(3.64) & 19.8(0.02) & 1.81(0.01) & 10.9 & \nodata & 2.29(2.13,2.61) \cr
       6 &        3.6172917 &       -30.368832 & 4.66(0.38) & 7.82(3.58) & 30.3(0.02) & 8.35(0.01) & 3.62 & \nodata & 3.50(2.46,3.50) \cr
       7 &        3.5938752 &       -30.356638 & 4.66(0.42) & 19.7(4.01) & 92.2(0.04) & 22.5(0.02) & 4.09 & \nodata & 1.40(1.36,1.87) \cr
       8 &        3.5720000 &       -30.382973 & 4.18(0.27) & 20.7(3.03) & 44.0(0.04) & 5.38(0.02) & 8.17 & ALMA6 & 1.83(1.55,1.89) \cr
       9 &        3.6341667 &       -30.437721 & 4.27(0.37) & 21.3(4.02) & 4.50(0.03) & 0.28(0.01) & 15.5 & \nodata & \nodata$^{\star}$ \cr
      10 &        3.5849168 &       -30.381777 & 3.27(0.28) & 9.30(2.97) & 20.2(0.04) & 4.40(0.02) & 4.60 & ALMA3$^{\dagger}$ & 3.058 \cr
      11 &        3.5990834 &       -30.359779 & 2.68(0.40) & 6.54(3.81) & 60.8(0.04) & 15.7(0.02) & 3.86 & \nodata & 1.26(1.22,1.35) \cr
      12 &        3.5404584 &       -30.359249 & 2.45(0.38) & 10.3(3.95) & 2.93(0.01) & 0.52(0.00) & 5.64 & \nodata & 2.91(2.13,3.09) \cr
      13 &        3.5332916 &       -30.358639 & 3.54(0.38) & -11.0(4.07) & \nodata &  \nodata &  \nodata & \nodata & \nodata \cr
      14 &        3.5796666 &       -30.378389 & 1.67(0.29) & 5.38(3.04) & 19.5(0.03) & 1.87(0.01) & 10.4 & ALMA5 & 2.409 \cr
      15 &        3.5579166 &       -30.377140 & 2.46(0.31) & 6.54(3.40) & 28.3(0.02) & 9.34(0.02) & 3.03 & \nodata & 2.62(2.56,2.65) \cr
      16 &        3.5863333 &       -30.425028 & 1.54(0.31) & 5.97(3.44) & 9.83(0.03) & 1.93(0.01) & 5.08 & \nodata & 3.62(3.52,3.67) \cr
      17 &        3.5809166 &       -30.386139 & 1.10(0.27) & 1.36(2.92) & \nodata &  \nodata &  \nodata & \nodata & \nodata \cr
      18 &        3.5600834 &       -30.417528 & 2.07(0.28) & 6.02(3.27) & 21.8(0.04) & 3.43(0.01) & 6.35 & \nodata & 1.48(1.47,1.63) \cr
      19 &        3.5850000 &       -30.381748 & 3.27(0.28) & 9.30(2.97) & 20.2(0.04) & 4.40(0.02) & 4.60 & ALMA3$^{\dagger}$ & 3.058 \cr
\enddata
\tablecomments{The columns are (1) SCUBA-2 source number,
(2) and (3) R.A.  and decl.\ of the JWST NIRCam counterpart, if there is one, or the SCUBA-2 \afluxb\ position
otherwise, (4) and (5) SCUBA-2 \afluxb\ and \afluxa\ fluxes (uncertainties in parentheses)
measured at each SCUBA-2 position, (6), (7), and (8) F444W and F150W fluxes (uncertainties in parentheses)
and their ratio for the sources with NIRCam counterparts (these fluxes are from the \citealt{paris23} catalog),
(9) ALMA source match from Table~\ref{tabALMA}, if there is one, and (10) redshift
(see Section~\ref{secdata};
spectroscopic has three digits after the decimal point,
while photometric has two digits after the decimal point, with the 16th and 84th percentiles of the posterior
given in parentheses).
$^{\star}$This source lies off the photometric redshift catalog of \cite{weaver23}.
$^{\dagger}$Two SCUBA-2 sources are matched to the same ALMA counterpart 
(see Figure~\ref{scuba2compare}).
}
\end{deluxetable*}

%---------------------------------------------------------------------
\section{JWST NIRCam Properties of ALMA Sources}
\label{secalma}
%---------------------------------------------------------------------
We used the ALMA positions to find the counterparts
in the JWST NIRCam images of \citet{paris23}.
As we show with thumbnails in Figure~\ref{alma_images}, all 9 $>4.5\sigma$ ALMA sources
have bright red NIRCam counterparts, which closely match
in position, and in some cases shape, to the ALMA images (white contours).
In Table~\ref{tabALMA}, we summarize the \afluxb\ and \afluxa\ fluxes and uncertainties
measured from the SCUBA-2 images at the ALMA positions, along with the
NIRCam F444W and F150W (4.44~$\mu$m and 1.5~$\mu$m) fluxes and uncertainties
and their ratio.
We also give the spectroscopic redshifts, where available, and otherwise the 
photometric redshifts and uncertainties.

In Figure~\ref{alma_color_plot} (left), we plot \jwratio\ for the 9 $>4.5\sigma$ ALMA 
sources (red squares) and for the full JWST NIRCam sample in the area covered by
the combined ALMA mosaics (black dots). Throughout, we show the data above 
\fluxa\ $= 0.05~\mu$Jy. Down to this level, \jwratio\ is well defined,
as can be seen from Figure~\ref{alma_color_plot} (left).
All 9 ALMA sources lie in the upper right corner defined by
\fluxa\ $>1~\mu$Jy and \jwratio\ $>3.5$ (red lines). 
In what follows, we will use this flux and color selection to
identify our sample of {\em red galaxies}.

Some of the other sources lying in the red galaxy region
are detected at lower significance in the ALMA data.
After combining one galaxy (second thumbnail in top row of Figure~\ref{alma_images})
that is divided into 4 parts in the \citet{paris23} catalog,
there are 16 JWST NIRCam sources that satisfy our red galaxy selection criteria.
Of these, 10 have $>3\sigma$ detections in the
AFFS mosaic, which has a minimum 1.1~mm rms of 0.055~mJy.
One of the 9 $>4.5\sigma$ ALMA sources is not included in this number, since
it was detected in longer follow-up ALMA observations 
(ALMA program \#2017.1.01219.S; PI: F.~Bauer).
Thus, in total, we have 11 ALMA $>3\sigma$ sources, 
or 69\% of our red galaxy sample.

In determining our red galaxy selection criteria, 
we used \jwratio\ $>3.5$, which matches most 
closely to previous definitions of dark galaxies. 
(As we discussed in the Introduction, these previous works
often used the Spitzer 4.5~$\mu$m to HST 1.6~$\mu$m flux ratio.)
However, there may be other flux ratios we should consider.

The use of \fluxa\ as our long-wavelength anchor is clear, since it is the
reddest JWST NIRCam band, but as
we illustrate in Figure~\ref{alma_color_plot} (right),
another shorter wavelength band, such as F115W, could replace F150W.
In this case, \jwratioc $>6$ (horizontal line) contains all 
9 $>4.5\sigma$ ALMA sources. There is only a
slightly higher contamination level for \jwratioc\ $>6$ (i.e., non-ALMA sources
above the horizontal line) than for \jwratio\ $>3.5$
(i.e., non-ALMA sources to the right of the vertical line), 
but both selections are comparably effective.
There are 22 sources found by both selections. There are
2 additional sources that satisfy the \jwratioc\ selection
but not the \jwratio\ selection, and 1 additional source that satisfies
the \jwratio\ selection but not the \jwratioc\ selection. 
The figure also emphasizes that using multiple colors would only
marginally improve the selection.

Use of the F150W band also avoids any contamination by
$z\sim 10$ galaxies. \citet{castellano23} report the detection
of seven such galaxies in the A2744 JWST NIRCam field. While these
galaxies are fainter than our flux selection threshold in F444W with fluxes
in the 0.03 to 0.35~$\mu$Jy range, they
satisfy, by construction, our red color threshold in \jwratioc\ but not
in \jwratio. That is, they are flat at longer wavelengths
and break at F115W.

%---------------------------------------------------------------------
\section{JWST NIRCam Counterparts to SCUBA-2 Sources}
\label{secscuba2cat}
%---------------------------------------------------------------------
We next aim to see whether we can obtain accurate positions for low-resolution
single-dish submillimeter/millimeter sources by finding their red galaxy counterparts.
We use the SCUBA-2 imaging of A2744, which has been slightly deepened over
that presented in \citet{cowie22}.
The reduction, extraction,
and cataloging follow that of \citet{cowie22}, providing \afluxb\ and \afluxa\ imaging
with central rms noise of 0.26 and 2.8~mJy, respectively. The noise quoted here
is the white noise; we add a confusion noise of 0.33~mJy \citep{cowie17}
in quadrature when selecting sources from the \afluxb\ image. 

\begin{figure}
\center{
\includegraphics[width=2.5in]{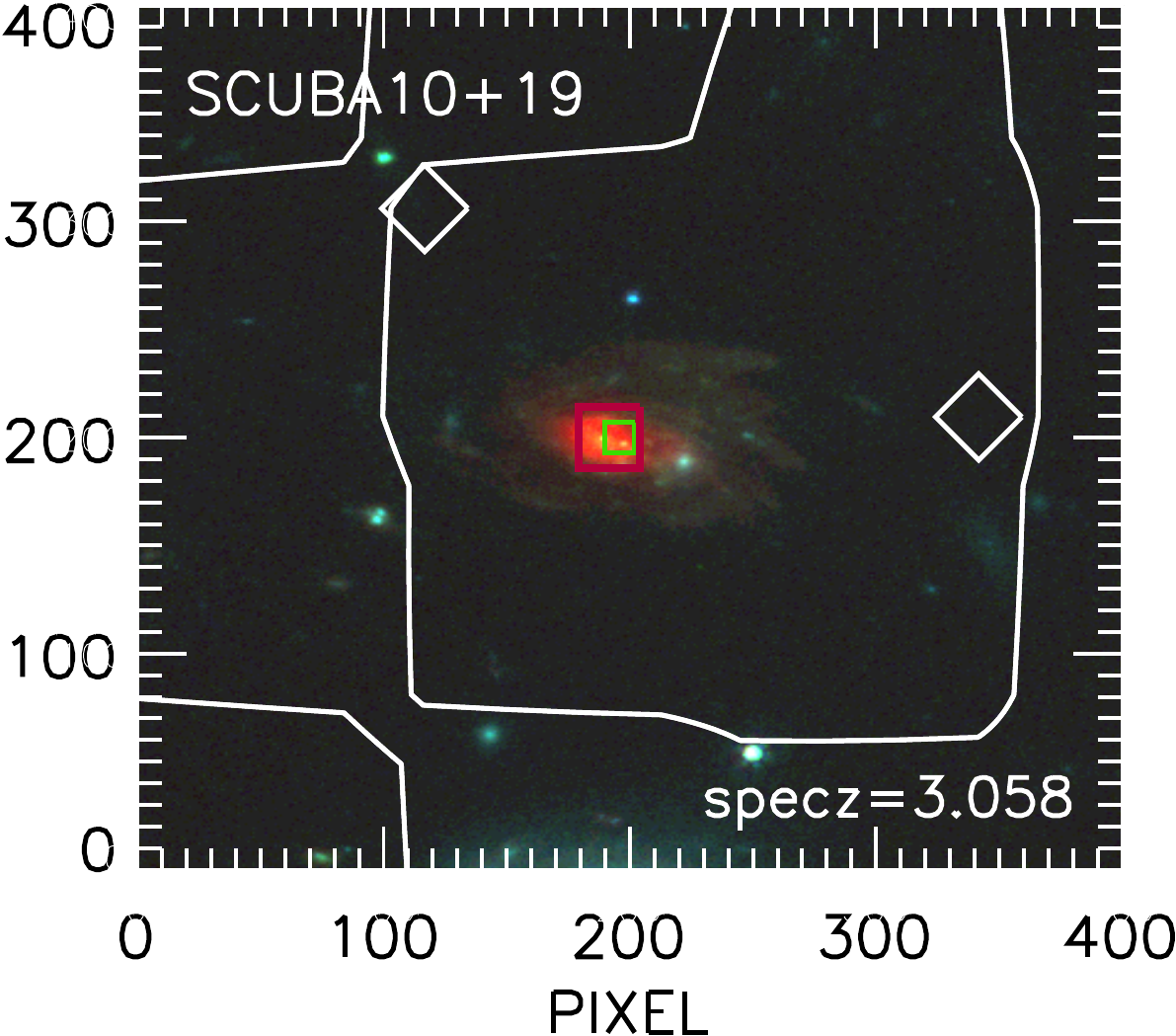}
\includegraphics[width=2.5in]{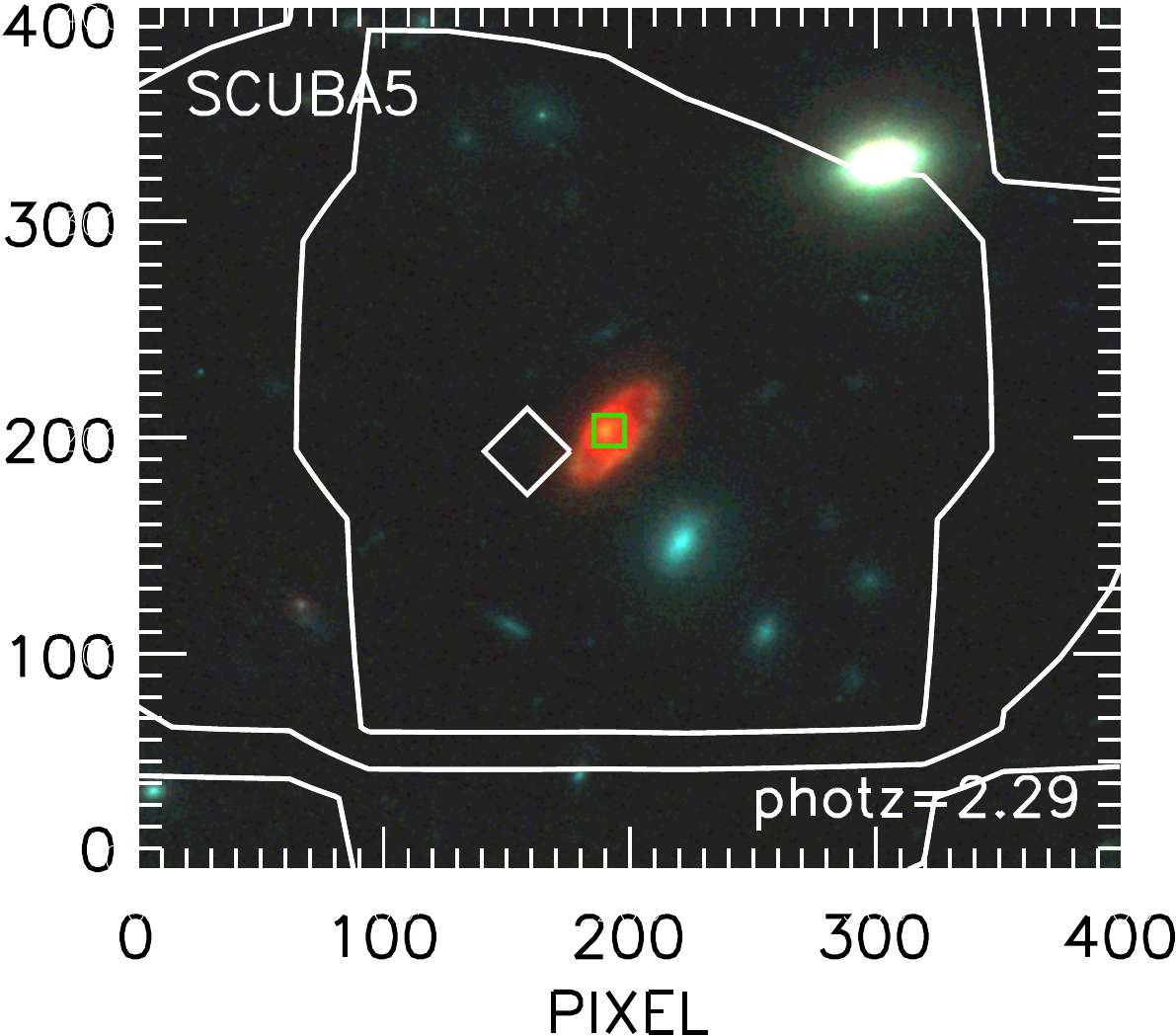}
}
\caption{
Two examples of SCUBA-2 \afluxb\ sources
(white contours: 1, 2, and 4~mJy per beam; diamonds: local peaks) 
and their assigned counterparts.
(Top) A red galaxy with an ALMA
counterpart (ALMA3; large red square) lying between two SCUBA-2 sources that are well
separated; this source has been assigned as the counterpart to both SCUBA-2 sources.
(Bottom) A red galaxy with no ALMA counterpart.
The underlying images are three-color JWST NIRCam (blue = F115W,
green = F150W, and red = F444W), centered on the assigned 
red galaxy counterpart (green squares).
The redshifts are marked as either spectroscopic
(specz) or photometric (photz).
The pixels are $0\farcs03$, and the fields are $12''$ on a side. 
\label{scuba2compare}
}
\end{figure}

As can be seen from Figure~\ref{footprint},
the JWST NIRCam observations are mostly well positioned on the SCUBA-2
image, and here we focus on the SCUBA-2 area also observed by NIRCam.
In this area, there are 19 \afluxb\ sources ($>5\sigma$)
stretching down to an \afluxb\ flux of 2.4~mJy.
For each \afluxb\ source, we determine the nearest
red galaxy.
We find that 17 of the 19 \afluxb\ sources have such
counterparts within a $4''$ match radius, the rough
uncertainty in the \afluxb\ position. We show these counterparts
in Figure~\ref{scuba2_images}, omitting the five that
are ALMA sources and hence already shown in Figure~\ref{alma_images}.  
We note that two of the SCUBA-2 sources
(SCUBA10 and SCUBA19) 
match to the same red galaxy and ALMA source (ALMA3).
We show this anomalous situation in Figure~\ref{scuba2compare}, alongside a
more typical match.
We list all 19 SCUBA-2 sources in Table~\ref{scuba2_acc}.

The full NIRCam area covers 171,900~arcsec$^2$, and there are 202 sources
satisfying our selection criteria, giving a surface density of 0.00118~arcsec$^{-2}$.
This corresponds to a probability of 0.057 of seeing such a source  in a $4''$ radius circle.
Thus, we expect one false positive in our sample of 19 sources.
Measurements of random positions in the field
give a similar contamination rate. Consequently, nearly
all 17 red galaxy counterparts in Table~\ref{scuba2_acc} are real.

\begin{figure}
\includegraphics[width=3.25in]{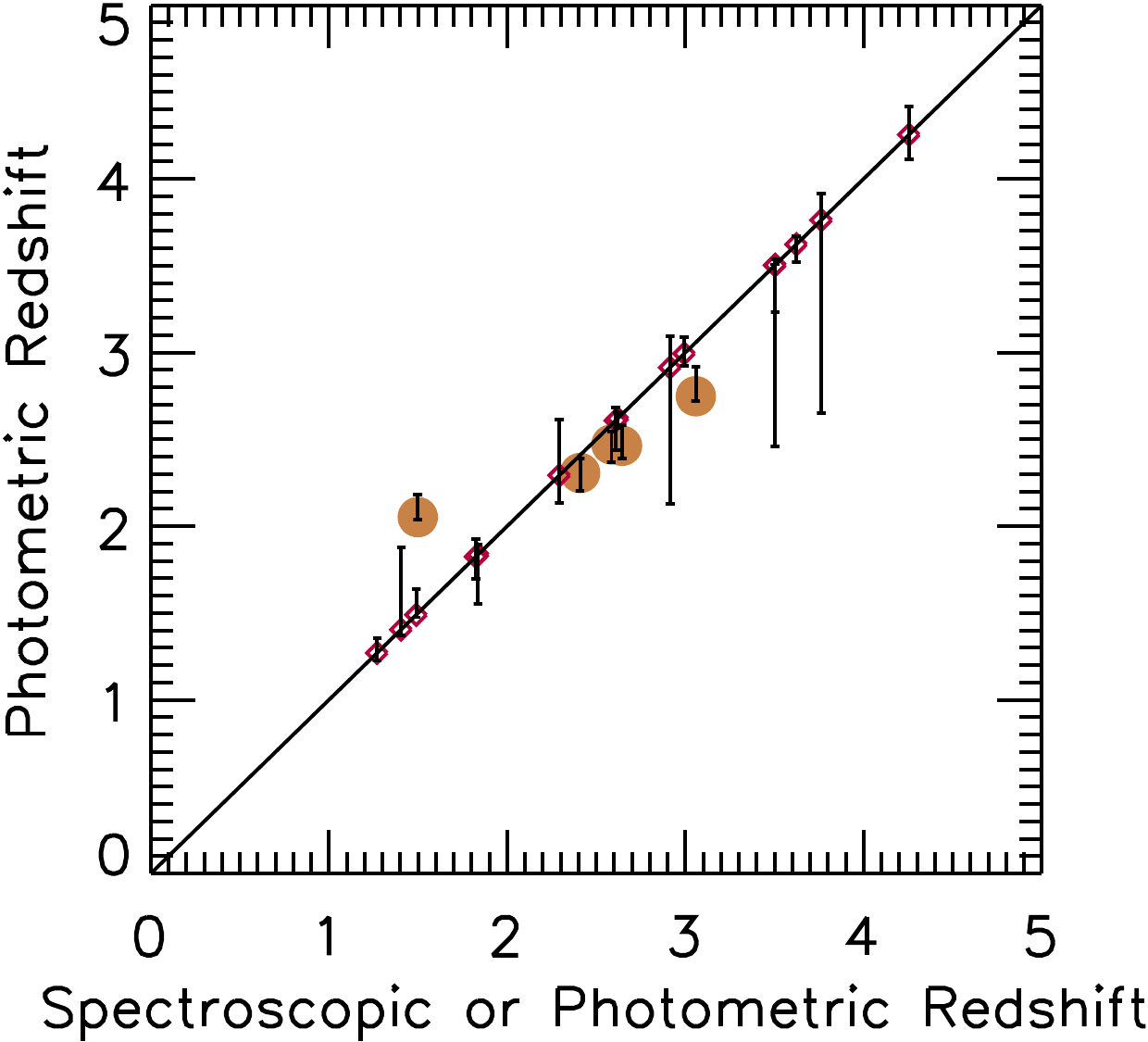}
\caption{
Photometric redshift with uncertainties (16th 
and 84th percentiles of the posterior) vs. either spectroscopic redshift
(5 sources; orange circles) or photometric redshift
(15 sources; red diamonds).
\label{redshift}
}
\end{figure}

In combination,
Figures~\ref{alma_images} and \ref{scuba2_images}
give 21 directly detected submillimeter/millimeter sources with accurate positions
in the field. Twenty of these have either spectroscopic or photometric redshifts
(the remaining source, SCUBA9, lies outside the \citealt{weaver23} 
catalog). In Figure~\ref{redshift},
we show for these sources the photometric redshifts and uncertainties versus the
spectroscopic redshifts, where available, or the photometric redshifts otherwise.
The photometric redshifts do a reasonable job of estimating the spectroscopic redshifts
for those sources that have both.

%-----------------------------------------------------------------------------
% FIGURE 5: SCUBA10 and SCUBA5
%-----------------------------------------------------------------------------
%-----------------------------------------------------------------------------

%-----------------------------------------------------------------------------
% FIGURE 6:  compare_specz 
%-----------------------------------------------------------------------------
%-----------------------------------------------------------------------------

\newpage
%---------------------------------------------------------------------
\section{JWST NIRCam Selection of SCUBA-2 Sources}
\label{secDSFG}
%---------------------------------------------------------------------
We can now invert the procedure of the
previous two sections and use the color selected JWST NIRCam sources
as priors to probe deeper in the SCUBA-2 image and to avoid the effects
of confusion present in a direct search.

We restrict to the portion of
the SCUBA-2 image where the \afluxb\ rms white noise
is $<0.5$~mJy (twice the central noise) and where the 
area is covered by the NIRCam footprint (see Figure~\ref{footprint}). 
There are 11,200 NIRCam sources in this region
with \fluxa\ $>0.05$~$\mu$Jy.
Of these, 156 have \fluxa\ $>1~\mu$Jy and 
\jwratio\ $>3.5$, satisfying our red galaxy selection criteria.

\begin{figure}
\hspace{-0.5cm}\includegraphics[width=3.5in]{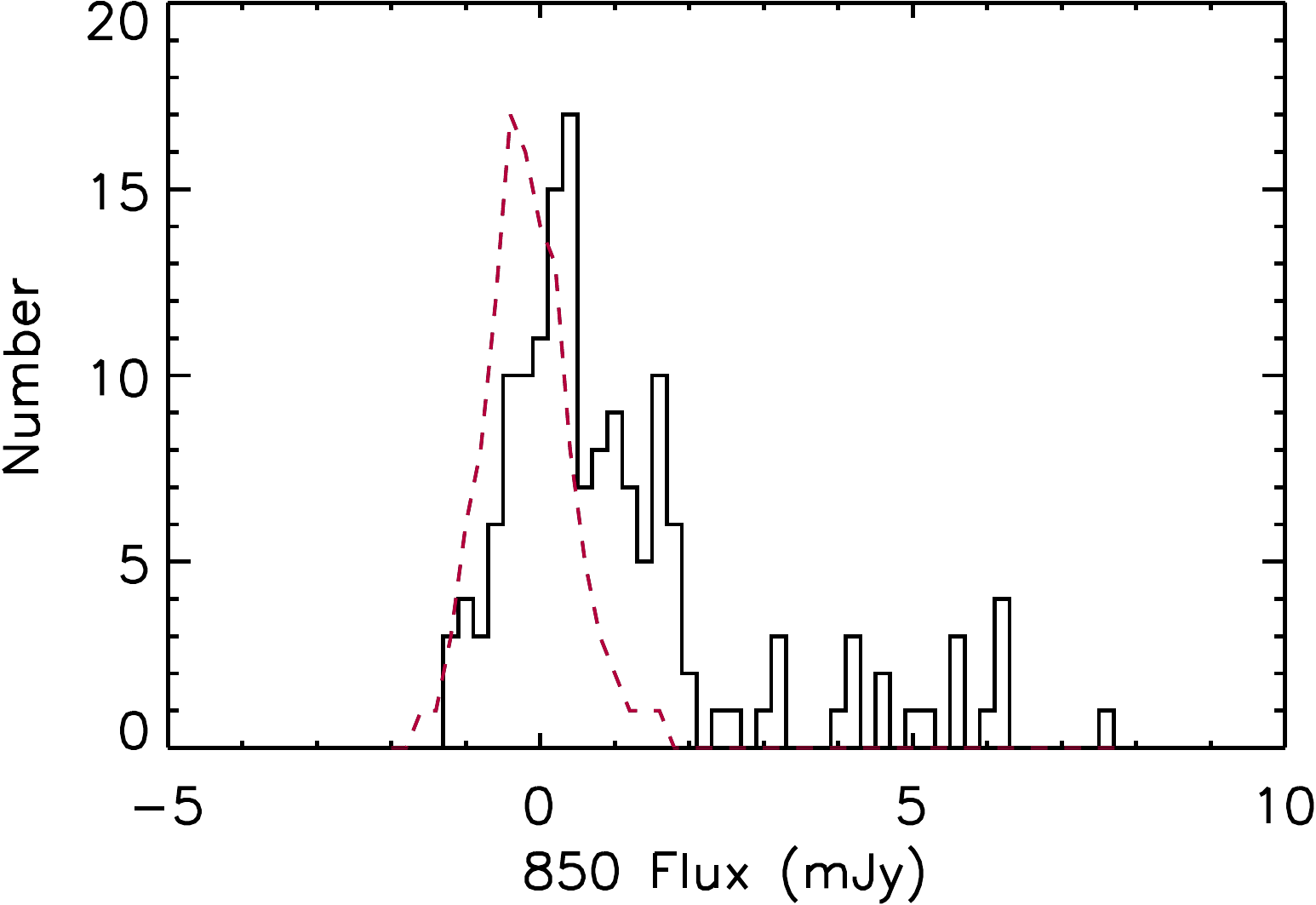}
\caption{Distribution of the measured \afluxb\ flux for the 
156 red galaxy priors (black histogram) and for
the full F444W $>0.05~\mu$Jy sample (red dashed histogram.
The latter excludes the priors and is
renormalized to match the peak of the priors histogram.
\label{jwst_scuba2_histogram}
}
\end{figure}

We take these 156 sources as our priors and measure the \afluxb\ flux and error at each
NIRCam position in the SCUBA-2 image. We make the same measurement for all
of the \fluxa\ $>0.05~\mu$Jy sources in the region, excluding the priors. 
In Figure~\ref{jwst_scuba2_histogram},
we show the distribution of measured \afluxb\ flux for the two populations. 

It is clear from Figure~\ref{jwst_scuba2_histogram}
that there is a significant \afluxb\ flux associated with the priors
(mean \afluxb\ flux of $1.15\pm0.13$~mJy). Meanwhile,
the full F444W $>0.05~\mu$Jy population has 
a mean \afluxb\ flux of $0.02\pm0.008$~mJy. In both cases,
we estimated the uncertainties using the bootstrap method.

\begin{figure*}
\includegraphics[width=3.6in]{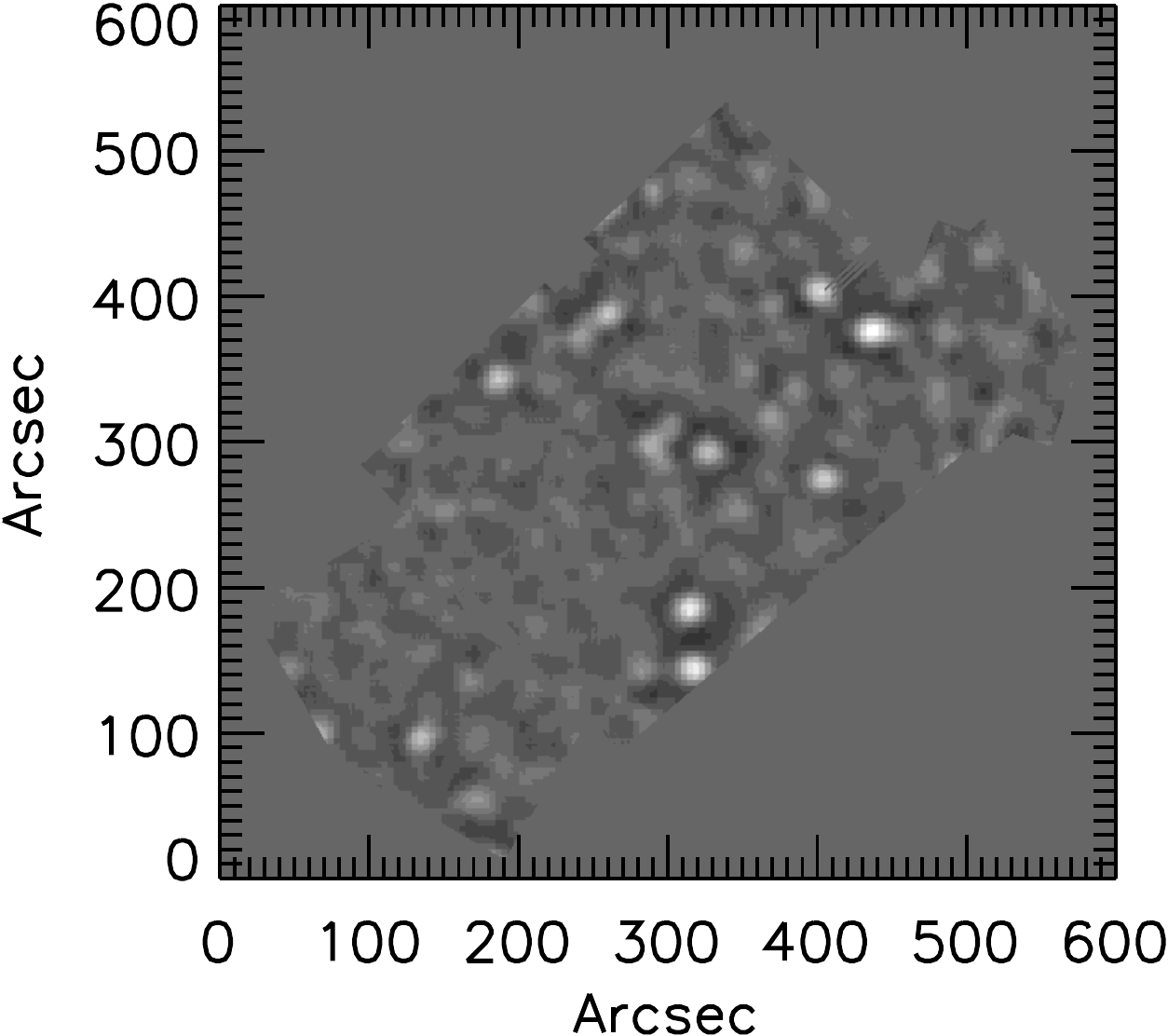}
\includegraphics[width=3.6in]{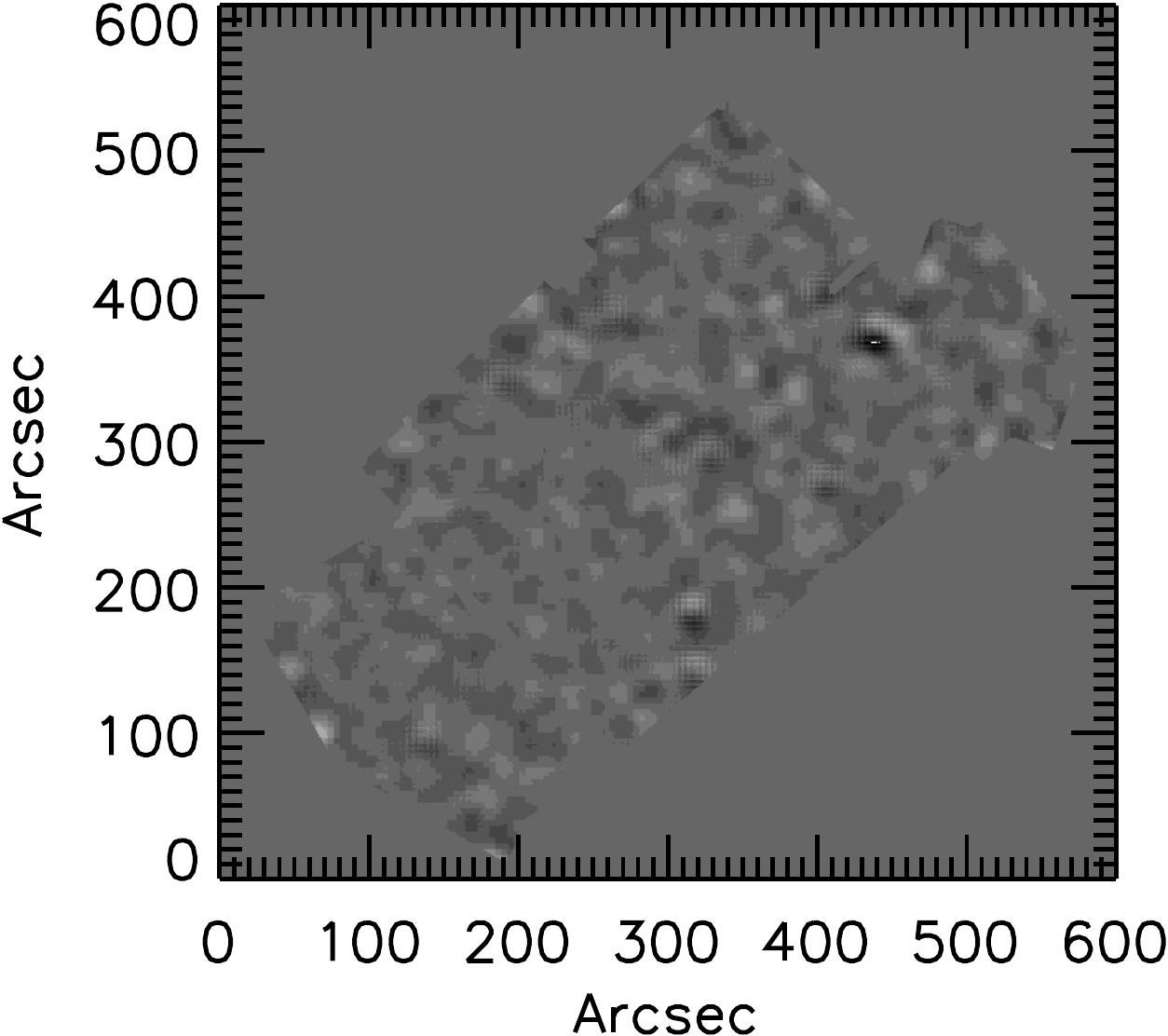}
\caption{(Left) Deep portion of the SCUBA-2 \afluxb\ image in the JWST NIRCam footprint.
(Right) SCUBA-2 \afluxb\ image after removing the \afluxb\ fluxes
corresponding to the red galaxy priors.
\label{SCUBA_2_clean}
}
\end{figure*}

There are 58 $>3\sigma$ \afluxb\ sources.  Nearly all of these are 
above 1.1~mJy. However, not all are real, as some are contaminated 
by the wings of neighboring \afluxb\ sources.
In order to deal with contamination and eliminate any double-counting,
we adopt the following procedure: 
We identify the brightest \afluxb\ peak within $4''$ from a prior.
We then measure the flux at the position of this prior,
convolve it with the SCUBA-2 matched-filter PSF,
and subtract it to form a new cleaned image. 
We repeat this procedure  in order of decreasing \afluxb\ flux, 
using the cleaned SCUBA-2 image 
until all of the priors are used.  
We compare the actual SCUBA-2 \afluxb\ image in 
Figure~\ref{SCUBA_2_clean} (left) with the
final cleaned image in Figure~\ref{SCUBA_2_clean} (right).

\begin{figure}
\center{
\includegraphics[width=2.5in]{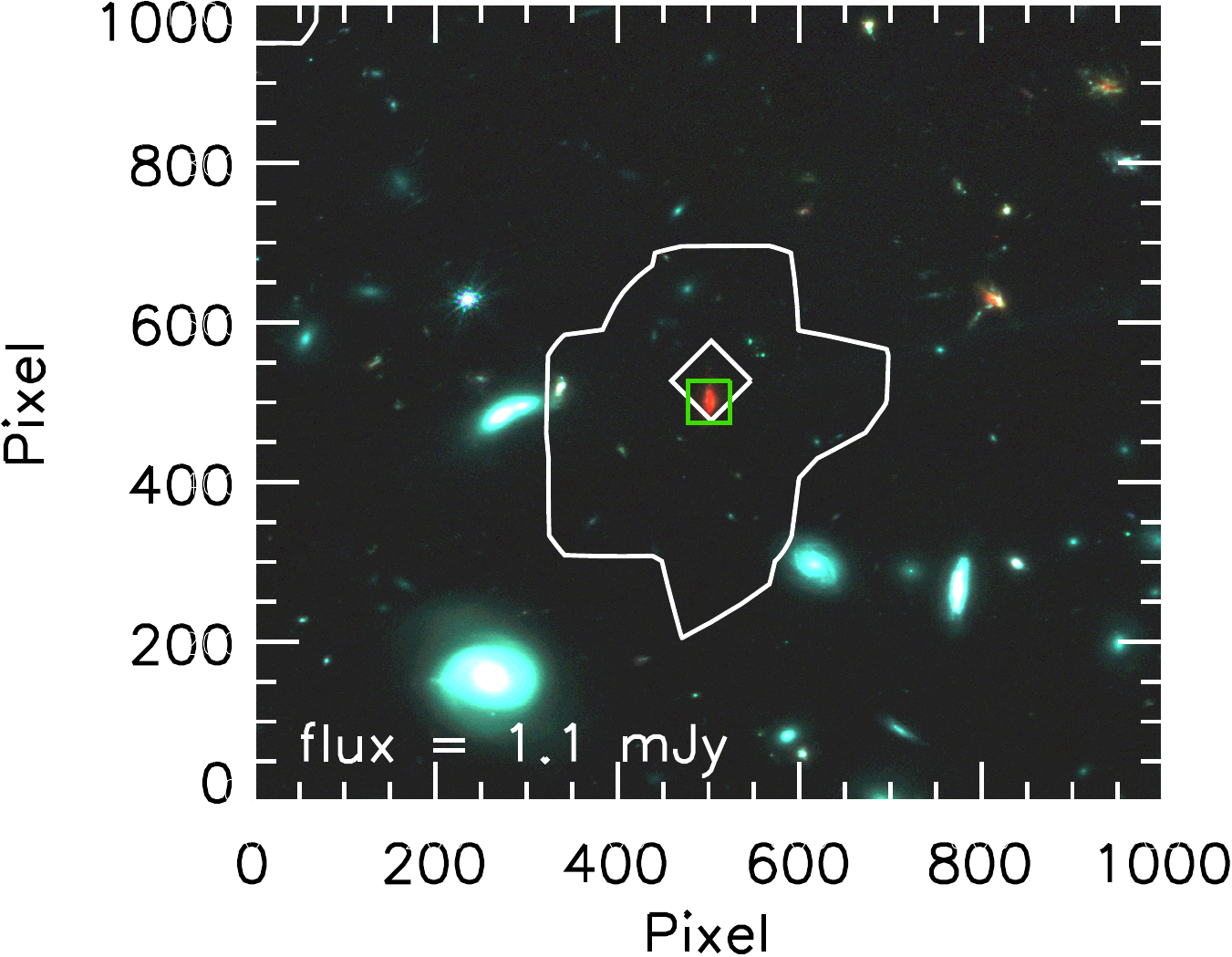}
\includegraphics[width=2.5in]{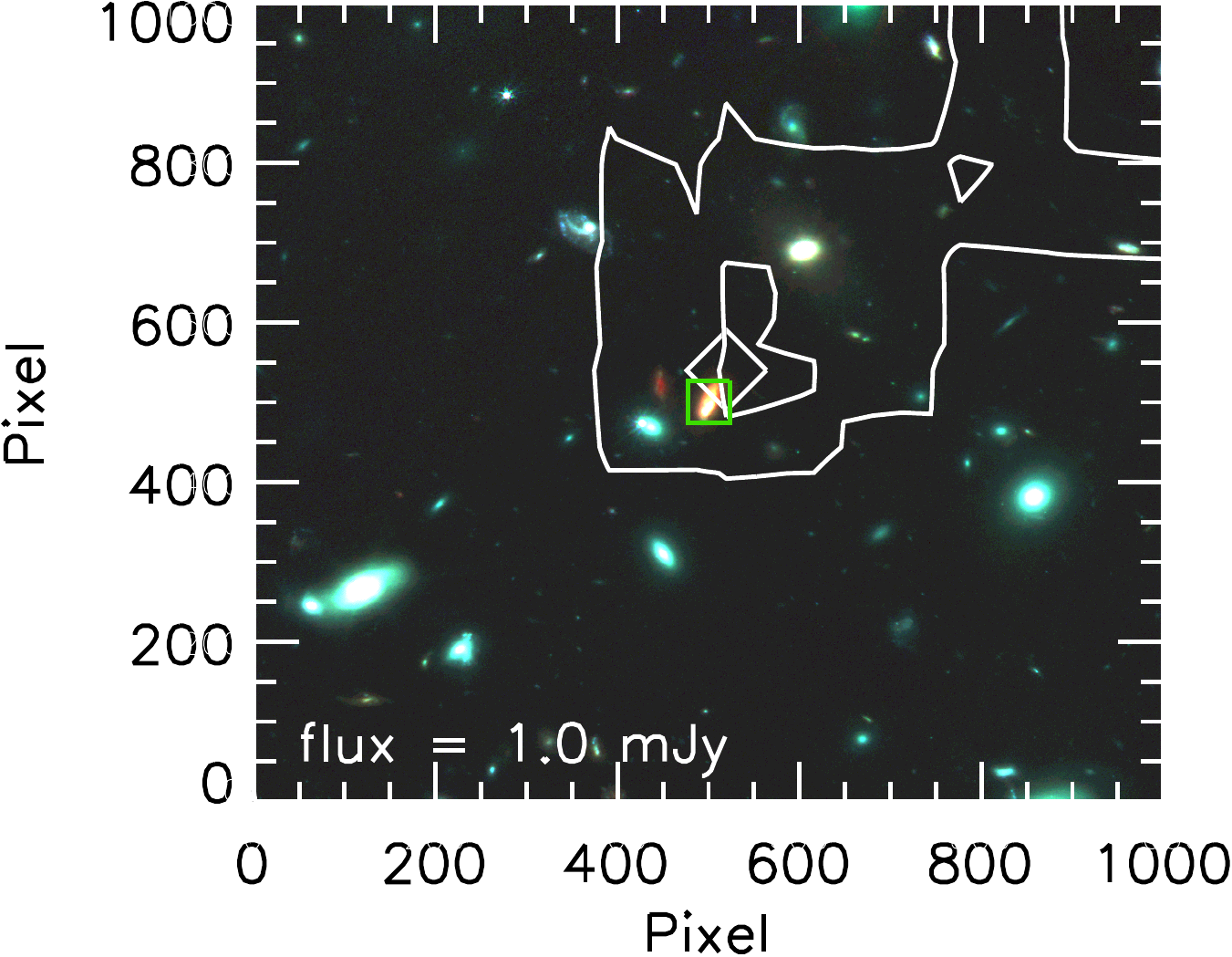}
}
\caption{Two examples of faint SCUBA-2 \afluxb\ sources 
(white contours: 0.6 and 1.2~mJy per beam; diamonds: local peaks; 
fluxes shown in lower-left corner) 
found by using the red galaxy priors.
Both sources lie in the JWST NIRCam footprint but not in the combined
ALMA mosaics from AFFS and ALCS.
The underlying images are three-color JWST (blue = F115W, green = F150W,
and red = F444W). The priors are shown with green squares.
The pixels are $0\farcs03$, and the fields are $30''$ on a side. 
\label{prior_sample}
}
\end{figure}

This procedure reduces the number of priors
with $>3\sigma$ \afluxb\ fluxes above 1.1~mJy 
to 43.  It recovers all 17 of the directly detected SCUBA-2
sources with JWST NIRCam counterparts in Table~\ref{scuba2_acc}.
These 43 sources contain an \afluxb\ extragalactic background
light (EBL) of 10.2~Jy~deg$^{-2}$, which is about a quarter
of the total EBL \citep{fixsen98}.
In Figure~\ref{prior_sample}, we show two examples of faint \afluxb\ 
sources found by using these priors.

%-----------------------------------------------------------------------------
% FIGURE 7:  scuba2_histogram
%-----------------------------------------------------------------------------
%-----------------------------------------------------------------------------

%-----------------------------------------------------------------------------
% FIGURE 8 scuba2_unclean and final_figure_clean_image
%-----------------------------------------------------------------------------
%-----------------------------------------------------------------------------

%-----------------------------------------------------------------------------
% FIGURE 9 red41 and red44
%-----------------------------------------------------------------------------
%-----------------------------------------------------------------------------

Using the same cleaning procedure that \citet{cowie22}
used for the direct SCUBA-2 search,
we now search the cleaned image for additional \afluxb\ sources without
priors, finding 21 with $>3\sigma$ \afluxb\ fluxes above 1.1~mJy.
We note that both sources~13 and 17 from Table~\ref{scuba2_acc} are contained in
the 21 sources detected in the residual image.
While not all of these 21 sources are necessarily real, 
combining them with the 43 found with priors gives a fraction of 67\% that are 
picked out by our red galaxy priors.

Before proceeding, we again note that the \citet{paris23} catalog
contains a small number of cases where a single object
is split into multiple components.
Replacing these with single objects
reduces the priors sample to 148. There are a further 4 objects
that appear to be red stars \citep[e.g.,][]{nonino23}
based on the SExtractor \citep{bertin96}
star classifier and visual inspection.
As expected, none of the stars are detected at \afluxb.

In summary, of our 144 non-star red 
galaxy priors, we find 43 with $>3\sigma$ \afluxb\ fluxes above 1.1~mJy.
There is one additional prior with a $>3\sigma$ \afluxb\ detection
whose flux is below 1.1~mJy. Thus, 30\% of the
priors have $>3\sigma$ \afluxb\ counterparts.
We list these 44 sources in Table~\ref{fintab} in the Appendix.

In $\sim20$\% of these, the SCUBA-2 \afluxb\ flux could be
associated with two priors. In the ALMA covered area,
sources~4 and 6 in Table~\ref{tabALMA} (see Figure~\ref{alma_images}
for their thumbnails) provide such an example.
This percentage is slightly higher than the 13\% (68\% confidence range 7\%--19\%) of
SCUBA-2 sources with $>4\sigma$ \afluxb\ flux above 2.25~mJy in the GOODS-S that have
multiple ALMA counterparts \citep{cowie18}.
While we assign all of the \afluxb\ flux to the nearest 
prior, the other prior could be partially contributing. 
Allowing for this possibility could increase the percentage of the
non-star red galaxy priors with $>3\sigma$ \afluxb\ counterparts to 37\%.

%-----------------------------------------------------------------------------
\section{Discussion}
\label{secdisc}
%-----------------------------------------------------------------------------
In order to see whether we are missing fainter 
\fluxa\ sources in our red galaxy selection, 
we now extend our \jwratio\ $>3.5$ sample down to \jwratio\ $= 0.05~\mu$Jy.
We combine the small number of multiple 
component objects from the \citet{paris23} catalog into single objects.
We also exclude the four red stars in the region.
Finally, we exclude any source that lies closer
than $4''$ to a \fluxa-brighter red galaxy
to avoid multiple-counting in the submillimeter.
However, we note that none of our subsequent results 
are significantly affected if we remove this condition.

\begin{figure}
\includegraphics[width=3.5in]{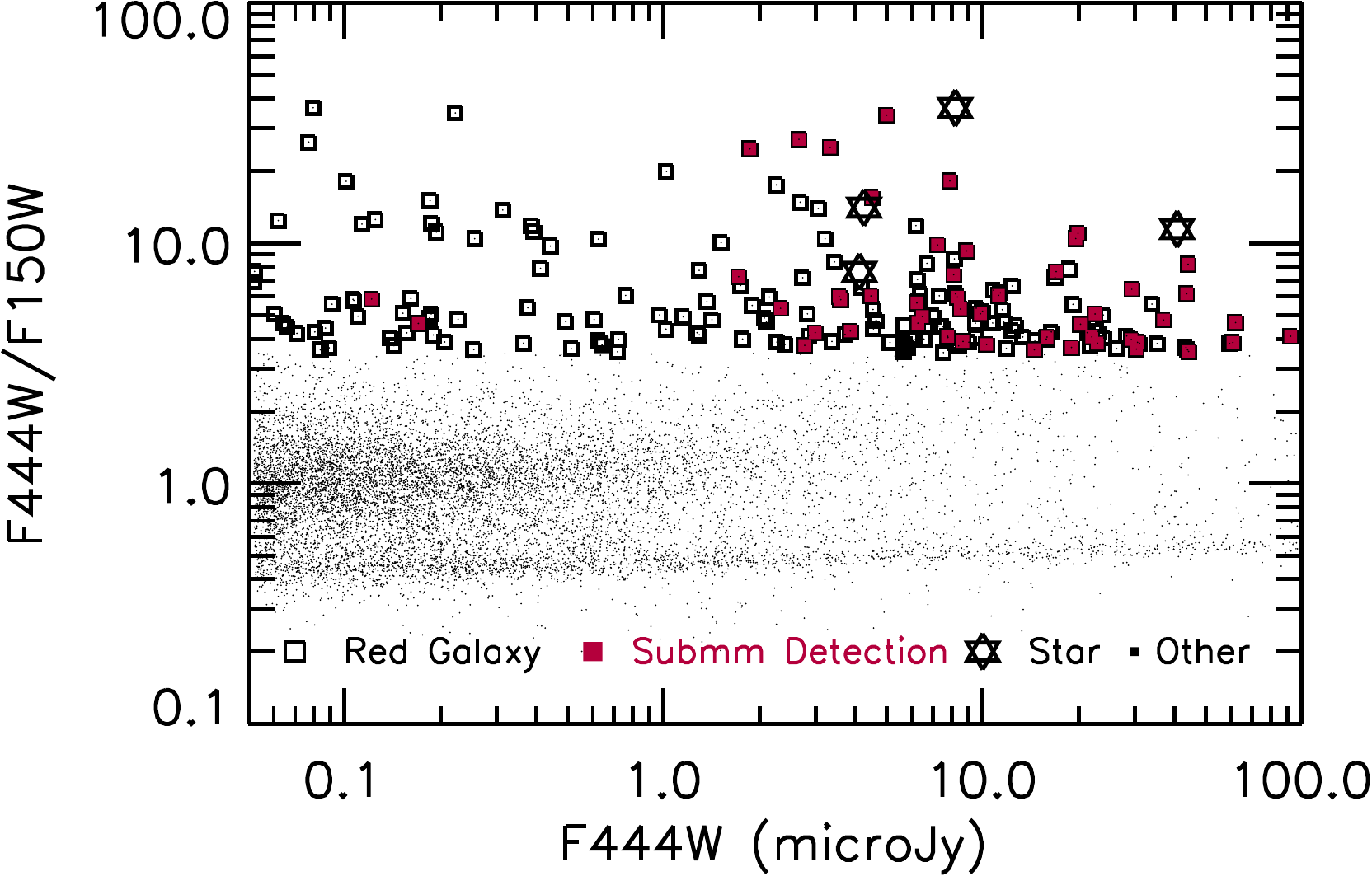}
\caption{
\jwratio\ vs. \fluxa\
for the full sample in the JWST NIRCam plus SCUBA-2 footprint
with \fluxa\ $>0.05~\mu$Jy (black dots).
Squares show the red galaxies with \jwratio\ $>3.5$, 
with those having $>3\sigma$ \afluxb\ detections
marked in red. For these red galaxies:
the small number of objects in the \citet{paris23} catalog with 
multiple parts were combined into single objects;
the four stars are shown with 
star symbols (none are detected at \afluxb); and
no source that lies closer
than $4''$ to a \fluxa-brighter red galaxy is shown
to avoid multiple-counting in the submillimeter.
\label{final_selection}
}
\end{figure}

In Figure~\ref{final_selection},
we show \jwratio\ versus \fluxa\ for the full sample in the
JWST NIRCam plus SCUBA-2 footprint with \fluxa\ $> 0.05~\mu$Jy
(black dots). We denote those with \jwratio\ $> 3.5$ as squares 
(red if they also have $>3\sigma$ \afluxb\ detections).
Consistent with our previous selection,
only 2 of the 55 galaxies with \fluxa\ between $0.05~\mu$Jy and $1~\mu$Jy
have a $>3\sigma$ \afluxb\ detection
(both are $3.2\sigma$, and their \afluxb\ fluxes are 
1.48~mJy and 1.17~mJy).

%-----------------------------------------------------------------------------
% FIGURE 10 dark_selection
%-----------------------------------------------------------------------------
%-----------------------------------------------------------------------------

We see from Figure~\ref{final_selection} a strong preference for the 
\fluxa-brightest of the red galaxies
to be detected at \afluxb. There are 47 red galaxies with \fluxa\ $>10~\mu$Jy. 
Of these, 26 (55\%) have $>3\sigma$ \afluxb\ detections.
We also see from Figure~\ref{final_selection} a strong preference for the reddest of the
red galaxies to be detected at \afluxb, as might be expected 
if the extinction is higher in these galaxies. 
There are 17 with \fluxa\ $>1~\mu$Jy and \jwratio\ $>8.3$.
Of these, 10 (59\%) have $>3\sigma$ \afluxb\ detections.

While these general preferences
are clear, we would like to understand in more detail
how red galaxies that are not \afluxb\ detected
are related to those that are \afluxb\ detected.
In order to carry out this analysis, we need redshift
estimates and magnifications. 
Given the extreme colors of these galaxies, only a very small
number of them have spectroscopic redshifts, so we must 
instead rely on photometric redshifts. We use the catalog 
from \citet{weaver23}, which includes both photometric 
redshifts and magnifications. However,
as we mentioned in Section~\ref{secdata}, 
it does not fully cover the JWST NIRCam image from \citet{paris23}.

Our {\em extended sample} contains
167 \fluxa\ $>0.05~\mu$Jy and \jwratio\ $>3.5$ galaxies
with both photometric redshifts and magnifications.
All but one have positive measurements
in all the NIR bands at and above $1.15~\mu$m, providing
six NIR bands for the photometric redshift determinations.
All but five have $>2\sigma$ detections in the $1.15~\mu$m 
and 1.5~$\mu$m bands, providing well-determined colors.
Seven are X-ray
sources in the Chandra catalog of \citet{swang16}
and appear to be luminous active galactic nuclei (AGNs).

\cite{weaver23} provide an extensive discussion of the
quality of the photometric redshifts for their entire sample, but
photometric redshifts are significantly
more uncertain for the red galaxies of the present sample due
to the degeneracy between reddening and redshift.
Reassuringly, our comparison of the
photometric redshifts with the spectroscopic redshifts for
the small number with both
(see Figure~\ref{redshift}) shows good agreement.
Although all of these spectroscopic redshifts lie in the $z=1-4$ range,
so, too, do 81\% of the photometric redshifts.
Thus, for sources in this redshift range, we
may have some confidence in the photometric redshifts. 
Moreover, the uncertainties assigned by the EAZY code are generally
small, as we illustrate in Figure~\ref{redshift} (see also the
uncertainties on the photometric redshifts for the $>3\sigma$ \afluxb\ detected
red galaxies given in Table~\ref{fintab} in the Appendix). 
Only six of the extended sample sources have $dz/(1+zp) > 0.5$.
We therefore proceed with the photometric redshifts, while bearing in mind
the need to confirm them with spectroscopy.

\citet{weaver23} do not provide the extinctions calculated
from their photometric redshift fits. We therefore reran EAZY on the 
$>3\sigma$ \afluxb\ detected red galaxies given in Table~\ref{fintab} in the Appendix. 
The measured median
$A_V$ for these sources is 2.9, which is consistent with them being dusty 
star-forming galaxies.

We next compute the demagnified fluxes at a rest-frame wavelength of
5000~\AA\ (the longest rest-frame wavelength observed by
JWST NIRCam at the highest redshifts), along with the corresponding
demagnified luminosities, \lnu.

\begin{figure}
\includegraphics[width=3.25in]{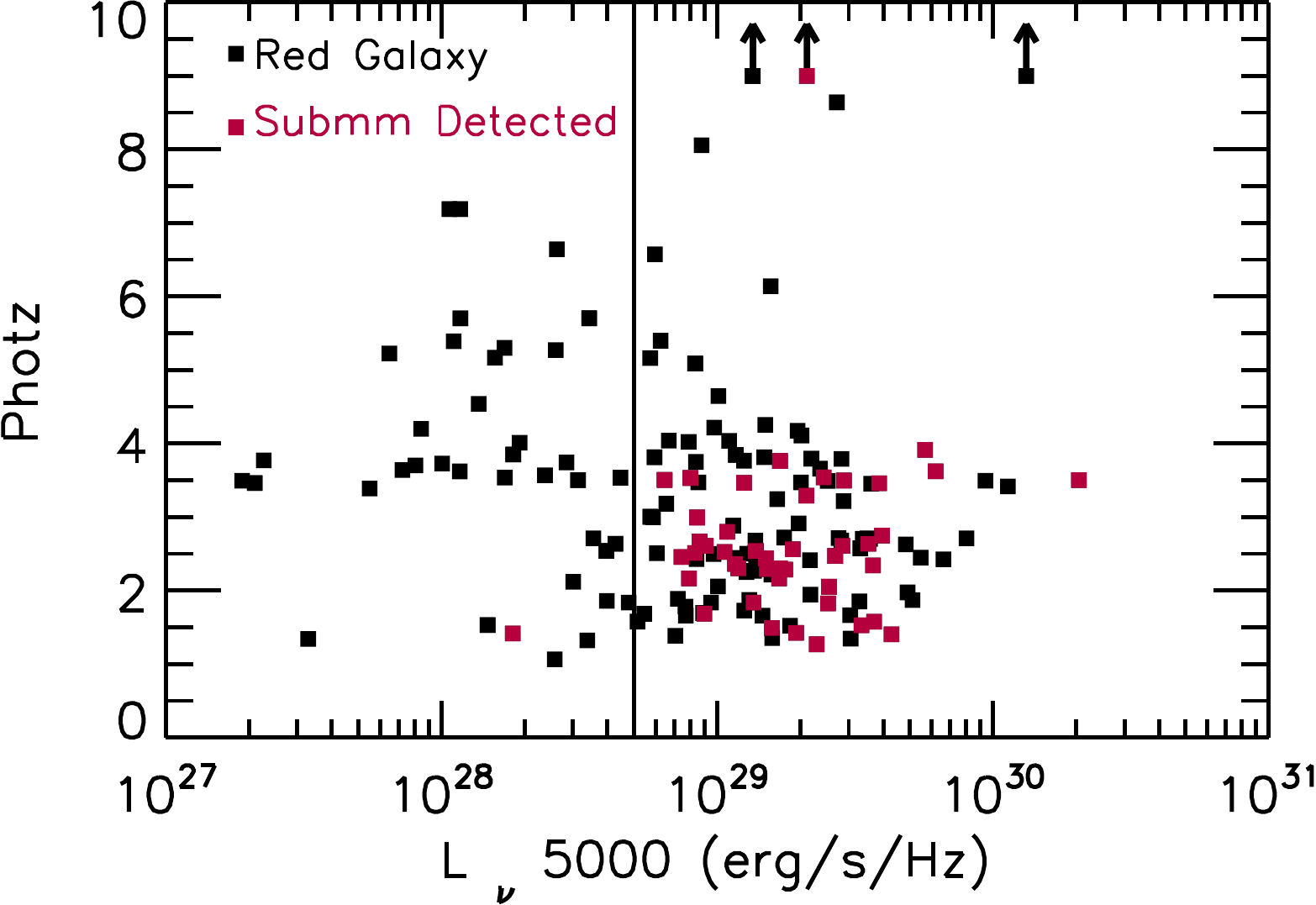}
\caption{
Photometric redshift vs. \lnu\ for our extended sample.
Sources with $>3\sigma$ \afluxb\ detections are marked in red. 
We show sources whose photometric redshifts are $>9$ at $z=9$
with upward pointing arrows for illustrative purposes only.
\label{dark_lum_zp}
}
\end{figure}

In Figure~\ref{dark_lum_zp}, we plot photometric redshift versus \lnu\
for our extended sample.
Nearly all of the $>3\sigma$ \afluxb\ detected sources (marked in red)
have \lnu\ $> 5\times10^{28}$~erg~s$^{-1}$~Hz$^{-1}$
(vertical black line), which corresponds to $\nu$\lnu\
$>3\times10^{43}$~erg~s$^{-1}$. 
There are 37 sources with 
\lnu\ $< 5\times10^{28}$~erg~s$^{-1}$~Hz$^{-1}$,
of which only one has a $>3\sigma$ \afluxb\ detection.

In contrast, there are 129 sources with
\lnu\ $>5\times10^{28}$~erg~s$^{-1}$~Hz$^{-1}$,
44 of which have $>3\sigma$ \afluxb\ detections. 
The mean demagnified \afluxb\ flux, \fafluxbb,
of the 129 sources is $0.57\pm0.09$~mJy, where we
estimated the uncertainties using the bootstrap method.

However, nearly all
of this is concentrated in the low-redshift population.
For $z<4$, the mean \fafluxbb\ is $0.66\pm0.08$~mJy,
while for $z>4$, it has dropped to $0.08\pm0.08$~mJy.
This result is not sensitive to the uncertainties
in the photometric redshifts. Even placing all of the photometric
redshifts at their $1\sigma$ upper limits only changes
the mean \fafluxbb\
to $0.63\pm0.08$~mJy at $z<4$ and $0.11\pm0.08$ at $z>4$.

%-----------------------------------------------------------------------------
% FIGURE 11 dark_lum_zp
%-----------------------------------------------------------------------------
%-----------------------------------------------------------------------------

%-----------------------------------------------------------------------------
% FIGURE 12 dark_lum_fl and dark_lum_fl_hiz
%-----------------------------------------------------------------------------
\begin{figure*}[t]
\includegraphics[width=3.25in]{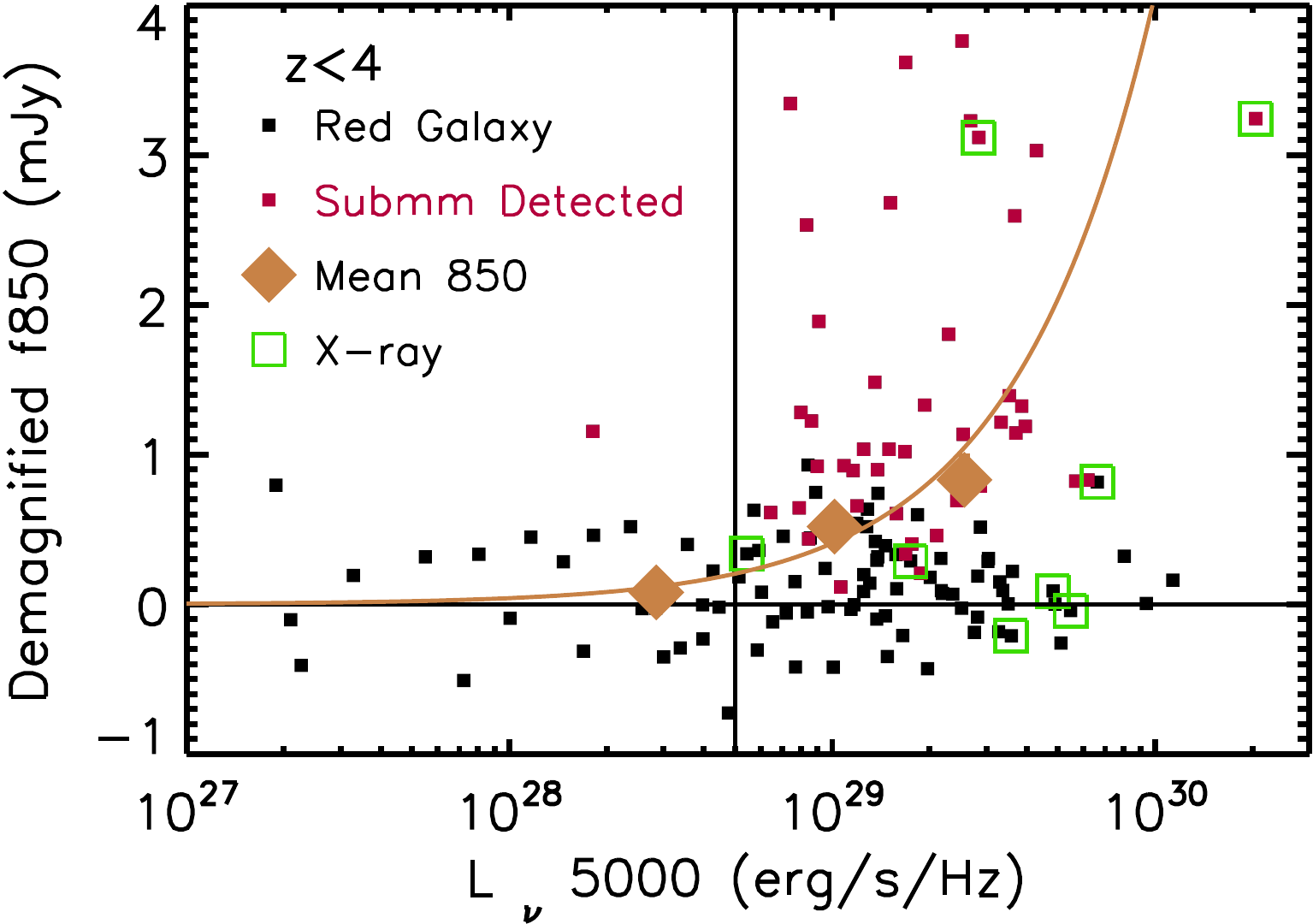}
\hskip 0.75cm
\includegraphics[width=3.25in]{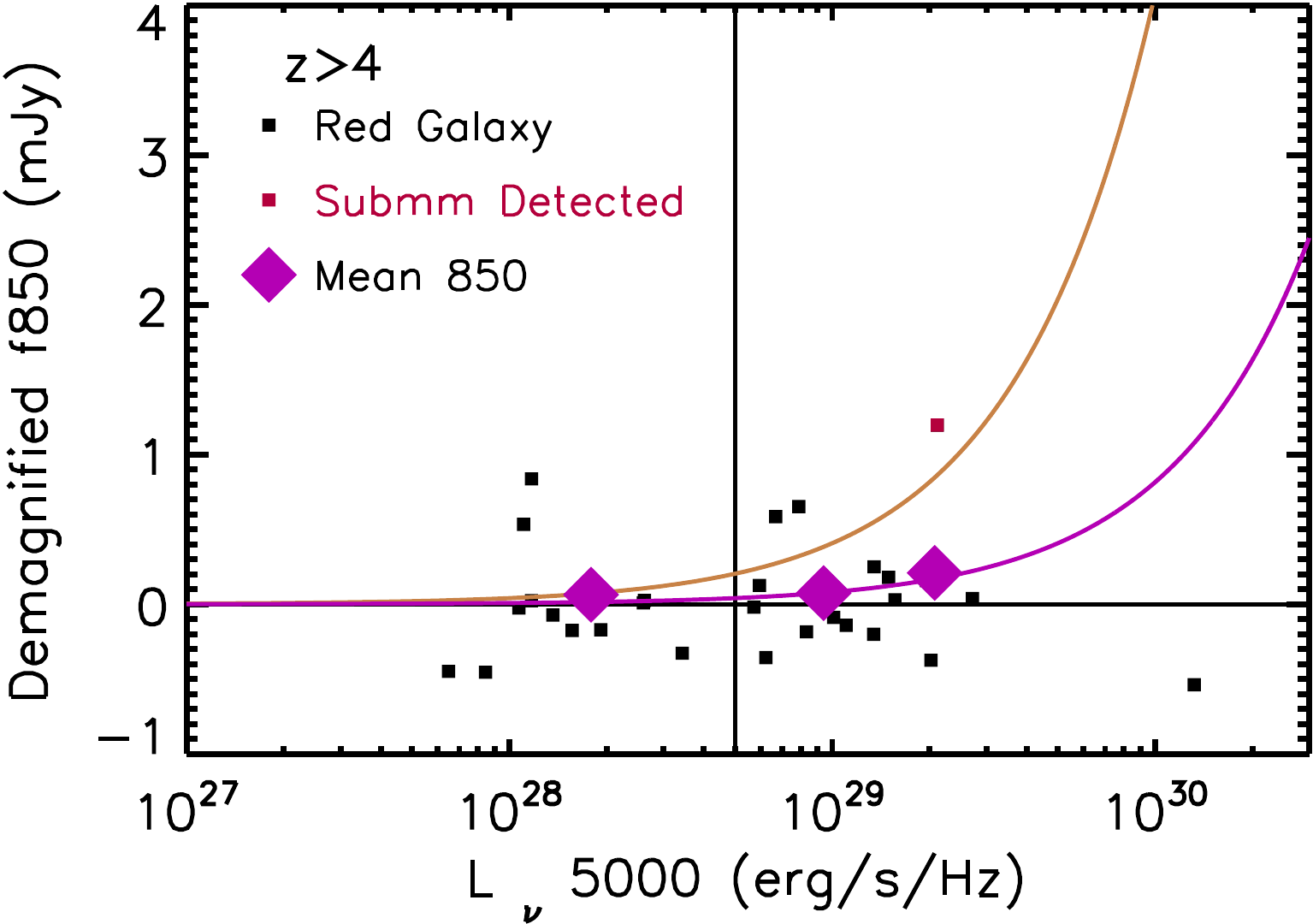}
\caption{
\fafluxbb\ vs. \lnu\ for the (a) $z<4$ and (b) $z>4$ sources in
our extended sample.
Sources with $>3\sigma$ \afluxb\ detections are marked in red. 
X-ray sources are enclosed in large green squares. 
The solid diamonds show the mean values
in the \lnu\ ranges $10^{28}$ to $5\times 10^{28},
5\times 10^{28}$ to $1.5\times 10^{29}$, 
and $1.5\times 10^{29}$ to $4\times 10^{29}$~erg~s$^{-1}$~Hz$^{-1}$.
The curves correspond to $R=30$ (gold) and $R=6$ (purple)
(see Equation~\ref{Requation}).
\label{dark_lum_fl}
}
\end{figure*}
%-----------------------------------------------------------------------------

Sources with \fafluxbb\ = 1~mJy have a demagnified FIR (8--1000~$\mu$m) luminosity of 
$L_{\rm FIR}^d\sim 4\times 10^{45}$~erg~s$^{-1}$ at $z>1$ \citep[e.g.,][]{cowie17}.
We now characterize the sources with the parameter
\begin{equation}
R=\frac{(4\times10^{45}) f_{\rm 850~\mu m}^d \, {\rm (in \, mJy)}}{(6\times 10^{14}) L_\nu^d (5000)} \,.
\label{Requation}
\end{equation}
This is the ratio of the FIR luminosity to $\nu$\lnu\
at rest-frame 5000~\AA.
The $R$ parameter will vary with extinction and SED, but
for a typical ultraluminous infrared galaxy such as Arp~220, $R$
computed with the \citet{silva98} SED is 20. 
In contrast, the ratio for the luminosity in the 1000--40000~\AA\ range to 
$\nu$\lnu\ at rest-frame 5000~\AA\ for Arp~220 is 2.4.
Thus, $\sim90$\% of its light emerges in the FIR.

For sources 
with \lnu\ $ > 5\times10^{28}$~erg~s$^{-1}$~Hz$^{-1}$,
the $R$ parameter drops from a mean of 54 at $z<4$ to 8 at $z>4$.
A Mann-Whitney test shows only a 0.0016 probability that
the two samples are consistent.
In Figure~\ref{dark_lum_fl}, we show \fafluxbb\
versus \lnu\ for the two redshift ranges.
We also show the mean \fafluxbb\ in
three \lnu\ ranges (diamonds). We overlay the
curves for $R=30$ (gold) and $R=6$ (purple), which provide
reasonable fits to these mean values at $z<4$ and $z>4$, respectively.

The rapid drop in $R$
as one moves to higher redshifts may suggest
that the high-redshift galaxies have much less dust.
However, ultimately, a full understanding of this result
requires a more detailed analysis. We leave this analysis
to a subsequent paper \citep{mckay23}, where we will
consider the changes in the physical properties of the sources
with redshift using size and structure measurements
and fits to the SEDs.

From Figure~\ref{dark_lum_fl}, we can also see
that the absence of $>3\sigma$ \afluxb\ detections at fainter \lnu\
is simply a selection effect---the \afluxb\ flux has likely become too faint to detect. 
At \lnu\ $ = 2.5\times 10^{28}$~erg~s$^{-1}$~Hz$^{-1}$, 
we expect a mean \fafluxbb\ of 0.1~mJy for $R=30$,  
comparable to the minimum $3\sigma$ value reached
in our measurements.

%---------------------------------------------------------------------
\section{Summary}
\label{secsummary}
%---------------------------------------------------------------------
We showed that a JWST NIRCam red selection criteria of \fluxa\ $>1~\mu$Jy
and \jwratio\ $>3.5$ locates all of the known ALMA 1.1~mm and 1.2~mm sources
and 17 of the 19 SCUBA-2 \afluxb\ ($>5\sigma$) sources in
the A2744 cluster field in the JWST NIRCam covered areas 
based on the images of \citet{paris23}.
Using these red galaxies as priors, we were able to probe
deeper in the SCUBA-2 data, finding
44 $>3\sigma$ \afluxb\ sources (this procedure recovers the
17 direct detections).

We analyzed an extended sample of 167 sources 
(\fluxa\ $>0.05~\mu$Jy and \jwratio\ $>3.5$,
where care was taken to combine the small
number of multiple component objects from the \citealt{paris23} 
catalog into single objects; to exclude
the four red stars in the region; and to exclude
any source that lies closer than $4''$ to a \fluxa-brighter red galaxy
to avoid multiple-counting in the
submillimeter) using photometric redshifts and gravitational lensing 
magnifications from the slightly smaller area UNCOVER catalog of \citet{weaver23}. 
We found that all but one of the $>3\sigma$ \afluxb\
detections lay at $z<4$, and all but one had a
demagnified luminosity at a rest-frame wavelength of 5000~\AA\ 
of \lnu\ $>5\times10^{28}$~erg~s$^{-1}$~Hz$^{-1}$. 

We concluded that the redshift dependence in the $>3\sigma$ \afluxb\ detections
may be a result of a significant decrease in the dust content of the galaxies
at the higher redshifts. Parameterizing this with the quantity $R$,
which is the ratio of the FIR luminosity estimated from the
\afluxb\ flux to $\nu$\lnu\ at rest-frame 5000~\AA, we found a drop of around 
5 between $z<4$ and $z>4$.

In contrast, we found that the \lnu\ dependence appeared to be a simple sensitivity
issue, with the sources 
$<5\times10^{28}$~erg~s$^{-1}$~Hz$^{-1}$ being too faint to be
detected in the SCUBA-2 \afluxb\ image.

%------------------------------------------------------------------------
\begin{acknowledgements}
%------------------------------------------------------------------------
We thank the anonymous referee for constructive comments that helped 
us to improve the manuscript.
We gratefully acknowledge support for this research from
a Kellett Mid-Career Award and a WARF Named Professorship from the 
University of Wisconsin-Madison Office of the 
Vice Chancellor for Research and Graduate Education with funding from the 
Wisconsin Alumni Research Foundation (A.~J.~B.) and
NASA grant NNX17AF45G (L.~L.~C.).

The National Radio Astronomy Observatory is a facility of the National Science
Foundation operated under cooperative agreement by Associated Universities, Inc.
This paper makes use of the following ALMA data: 
ADS/JAO.ALMA\#2013.1.00999.S, \\ %(FB) G-L et al. 2017
ADS/JAO.ALMA\#2015.1.01425.S, \\ %(FB) G.L. et al. 2017
ADS/JAO.ALMA\#2017.1.01219.S, \\ %(FB) 2 sources in A2744
and ADS/JAO.ALMA\#2018.1.00035.L. %(KK)
ALMA is a partnership of ESO (representing its member states), 
NSF (USA) ,and NINS (Japan), together with NRC (Canada), 
MOST and ASIAA (Taiwan), and KASI (Republic of Korea), in cooperation 
with the Republic of Chile. The Joint ALMA Observatory is operated by 
ESO, AUI/NRAO and NAOJ.

The James Clerk Maxwell Telescope is operated by the East Asian Observatory 
on behalf of The National Astronomical Observatory of Japan, Academia Sinica 
Institute of Astronomy and Astrophysics, the Korea Astronomy and Space 
Science Institute, the National Astronomical Observatories of China and the 
Chinese Academy of Sciences (Grant No. XDB09000000), with additional funding 
support from the Science and Technology Facilities Council of the United Kingdom 
and participating universities in the United Kingdom and Canada.

We wish to recognize and acknowledge the very significant 
cultural role and reverence that the summit of Maunakea has always 
had within the indigenous Hawaiian community. We are most fortunate 
to have the opportunity to conduct observations from this mountain.

\end{acknowledgements}

\facilities{ALMA, JCMT}

\bibliography{smmref}

\appendix

In Table~\ref{fintab},
we summarize the properties of the 44 $>3\sigma$ SCUBA-2 \afluxb\ detected JWST NIRCam 
\fluxa\ $>1~\mu$Jy and \jwratio\ $>3.5$ sources with photometric redshifts.

We show the corresponding three-color NIRCam images in 
Figure~\ref{all_scuba2_images}.
The morphological classification of DSFGs is a complex and uncertain process.
Based on HST data, some visual analyses
have found a high merger fraction \citep[e.g.,][]{chen15,cowie18}, while some 
quantitative analyses have instead found them to be massive disks \citep{targett13} and 
not preferentially major mergers \citep{swinbank10}.
\citet{chen15} argue that quantitative methods can miss many merging or disturbed 
sources, which are easily distinguished in a visual inspection. 

JWST NIRCam data are needed for more reliable analyses of the
morphologies of DSFGs \citep[e.g.,][]{chen22,cheng22,cheng23},
and, in particular, those of color selected or dark galaxies
(see, e.g., \citealt{kokorev23}, who classified source~5 in  
Figure~\ref{all_scuba2_images} as an edge-on spiral galaxy).
However, much larger sample sizes than have currently been analyzed 
are needed to make definitive statements.
We leave a more detailed discussion to \citet{mckay23}. Here we note only that
the morphologies of the galaxies in Figure~\ref{all_scuba2_images} are quite heterogeneous, ranging from
compact sources, including the quasar of source~4, all the way to large mergers,
such as sources~7, 15, and 25. Slightly more than a quarter
of the sources show clear evidence of merging.

%-----------------------------------------------------------------------------
%TABLE 3
%-----------------------------------------------------------------------------
\begin{deluxetable*}{rcccccccccccc}
\renewcommand\baselinestretch{1.0}
\tablewidth{0pt}
\tablecaption{JWST NIRCam \fluxa\ $>1~\mu$Jy and \jwratio\ $>3.5$ Selected $>3\sigma$ SCUBA-2 \afluxb\ Detections in the UNCOVER Area \label{fintab}}
\scriptsize
\tablehead{No.  &  R.A.  &  Decl. &  \fafluxbb\  & $\log \nu$\lnu\ &  Redshift & $A_V$ & SFR & $\mu$  &  \fluxa\  &  \fluxb\ &  Ratio  &  Match  \\  
&  \multicolumn{2}{c}{(J2000.0)} &  (mJy)   &  (erg~s$^{-1}$)  & & &  (M$_\odot$~yr$^{-1}$) & & ($\mu$Jy) &  ($\mu$Jy)   & &  \\ (1)  &  (2)  &  (3)  &  (4)  &  (5)  &  (6)  &  (7)  &  (8)  &  (9) & (10) & (11 & (12) & (13))}
\startdata
       1 &        3.5362501 &       -30.360361 & 3.76(0.18) & 44.18 & 1.82(1.69,1.92) & 2.99 & 1.17  & 2.0 & 43.5 & 7.06 & 6.16 & SCUBA1 \cr
       2 &        3.5755835 &       -30.424389 & 3.61(0.17) & 44.00 & 3.76(2.64,3.91) & 3.69 & 106.  & 1.7 & 5.00 & 0.14 & 34.0 & SCUBA3 \cr
       3 &        3.5476665 &       -30.352140 & 3.34(0.24) & 43.64 & 2.45(2.24,2.61) & 1.76 & 6.08  & 1.6 & 2.98 & 0.70 & 4.24 & \nodata \cr
       4 &        3.6172502 &       -30.368807 & 3.24(0.26) & 45.08 & 3.50(2.46,3.50) & 2.00 & 177.  & 1.4 & 30.3 & 8.35 & 3.62 & SCUBA6 \cr
       5 &        3.5761251 &       -30.413166 & 3.22(0.13) & 44.20 & 2.585 & 3.40 & 0.04  & 1.9 & 3.11 & 0.15 & 20.5 & ALMA2 \cr
       6 &        3.5491250 &       -30.352251 & 3.11(0.24) & 44.23 & 2.60(2.43,2.68) & 2.99 & 0.12  & 1.7 & 16.9 & 2.23 & 7.61 & SCUBA4 \cr
       7 &        3.5938334 &       -30.356609 & 3.03(0.27) & 44.41 & 1.40(1.36,1.87) & 2.50 & 78.7  & 1.5 & 92.2 & 22.5 & 4.09 & SCUBA7 \cr
       8 &        3.5474167 &       -30.388277 & 2.68(0.15) & 43.95 & 2.29(2.13,2.61) & 3.80 & 52.5  & 1.9 & 19.8 & 1.81 & 10.9 & SCUBA5 \cr
       9 &        3.5925832 &       -30.356140 & 2.59(0.25) & 44.34 & 2.34(2.13,2.54) & 2.91 & 531.  & 1.6 & 15.9 & 3.91 & 4.06 & \nodata \cr
      10 &        3.5482500 &       -30.353277 & 2.53(0.23) & 43.69 & 2.51(2.13,2.64) & 1.46 & 12.5  & 1.7 & 2.76 & 0.73 & 3.75 & \nodata \cr
      11 &        3.5464168 &       -30.353443 & 1.88(0.23) & 43.73 & 2.60(2.17,2.72) & 2.87 & 12.7  & 1.7 & 4.46 & 0.73 & 6.03 & \nodata \cr
      12 &        3.5990419 &       -30.359722 & 1.80(0.27) & 44.13 & 1.26(1.22,1.35) & 2.55 & 31.8  & 1.4 & 60.8 & 15.7 & 3.86 & SCUBA11 \cr
      13 &        3.5719585 &       -30.382973 & 1.48(0.09) & 43.90 & 1.83(1.55,1.89) & 3.66 & 137.  & 2.8 & 44.0 & 5.38 & 8.17 & ALMA6 \cr
      14 &        3.5847917 &       -30.333166 & 1.39(0.33) & 44.32 & 2.63(2.22,2.84) & 3.05 & 14.4  & 1.3 & 10.2 & 2.71 & 3.79 & \nodata \cr
      15 &        3.5610831 &       -30.330000 & 1.32(0.34) & 44.06 & 1.42(1.43,2.41) & 3.07 & 15.8  & 1.3 & 36.8 & 7.69 & 4.79 & \nodata \cr
      16 &        3.5641251 &       -30.344444 & 1.32(0.29) & 44.36 & 3.45(2.82,3.28) & 2.86 & 33.1  & 1.6 & 6.26 & 1.34 & 4.64 & \nodata \cr
      17 &        3.6276667 &       -30.394249 & 1.28(0.25) & 43.68 & 3.53(3.26,3.60) & 3.09 & 50.4  & 1.4 & 1.86 & 0.07 & 24.6 & \nodata \cr
      18 &        3.6006250 &       -30.362720 & 1.22(0.24) & 43.71 & 2.66(2.31,2.72) & 3.29 & 0.01  & 1.5 & 7.88 & 0.43 & 18.1 & \nodata \cr
      19 &        3.5211668 &       -30.360666 & 1.21(0.25) & 44.30 & 1.52(1.38,1.75) & 2.71 & 92.8  & 1.5 & 62.0 & 13.2 & 4.67 & \nodata \cr
      20 &        3.5849586 &       -30.381777 & 1.18(0.10) & 44.37 & 3.058 & 3.15 & 287.  & 2.7 & 20.2 & 4.40 & 4.60 & ALMA3 \cr
      21 &        3.6235831 &       -30.426777 & 1.15(0.25) & 43.03 & 1.41(1.42,2.52) & 2.97 & 7.00  & 1.4 & 3.84 & 0.89 & 4.31 & \nodata \cr
      22 &        3.5965416 &       -30.358555 & 1.14(0.26) & 44.34 & 1.57(1.49,1.70) & 1.40 & 5.68  & 1.5 & 44.1 & 12.4 & 3.54 & \nodata \cr
      23 &        3.5732501 &       -30.383501 & 1.13(0.09) & 44.18 & 1.498 & 2.81 & 51.3  & 2.9 & 29.2 & 7.34 & 3.97 & ALMA4 \cr
      24 &        3.5757084 &       -30.426556 & 1.03(0.17) & 43.95 & 2.43(2.14,2.50) & 2.53 & 13.4  & 1.7 & 8.51 & 1.59 & 5.32 & \nodata \cr
      25 &        3.5137918 &       -30.345861 & 1.03(0.32) & 43.87 & 3.46(3.23,3.51) & 2.15 & 35.9  & 1.4 & 1.71 & 0.23 & 7.25 & \nodata \cr
      26 &        3.6100416 &       -30.355389 & 1.01(0.29) & 44.00 & 2.16(2.02,2.69) & 3.28 & 81.7  & 1.3 & 11.2 & 1.84 & 6.06 & \nodata \cr
      27 &        3.5632915 &       -30.418694 & 0.92(0.19) & 43.81 & 2.79(1.67,3.20) & 3.73 & 8.20  & 1.5 & 3.58 & 0.61 & 5.82 & \nodata \cr
      28 &        3.5102501 &       -30.375473 & 0.92(0.27) & 43.73 & 1.68(1.68,1.99) & 2.64 & 23.3  & 1.5 & 11.8 & 2.42 & 4.88 & \nodata \cr
      29 &        3.5872500 &       -30.423056 & 0.89(0.15) & 43.91 & 2.53(2.44,2.63) & 1.59 & 16.2  & 1.9 & 5.52 & 1.31 & 4.21 & \nodata \cr
      30 &        3.5237501 &       -30.371445 & 0.89(0.18) & 43.84 & 2.36(2.09,2.67) & 2.54 & 36.9  & 2.0 & 8.32 & 1.39 & 5.96 & \nodata \cr
      31 &        3.5863333 &       -30.425083 & 0.82(0.16) & 44.57 & 3.62(3.52,3.67) & 2.42 & 142.  & 1.8 & 9.83 & 1.93 & 5.08 & SCUBA16 \cr
      32 &        3.5126665 &       -30.381001 & 0.82(0.26) & 44.53 & 3.91(3.87,3.96) & 2.39 & 131.  & 1.5 & 6.46 & 1.30 & 4.95 & \nodata \cr
      33 &        3.5668333 &       -30.394890 & 0.78(0.11) & 44.23 & 3.49(2.74,3.22) & 3.55 & 19.2  & 2.2 & 8.91 & 0.96 & 9.28 & \nodata \cr
      34 &        3.6214998 &       -30.393084 & 0.69(0.22) & 44.16 & 3.53(3.21,3.61) & 2.17 & 45.0  & 1.4 & 3.54 & 0.59 & 5.99 & \nodata \cr
      35 &        3.5797083 &       -30.378389 & 0.65(0.11) & 43.85 & 2.409 & 3.17 & 19.7  & 2.5 & 19.5 & 1.87 & 10.4 & ALMA5 \cr
      36 &        3.5996249 &       -30.374695 & 0.64(0.19) & 43.67 & 2.16(2.06,2.39) & 3.34 & 6.24  & 1.7 & 8.11 & 1.09 & 7.40 & \nodata \cr
      37 &        3.5812917 &       -30.380220 & 0.61(0.09) & 43.58 & 3.50(3.23,3.53) & 3.40 & 109.  & 2.9 & 3.32 & 0.13 & 25.0 & ALMA8 \cr
      38 &        3.5312083 &       -30.361279 & 0.60(0.18) & 43.97 & 1.49(1.50,1.71) & 3.55 & 0.45  & 2.0 & 30.5 & 7.90 & 3.86 & \nodata \cr
      39 &        3.5388751 &       -30.362278 & 0.45(0.14) & 44.10 & 3.28(3.21,3.38) & 2.15 & 51.9  & 2.5 & 6.21 & 1.10 & 5.64 & \nodata \cr
      40 &        3.5825000 &       -30.385473 & 0.43(0.06) & 43.70 & 2.99(2.92,3.09) & 2.42 & 8.97  & 4.1 & 7.21 & 0.73 & 9.84 & ALMA1 \cr
      41 &        3.5582082 &       -30.374500 & 0.40(0.07) & 44.02 & 2.28(2.24,2.39) & 2.17 & 70.5  & 4.3 & 21.9 & 5.39 & 4.07 & \nodata \cr
      42 &        3.5543332 &       -30.372002 & 0.33(0.08) & 44.00 & 2.30(2.25,2.68) & 3.20 & 10.8  & 3.9 & 22.5 & 4.42 & 5.08 & \nodata \cr
      43 &        3.5442917 &       -30.368055 & 0.20(0.06) & 44.05 & 2.56(2.40,2.66) & 2.02 & 0.60  & 5.6 & 29.3 & 4.55 & 6.43 & \nodata \cr
      44 &        3.5430834 &       -30.369110 & 0.11(0.03) & 43.80 & 2.52(2.44,2.65) & 1.00 & 0.00  & 9.7 & 18.9 & 5.15 & 3.68 & \nodata \cr
\enddata
\tablecomments{The columns are (1) source number, 
(2) and (3) R.A. and decl. of the JWST NIRCam \fluxa\ $>1~\mu$Jy and \jwratio\ $>3.5$ prior,
(4) demagnifed \afluxb\ flux, 
(5) logarithm of the demagnified luminosity,
(6) redshift (see Section~\ref{secdata};
spectroscopic has three digits after the decimal point,
while photometric has two digits after the decimal point, with the 16th and 84th percentiles of the posterior
given in parentheses),
(7) $A_V$ from EAZY,
(8) SFR from EAZY,
(9) magnification from the \citet{weaver23} catalog,
(10), (11), and (12) observed \fluxa\ and \fluxb\ (not demagnified; these fluxes are from the \citealt{paris23} catalog)
and their ratio,
(13) direct ALMA source match from Table~\ref{tabALMA}, or direct 
SCUBA-2 source match from Table~\ref{scuba2_acc},
when available.
}
\end{deluxetable*}

%---------------------------------------------------------------------
% FIGURE 13 ; final_images.pro
%---------------------------------------------------------------------
\begin{figure*}
\includegraphics[width=1.8in,angle=0]{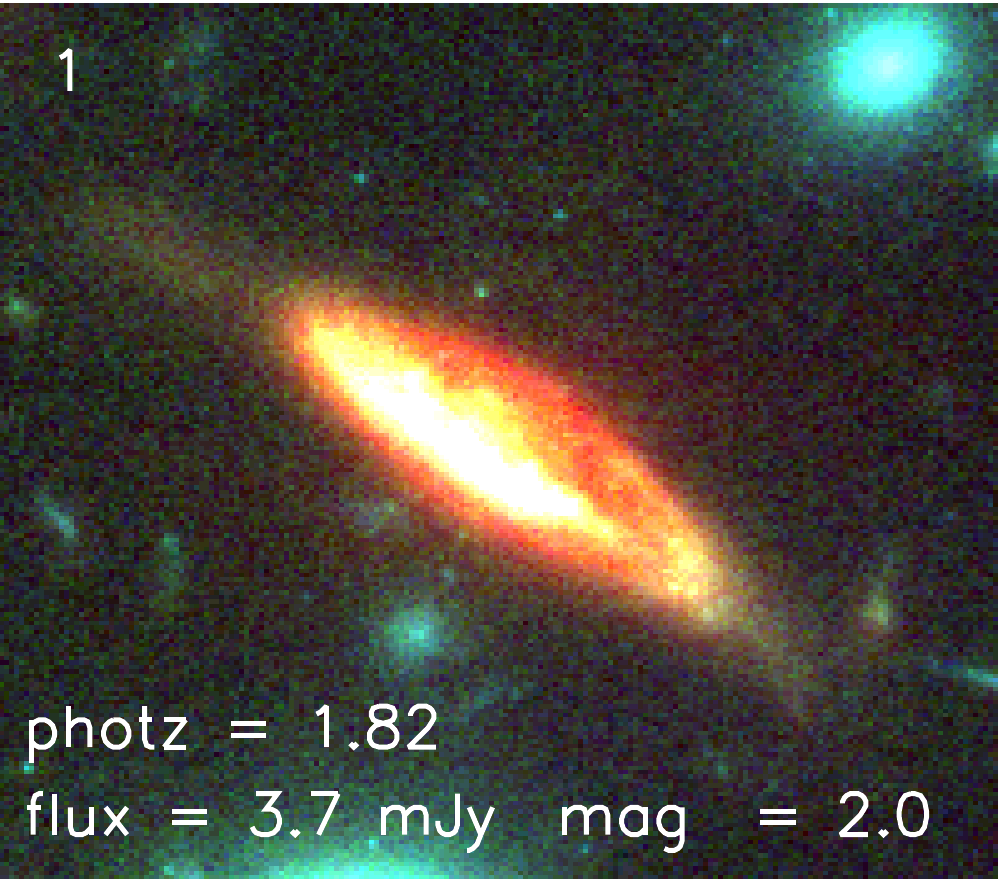}
\hskip -0.8cm
\includegraphics[width=1.8in,angle=0]{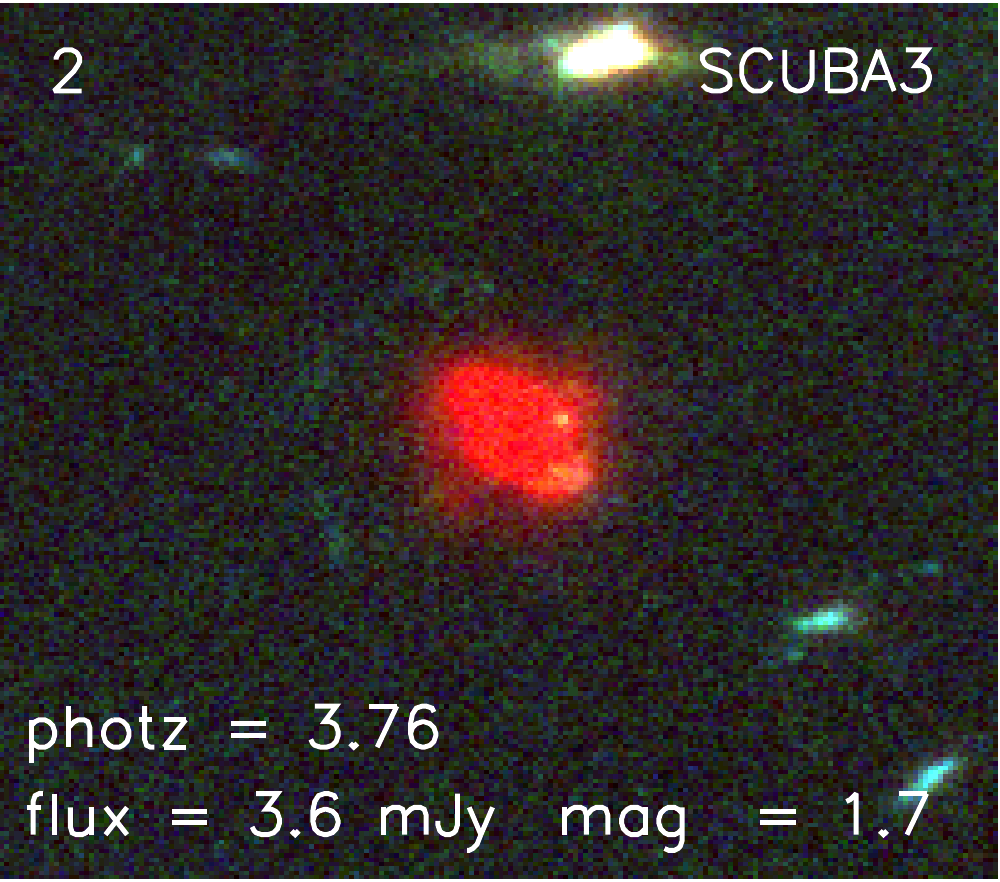}
\hskip -0.8cm
\includegraphics[width=1.8in,angle=0]{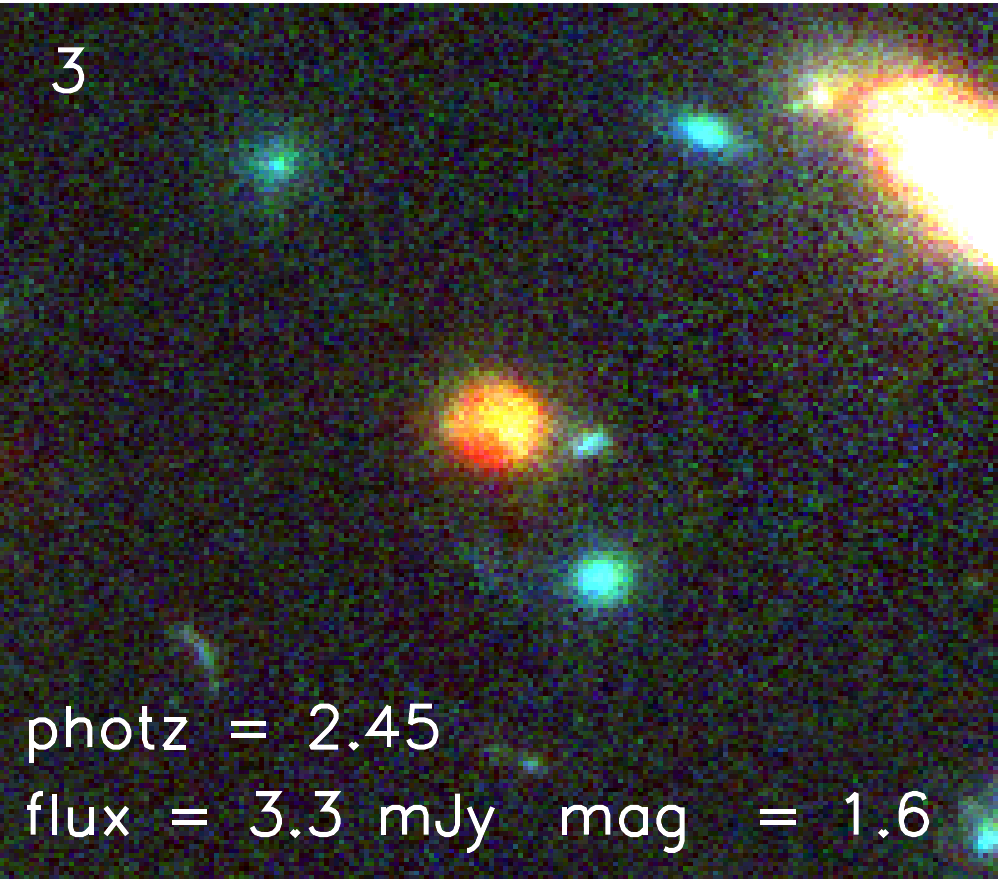}
\hskip -0.8cm
\includegraphics[width=1.8in,angle=0]{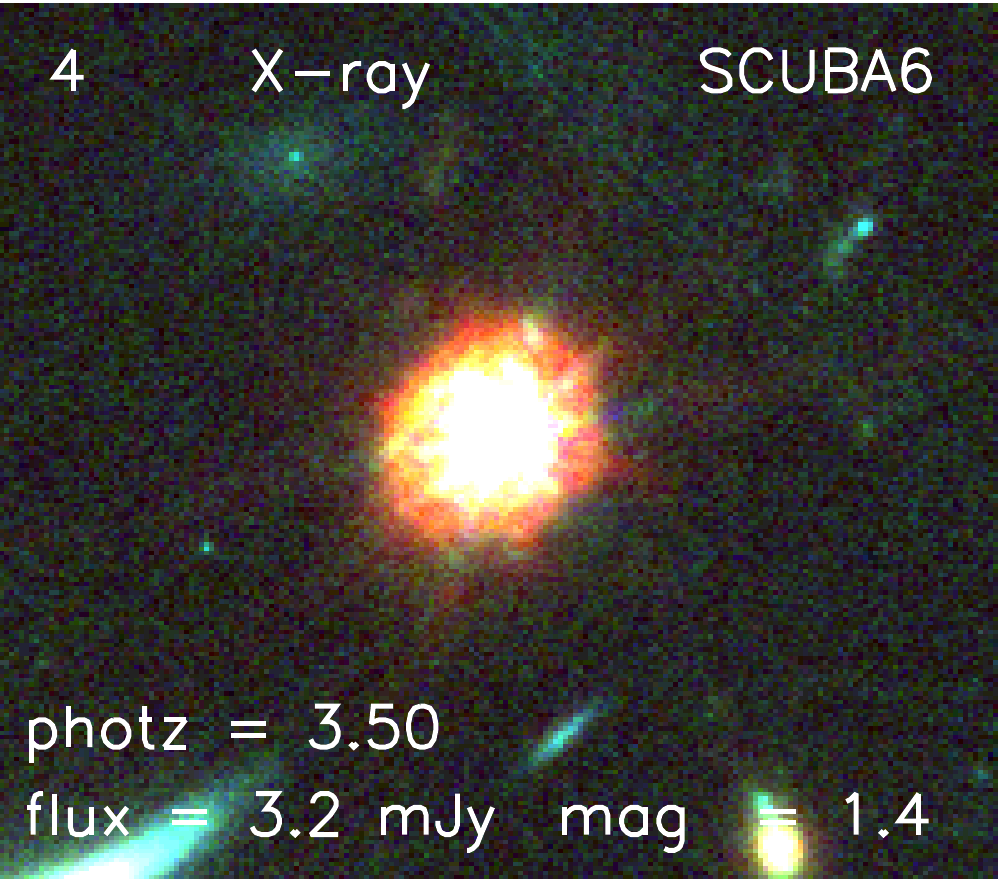}
\hskip -0.8cm
\includegraphics[width=1.8in,angle=0]{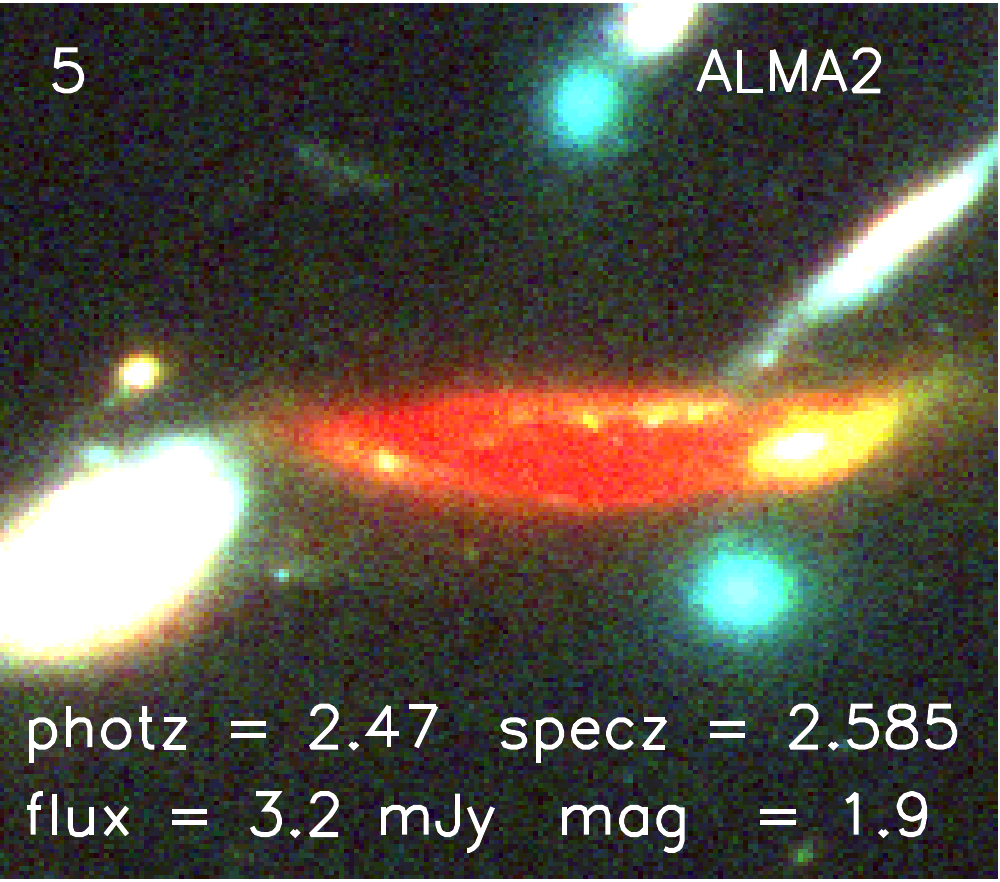}
\hskip -0.8cm
\includegraphics[width=1.8in,angle=0]{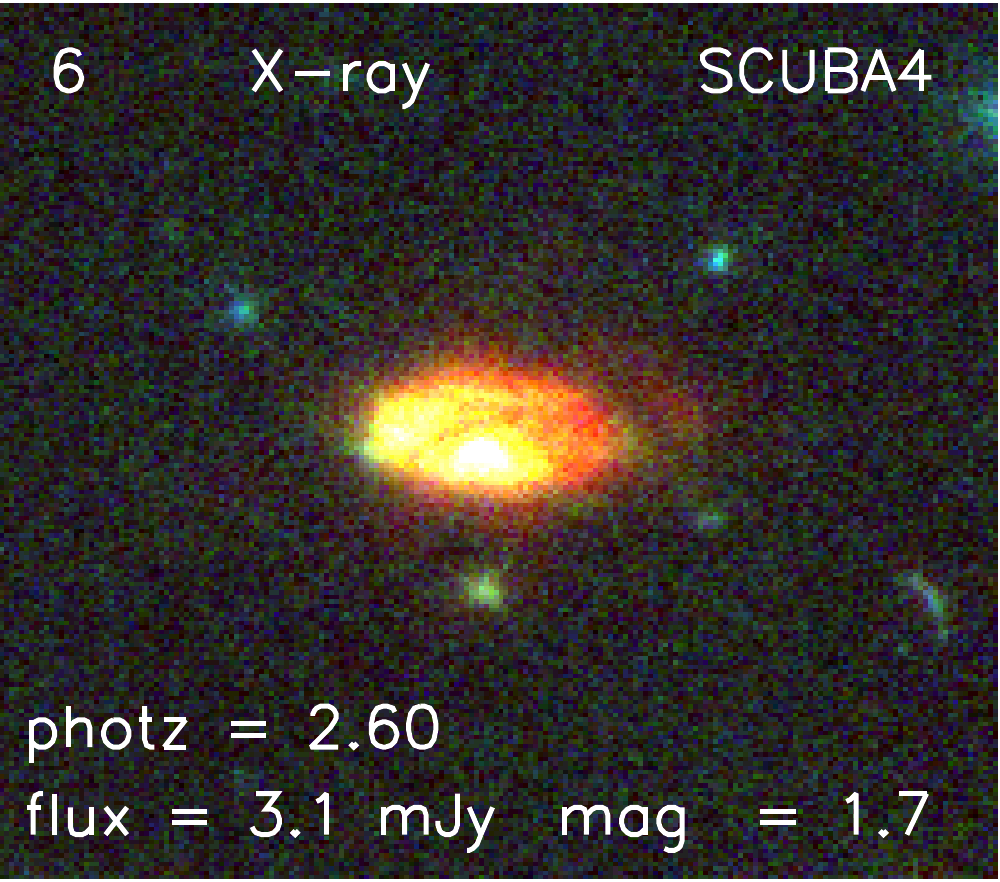}
\hskip -0.8cm
\includegraphics[width=1.8in,angle=0]{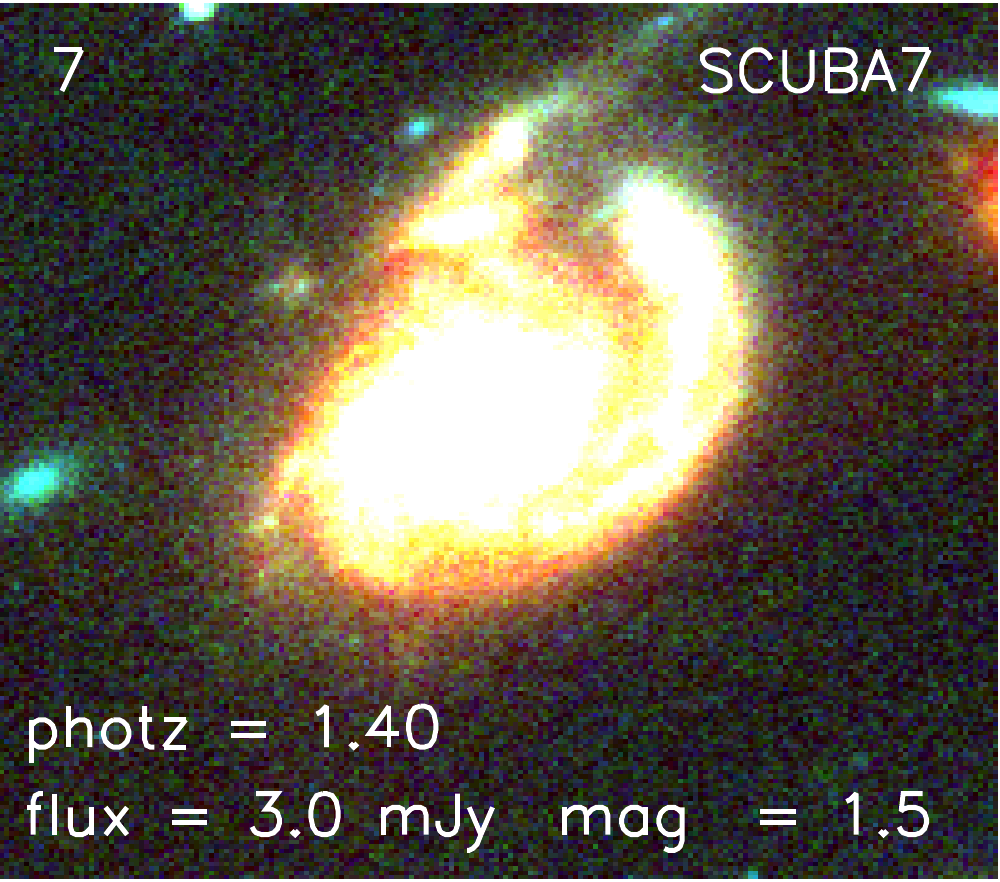}
\hskip -0.8cm
\includegraphics[width=1.8in,angle=0]{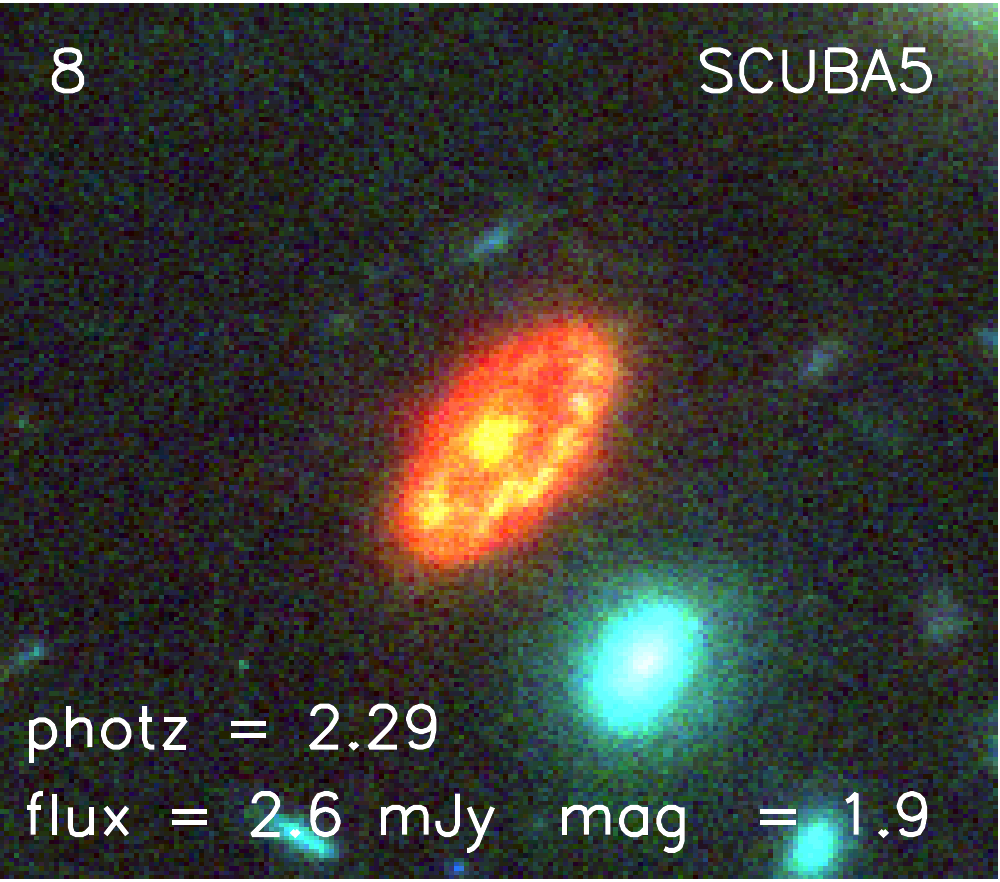}
\hskip -0.8cm
\includegraphics[width=1.8in,angle=0]{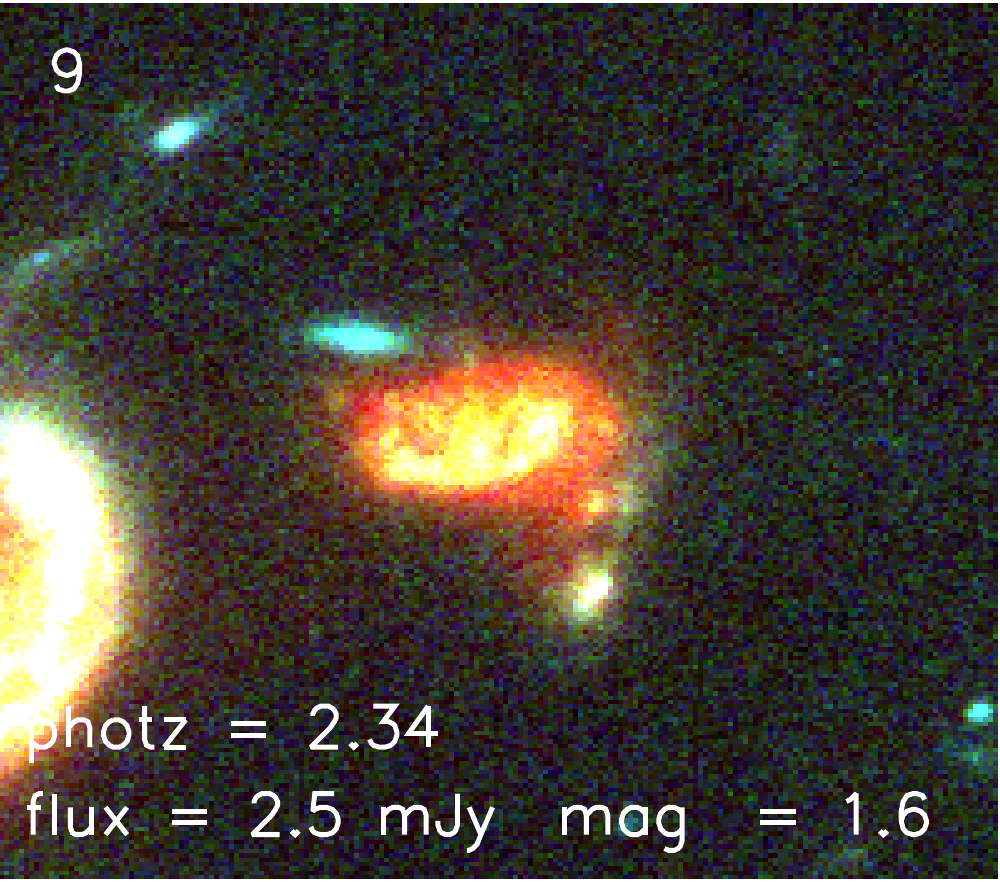}
\hskip -0.8cm
\includegraphics[width=1.8in,angle=0]{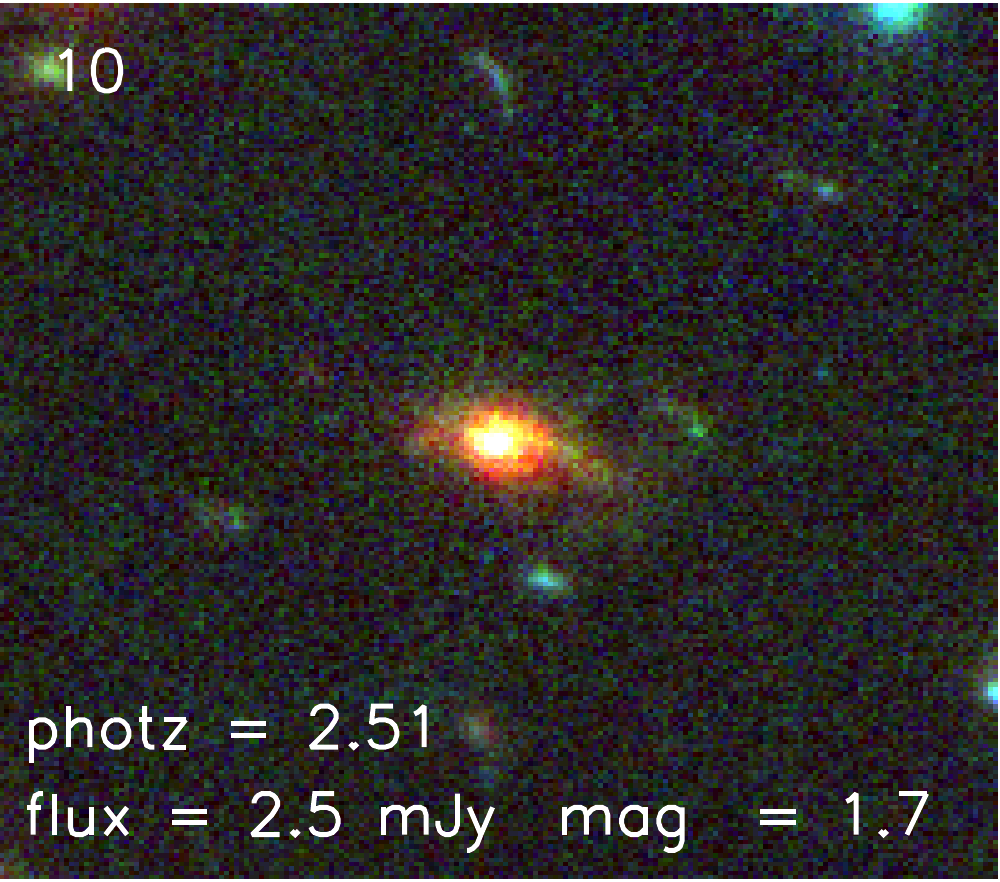}
\hskip -0.8cm
\includegraphics[width=1.8in,angle=0]{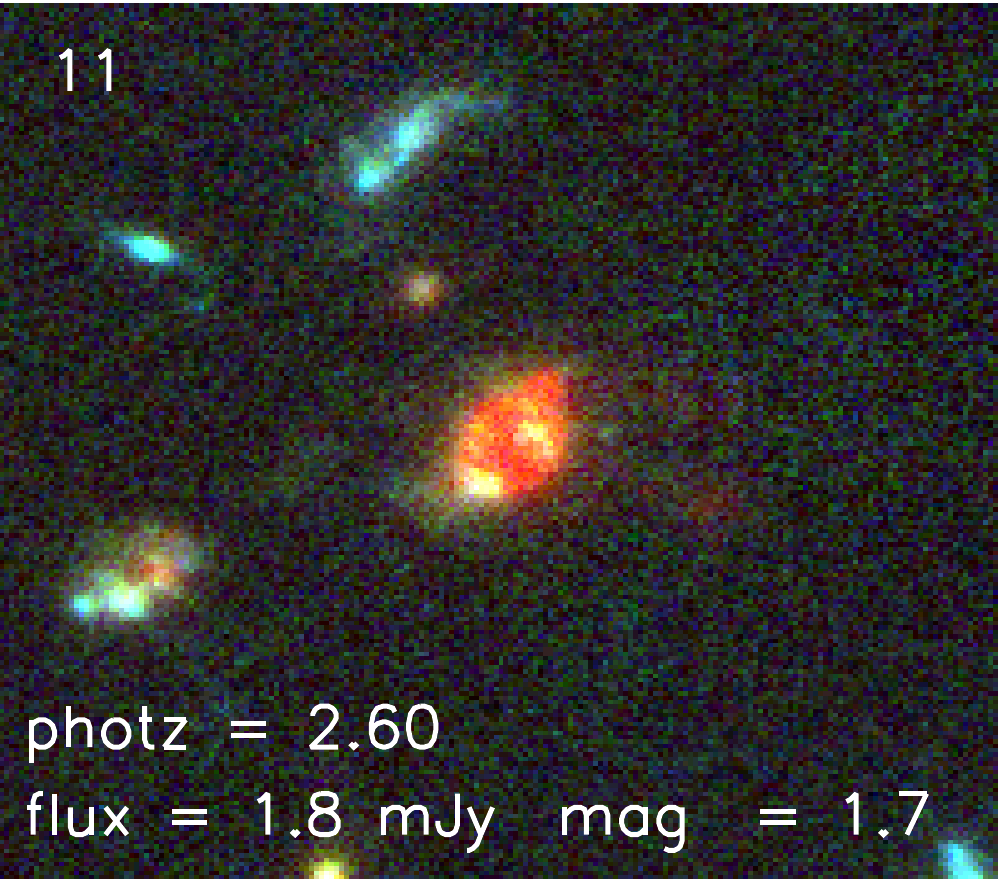}
\hskip -0.8cm
\includegraphics[width=1.8in,angle=0]{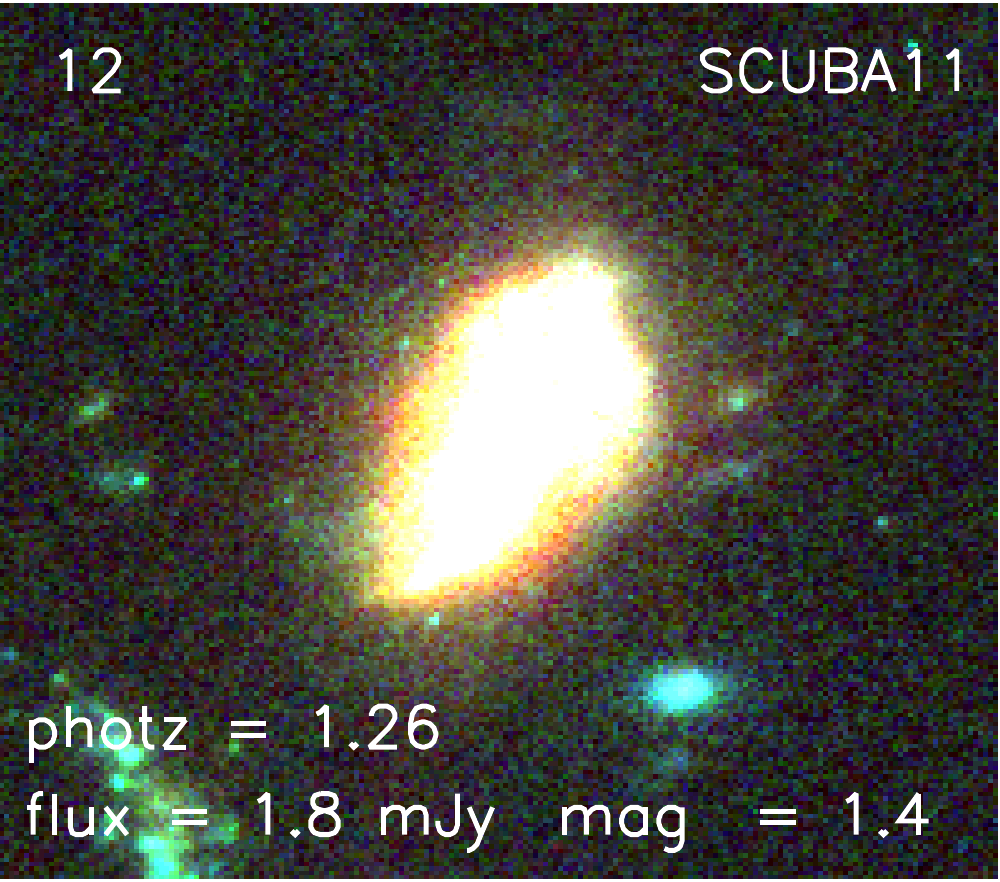}
\hskip -0.8cm
\includegraphics[width=1.8in,angle=0]{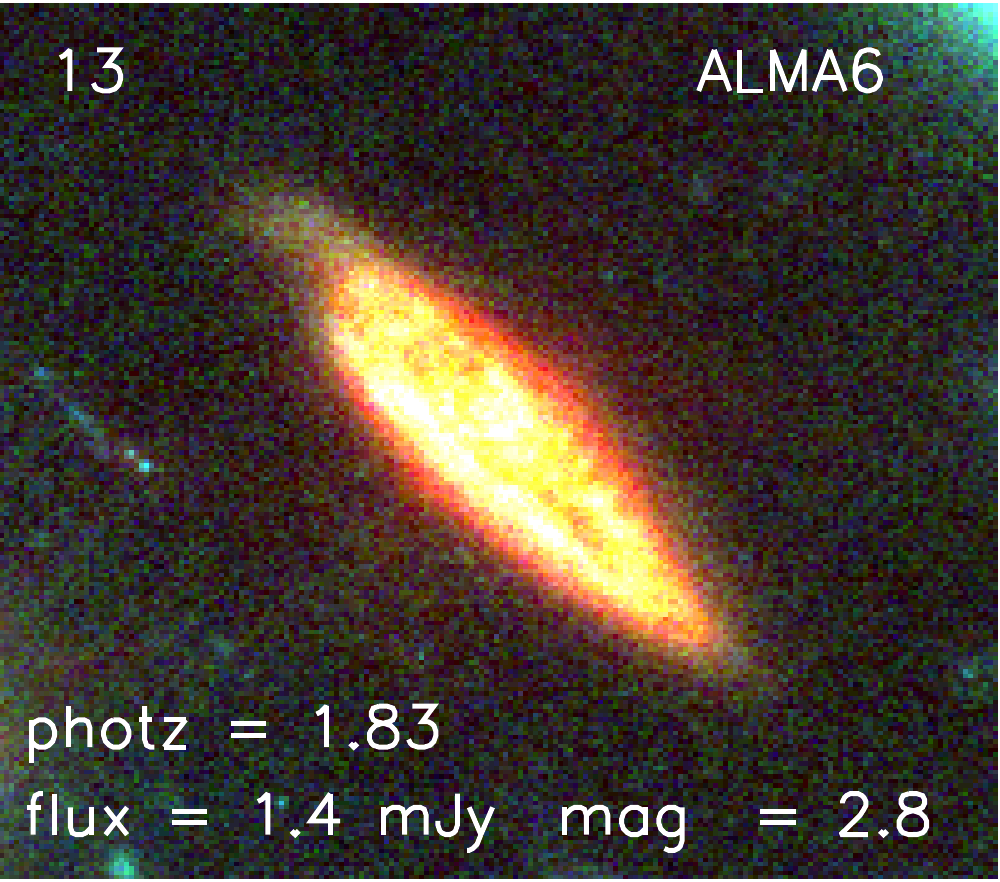}
\hskip -0.8cm
\includegraphics[width=1.8in,angle=0]{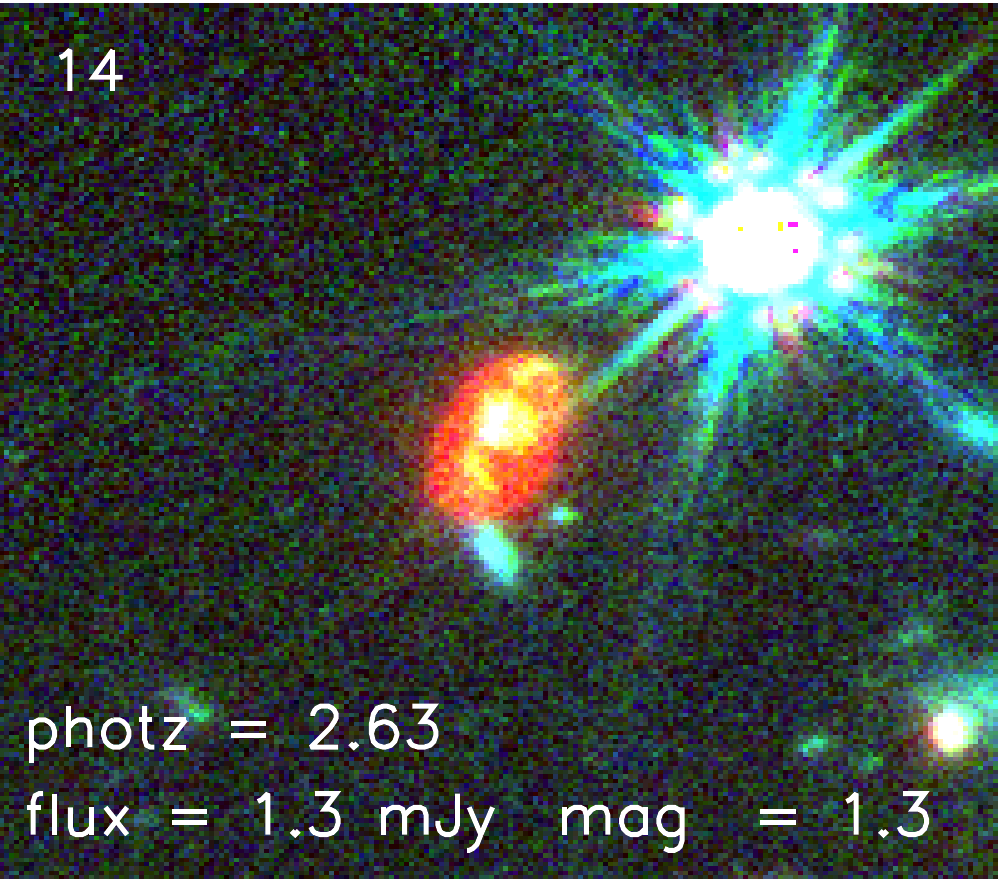}
\hskip -0.8cm
\includegraphics[width=1.8in,angle=0]{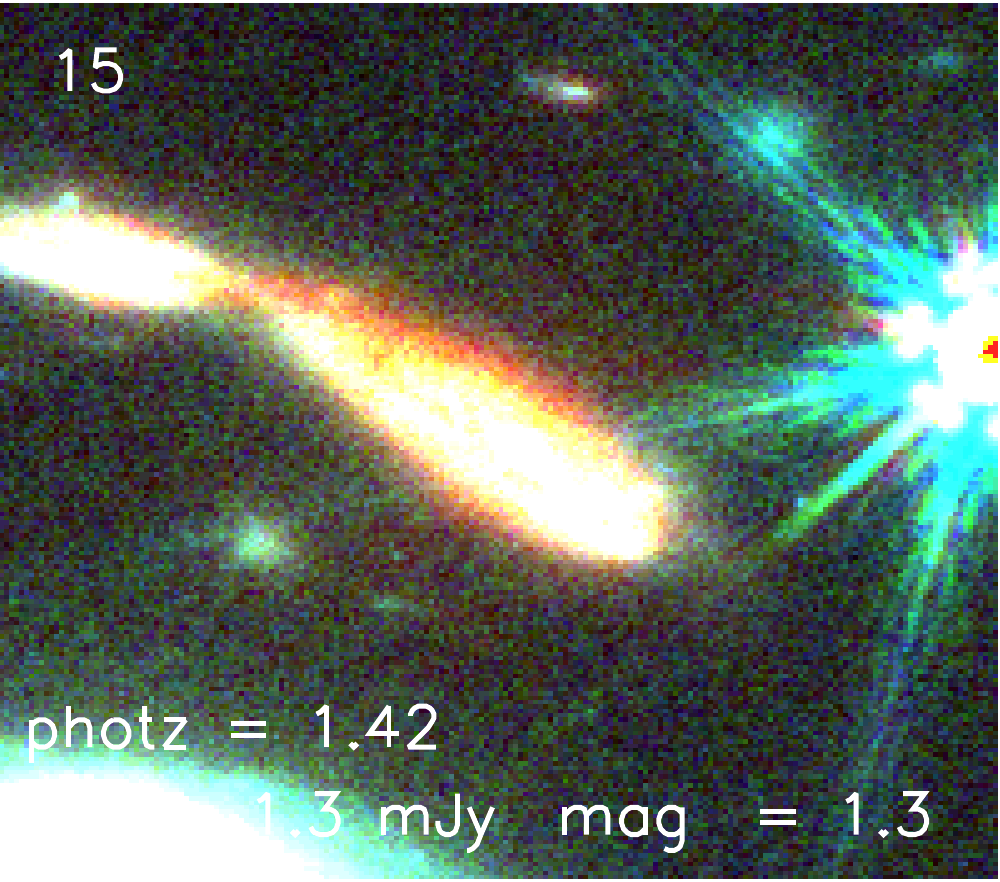}
\hskip -0.8cm
\includegraphics[width=1.8in,angle=0]{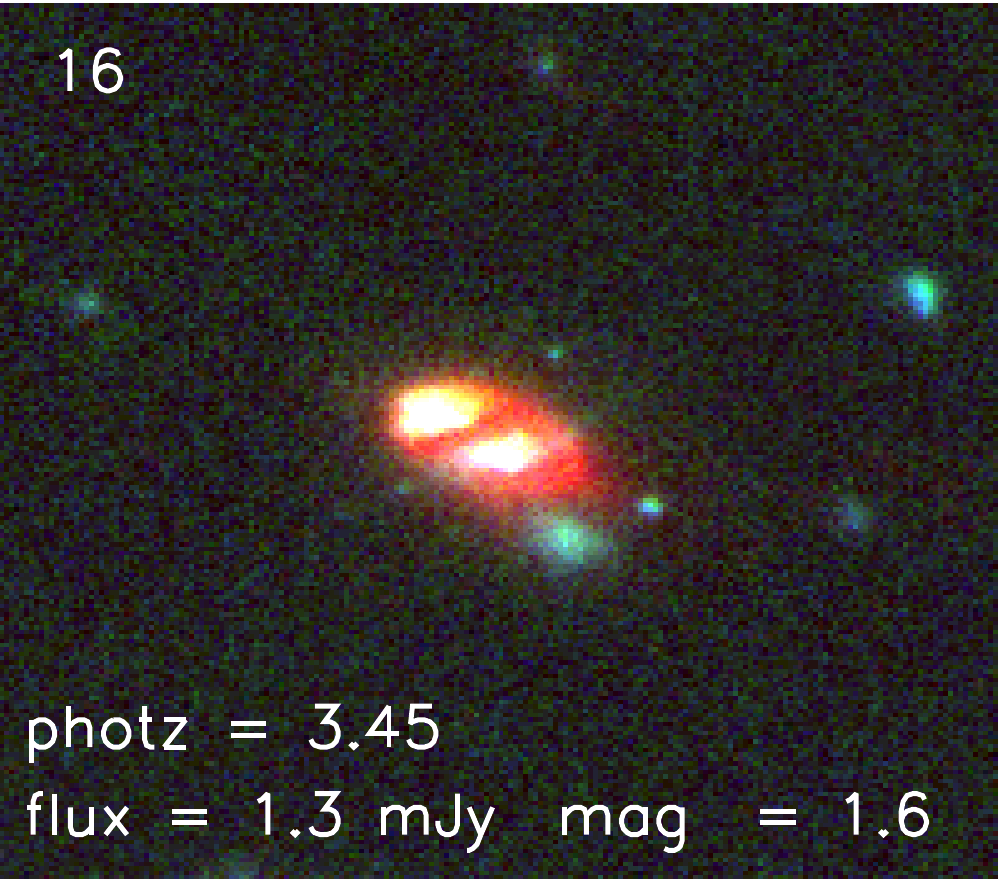}
\hskip -0.8cm
\includegraphics[width=1.8in,angle=0]{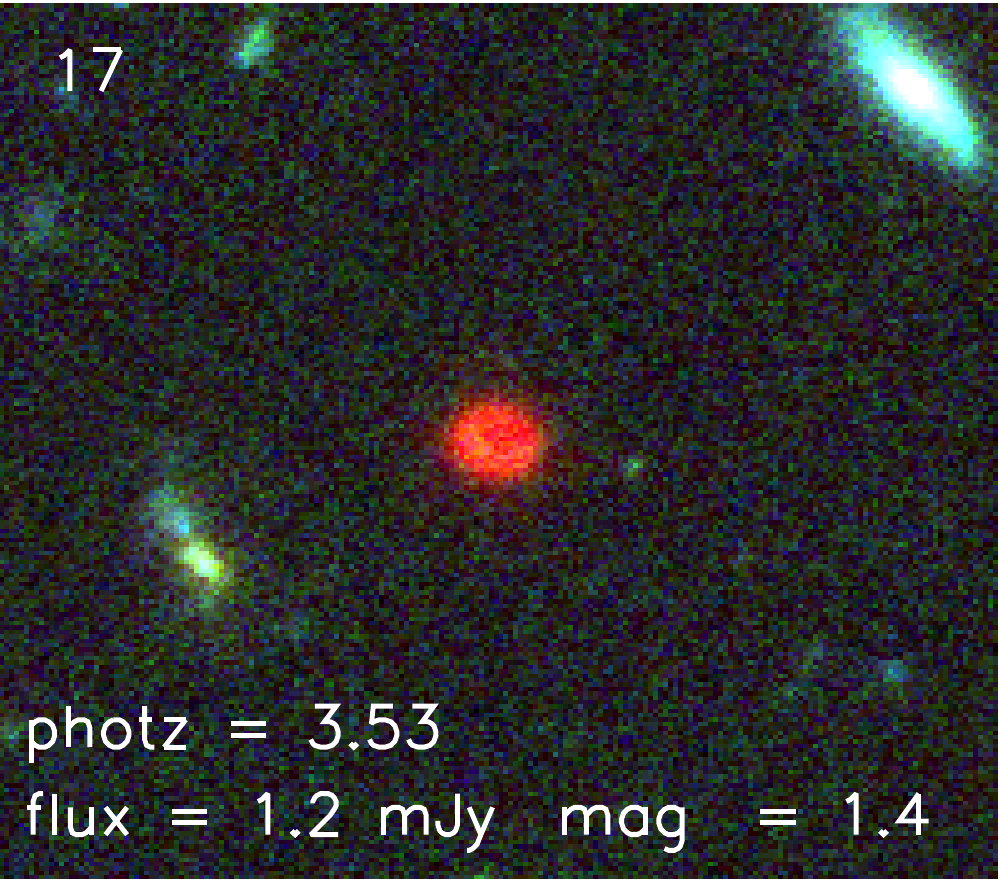}
\hskip -0.2cm
\includegraphics[width=1.8in,angle=0]{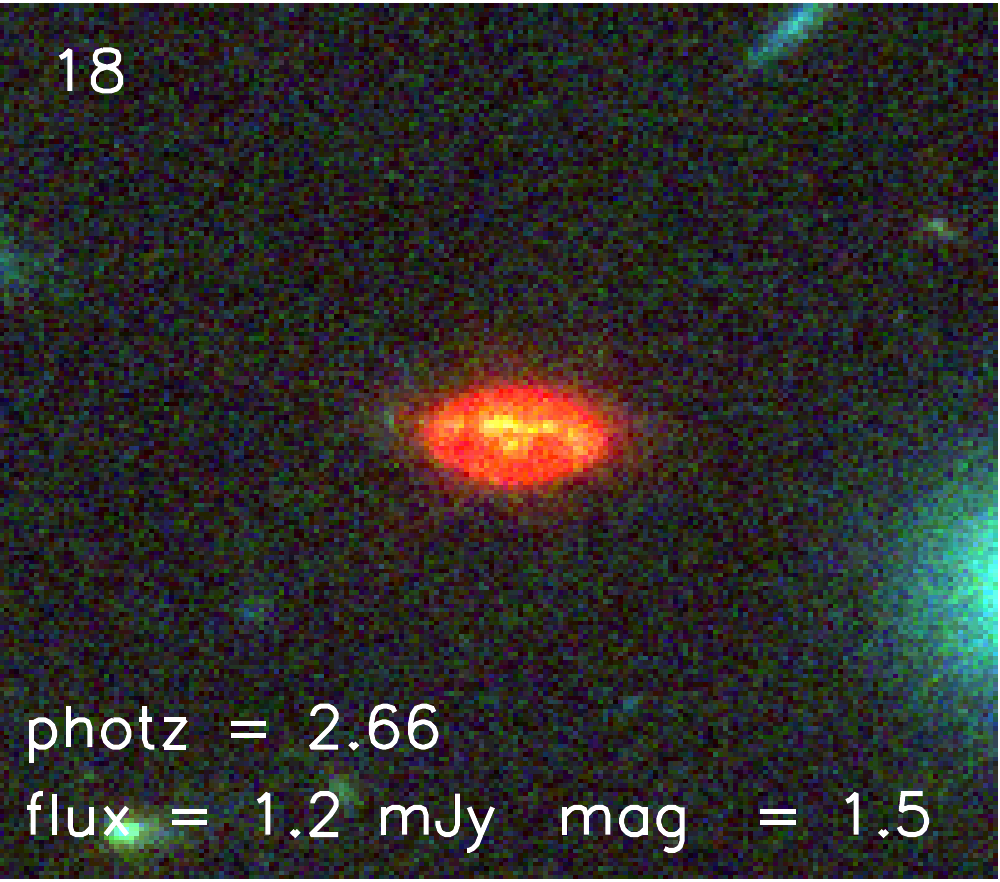}
\hskip -0.2cm
\includegraphics[width=1.8in,angle=0]{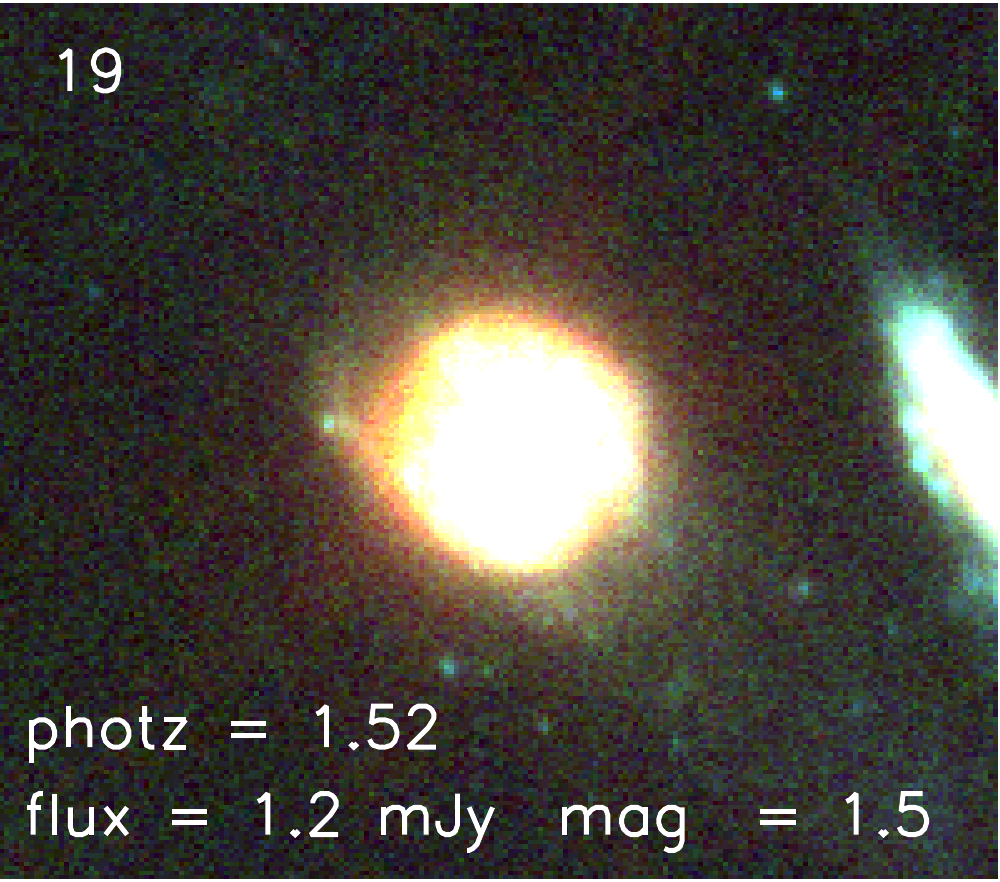}
\hskip -0.2cm
\includegraphics[width=1.8in,angle=0]{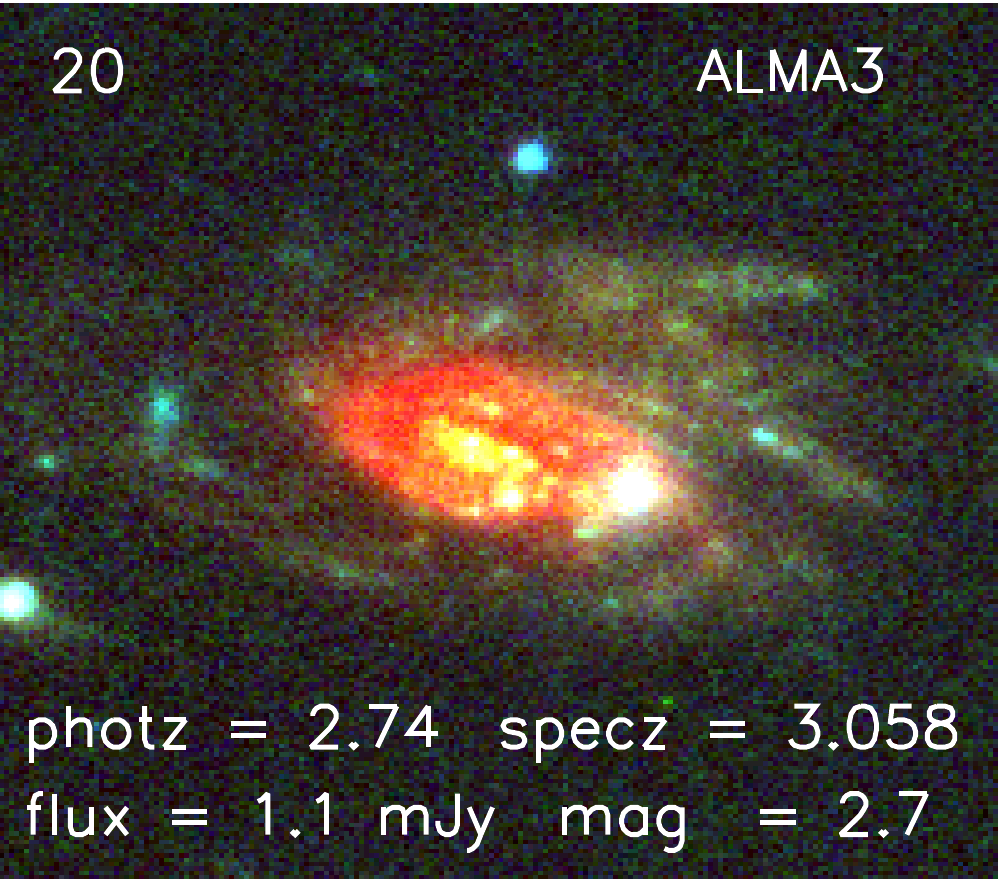}
\caption{Three-color JWST NIRCam images (blue = F115W, green = F150W, 
and red = F444W) for the 44 $>3\sigma$ SCUBA-2 \afluxb\ detected JWST NIRCam 
\fluxa\ $>1~\mu$Jy and
\jwratio\ $>3.5$ sources in the UNCOVER area (Table~\ref{fintab}). The demagnified \afluxb\ flux
is shown in the lower left, together with the photometric (photz) and spectroscopic (specz)
redshifts.
Sources with a Chandra detection are labeled X-ray in the upper left.
Sources with direct $>4.5\sigma$ ALMA detections (Table~\ref{tabALMA}) or
direct $>5\sigma$ SCUBA-2 detections (Table~\ref{scuba2_acc}) are labeled
with those numbers in the upper right.
The thumbnails are $6''$ on a side, or roughly 50~kpc at $z=2$.
\label{all_scuba2_images}
}
\end{figure*}

%---------------------------------------------------------------------
% FIGURE 13 (cont)
%---------------------------------------------------------------------
\begin{figure*}
\setcounter{figure}{12}
\includegraphics[width=1.8in,angle=0]{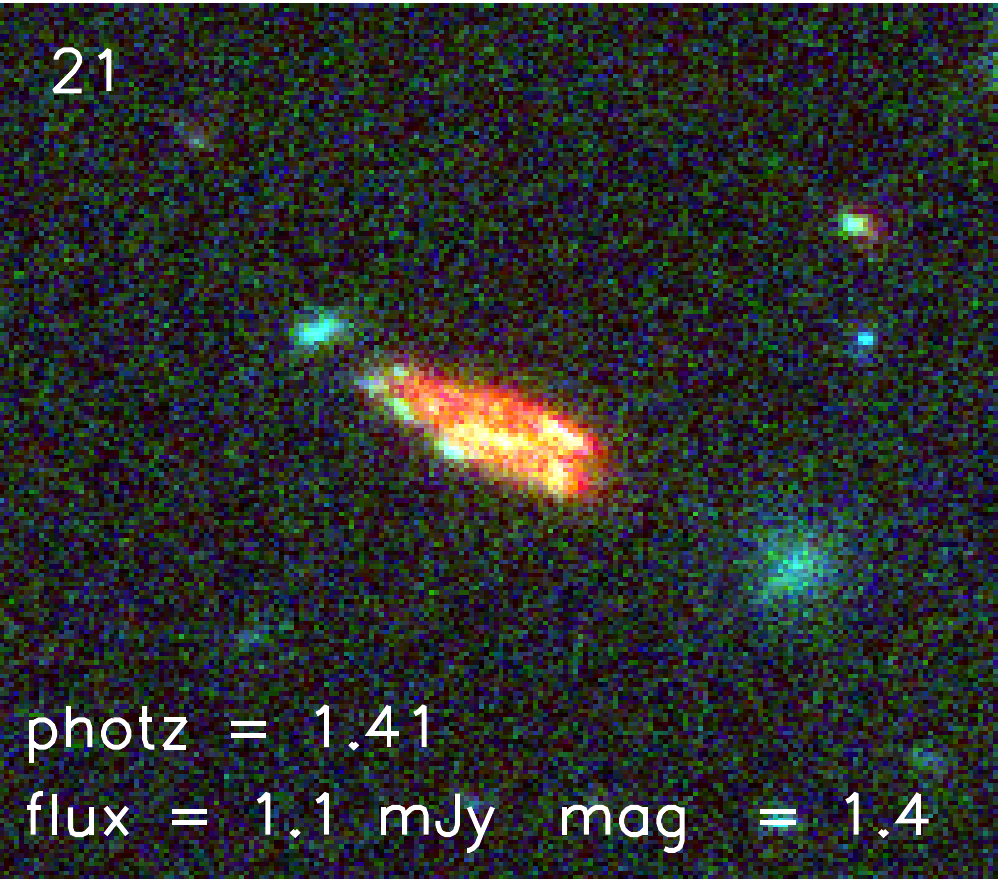}
\hskip -0.8cm
\includegraphics[width=1.8in,angle=0]{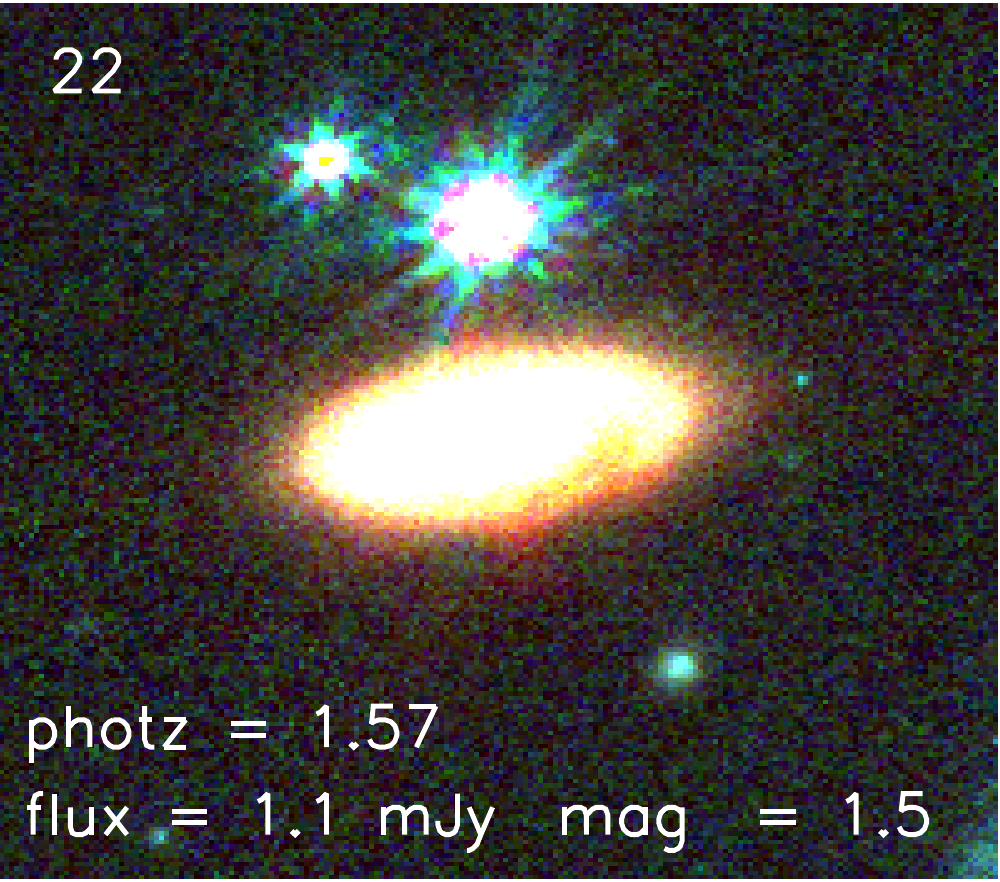}
\hskip -0.8cm
\includegraphics[width=1.8in,angle=0]{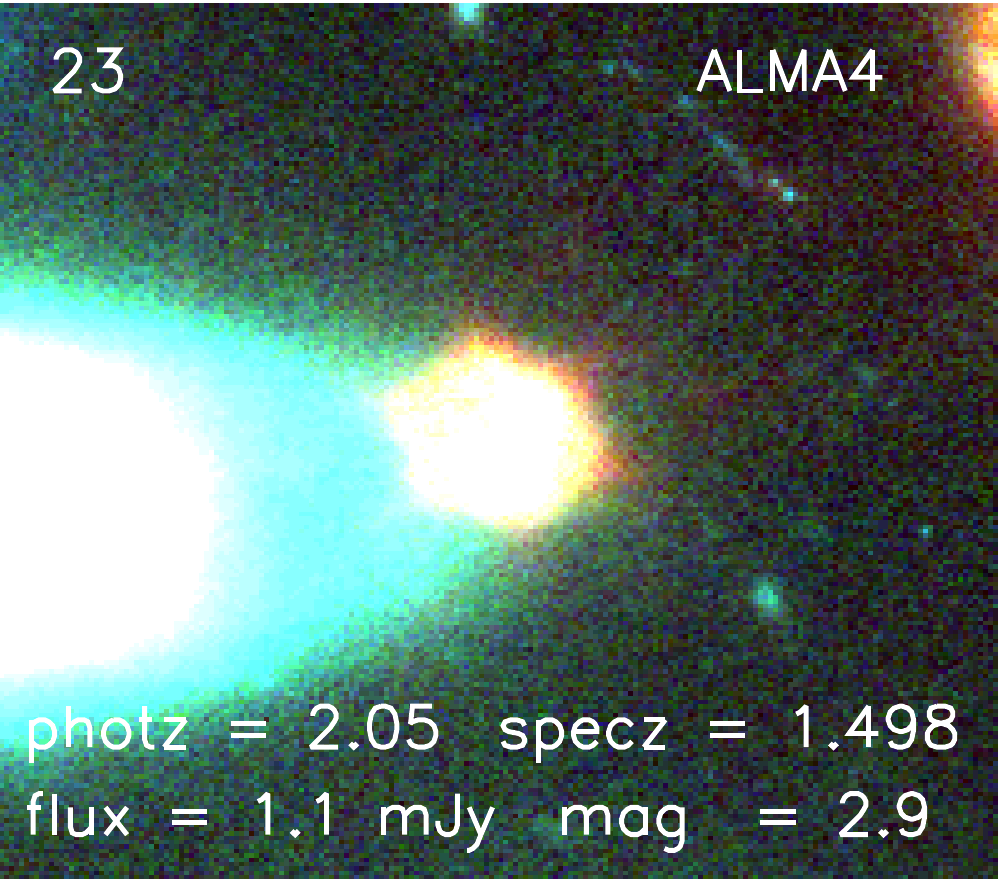}
\hskip -0.8cm
\includegraphics[width=1.8in,angle=0]{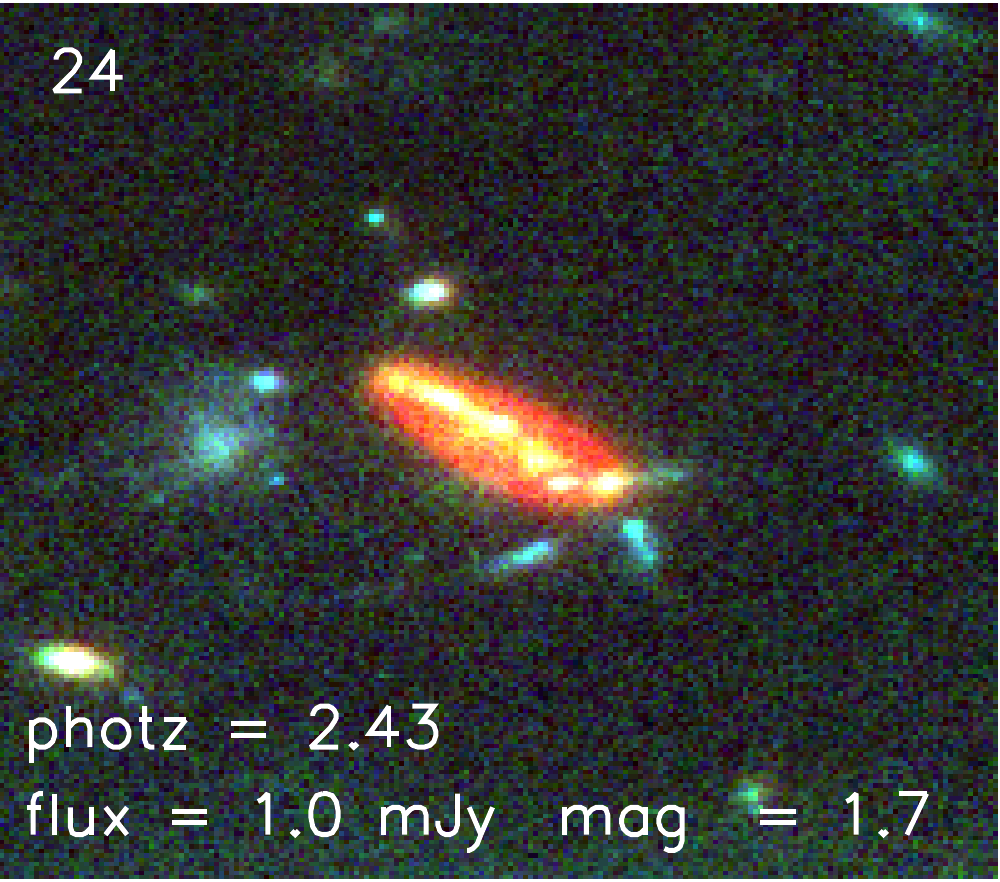}
\hskip -0.8cm
\includegraphics[width=1.8in,angle=0]{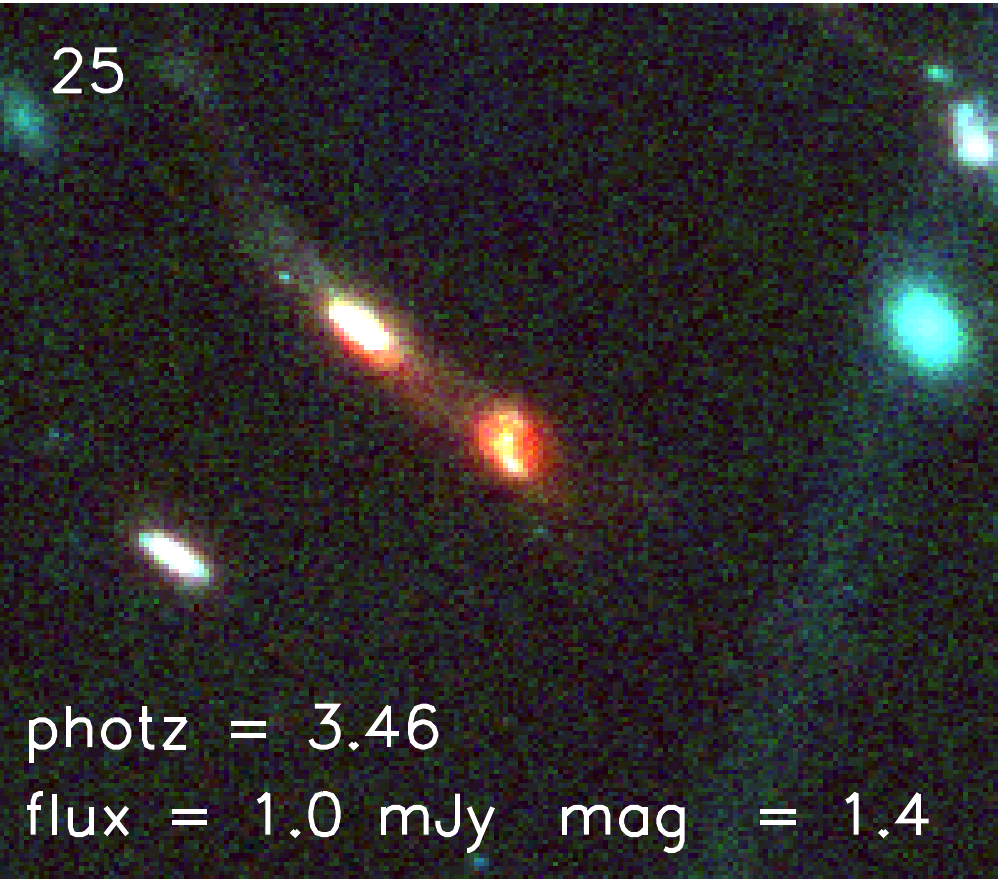}
\hskip -0.8cm
\includegraphics[width=1.8in,angle=0]{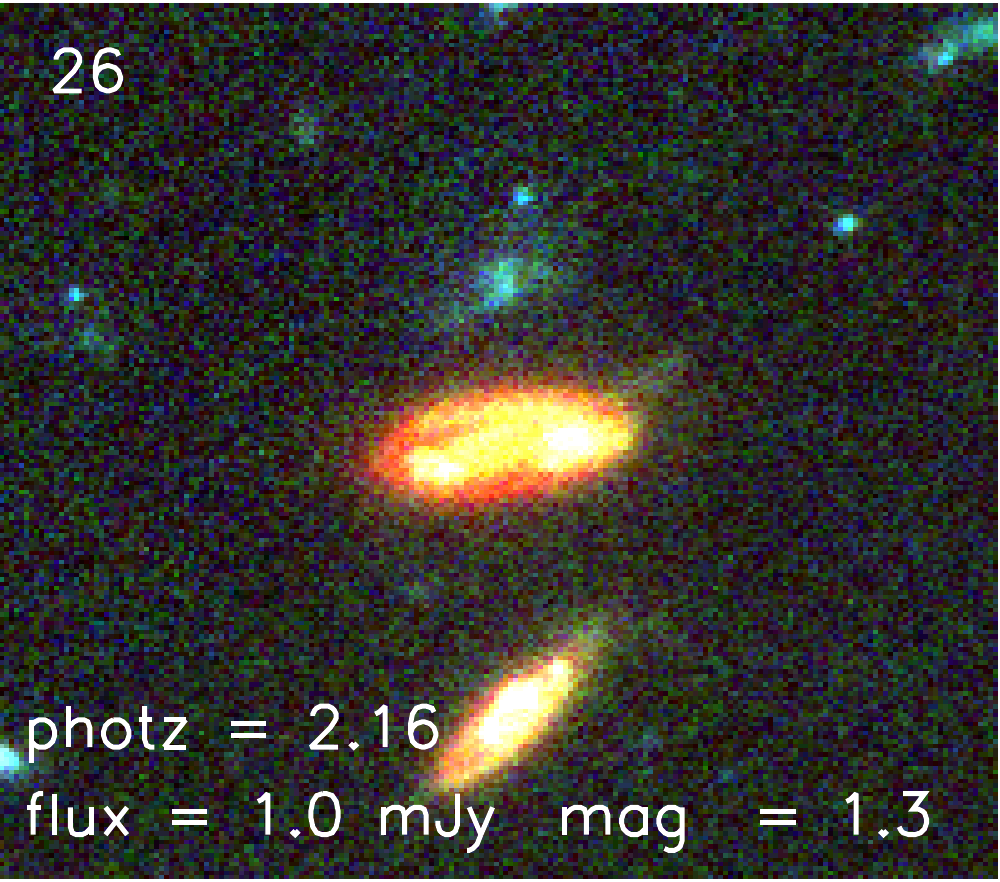}
\hskip -0.8cm
\includegraphics[width=1.8in,angle=0]{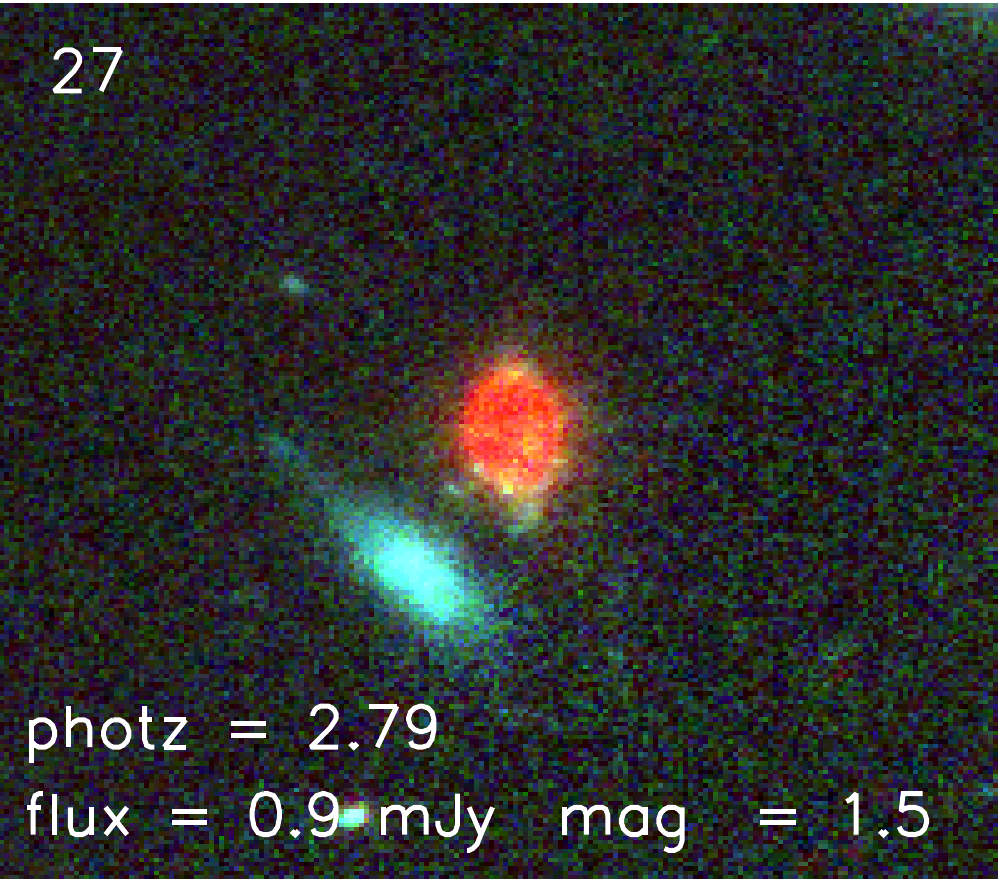}
\hskip -0.8cm
\includegraphics[width=1.8in,angle=0]{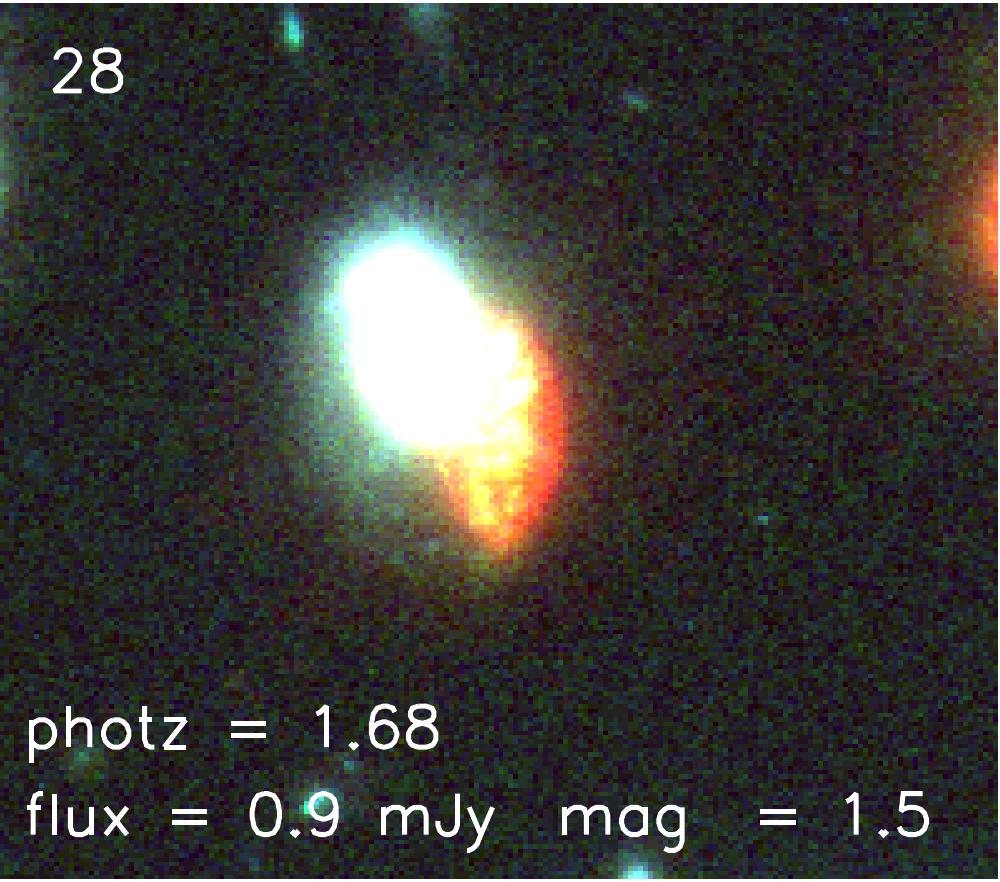}
\hskip -0.8cm
\includegraphics[width=1.8in,angle=0]{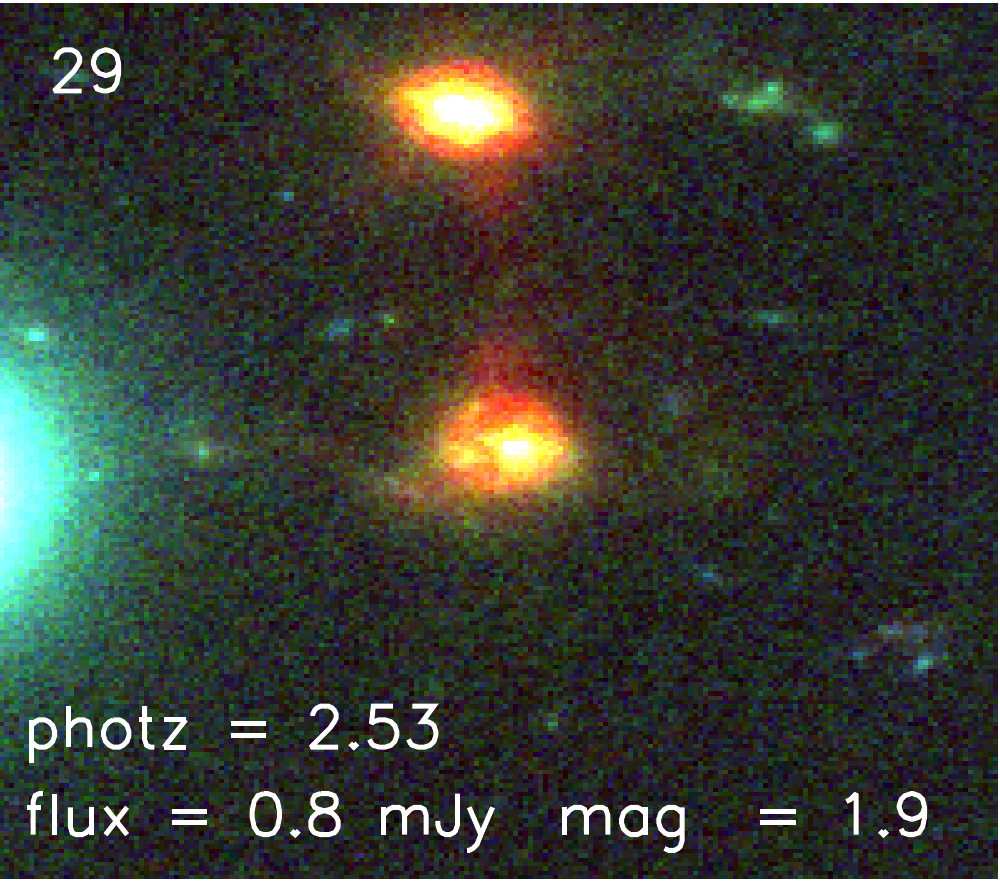}
\hskip -0.8cm
\includegraphics[width=1.8in,angle=0]{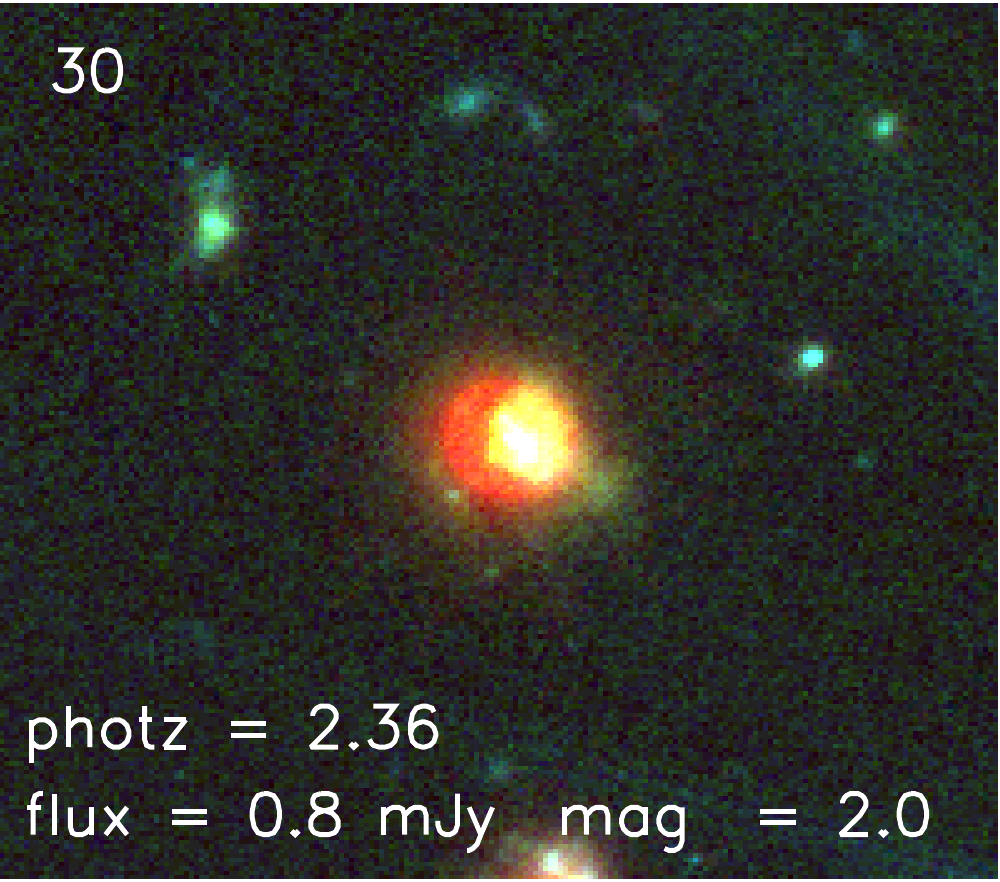}
\hskip -0.8cm
\includegraphics[width=1.8in,angle=0]{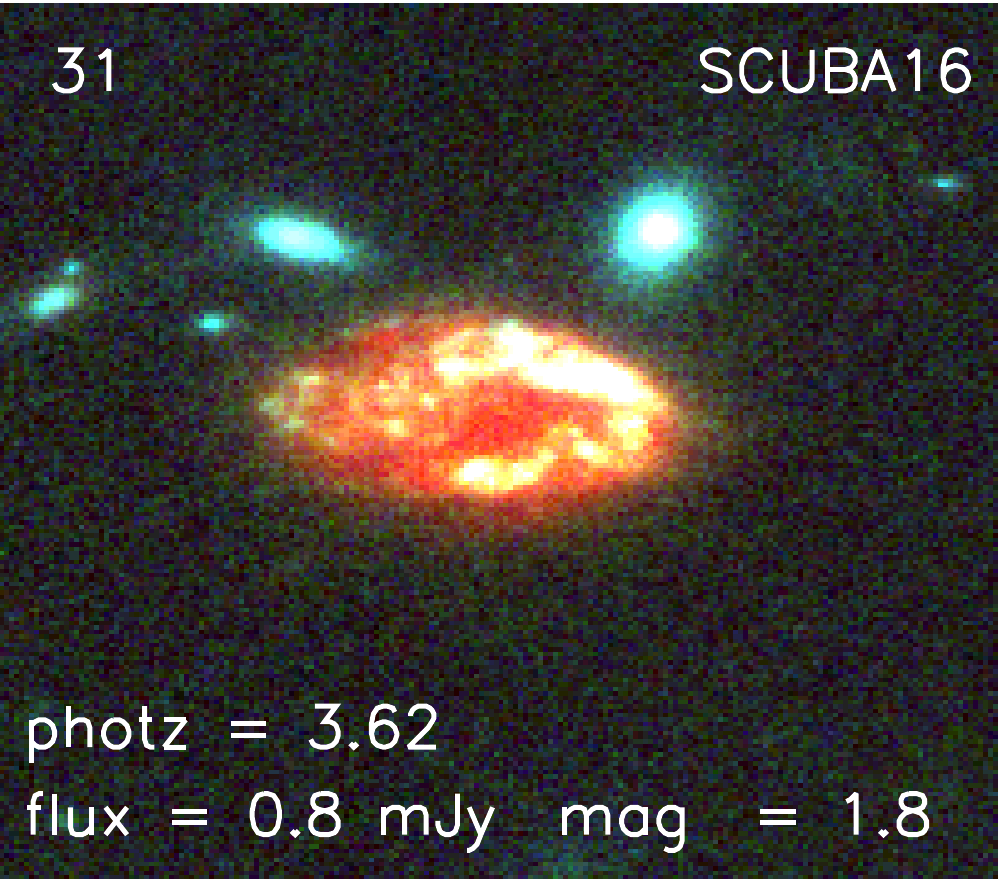}
\hskip -0.8cm
\includegraphics[width=1.8in,angle=0]{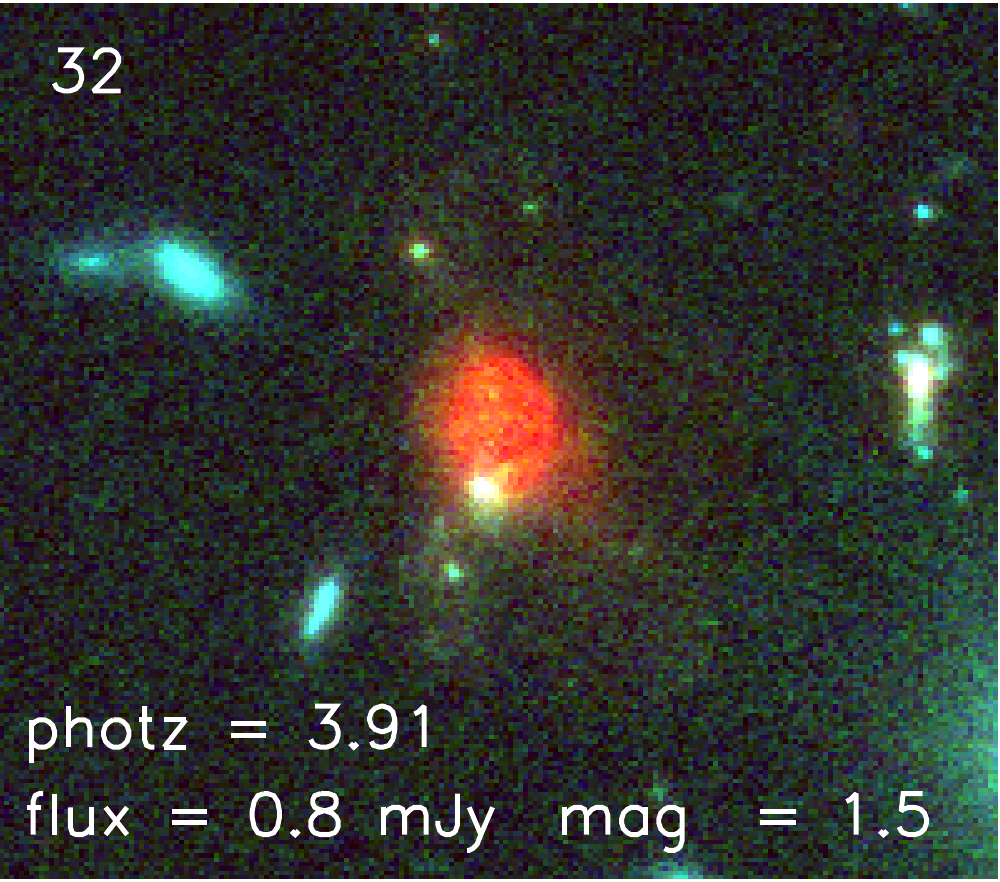}
\hskip -0.8cm
\includegraphics[width=1.8in,angle=0]{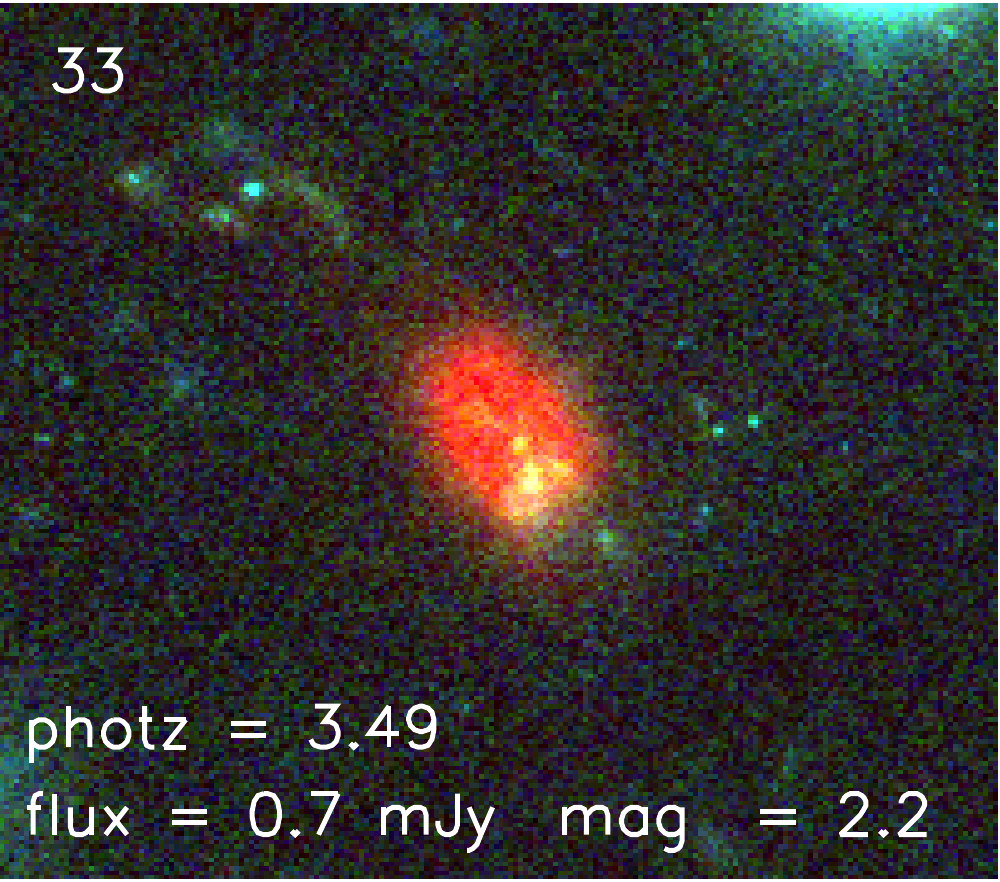}
\hskip -0.8cm
\includegraphics[width=1.8in,angle=0]{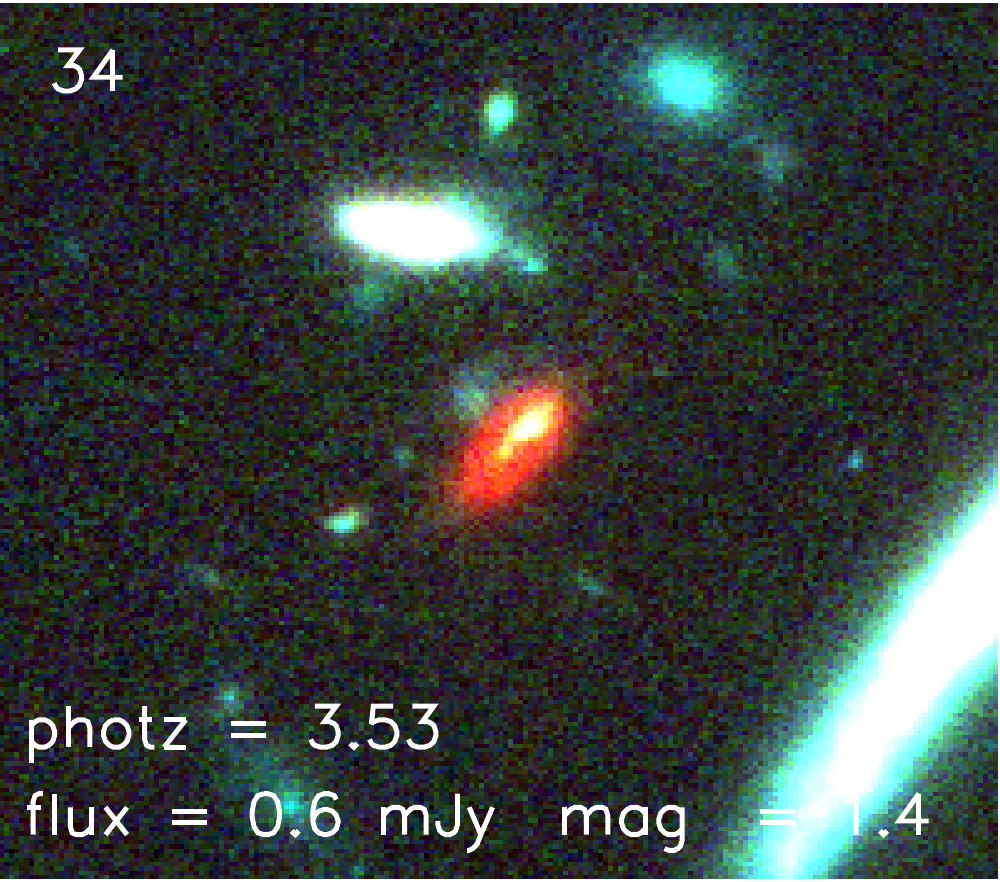}
\hskip -0.8cm
\includegraphics[width=1.8in,angle=0]{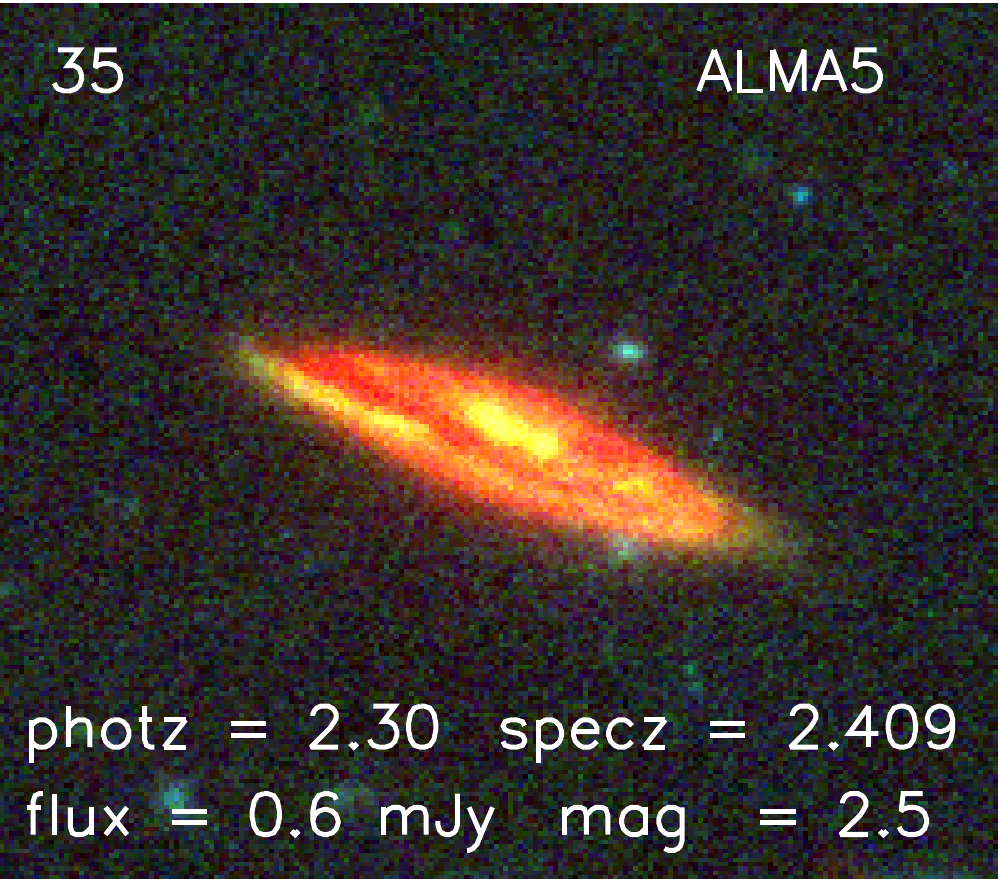}
\hskip -0.8cm
\includegraphics[width=1.8in,angle=0]{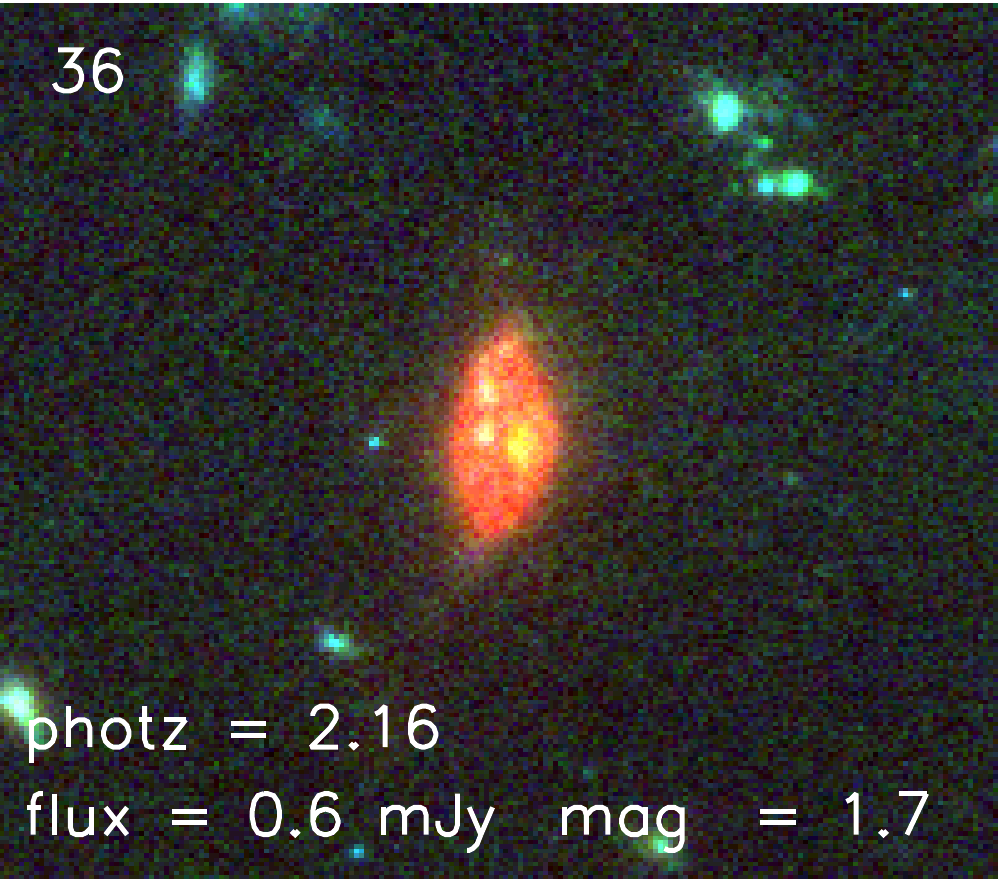}
\hskip -0.8cm
\includegraphics[width=1.8in,angle=0]{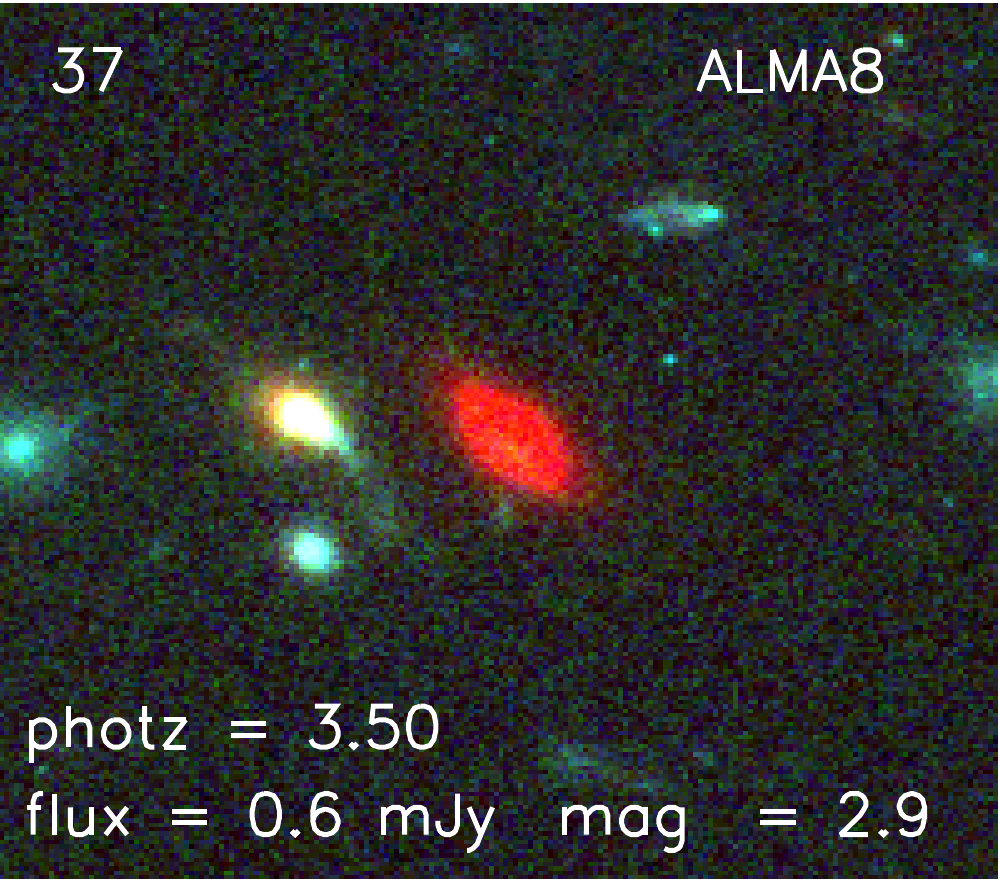}
\hskip -0.2cm
\includegraphics[width=1.8in,angle=0]{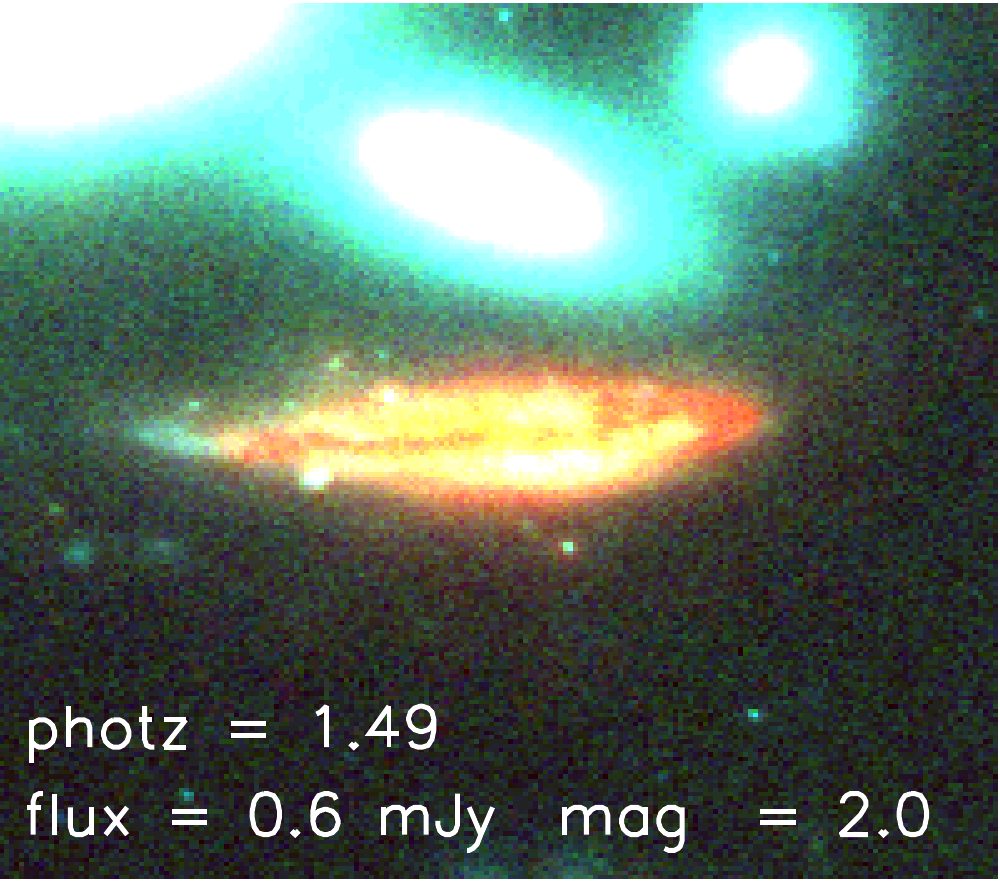}
\hskip -0.2cm
\includegraphics[width=1.8in,angle=0]{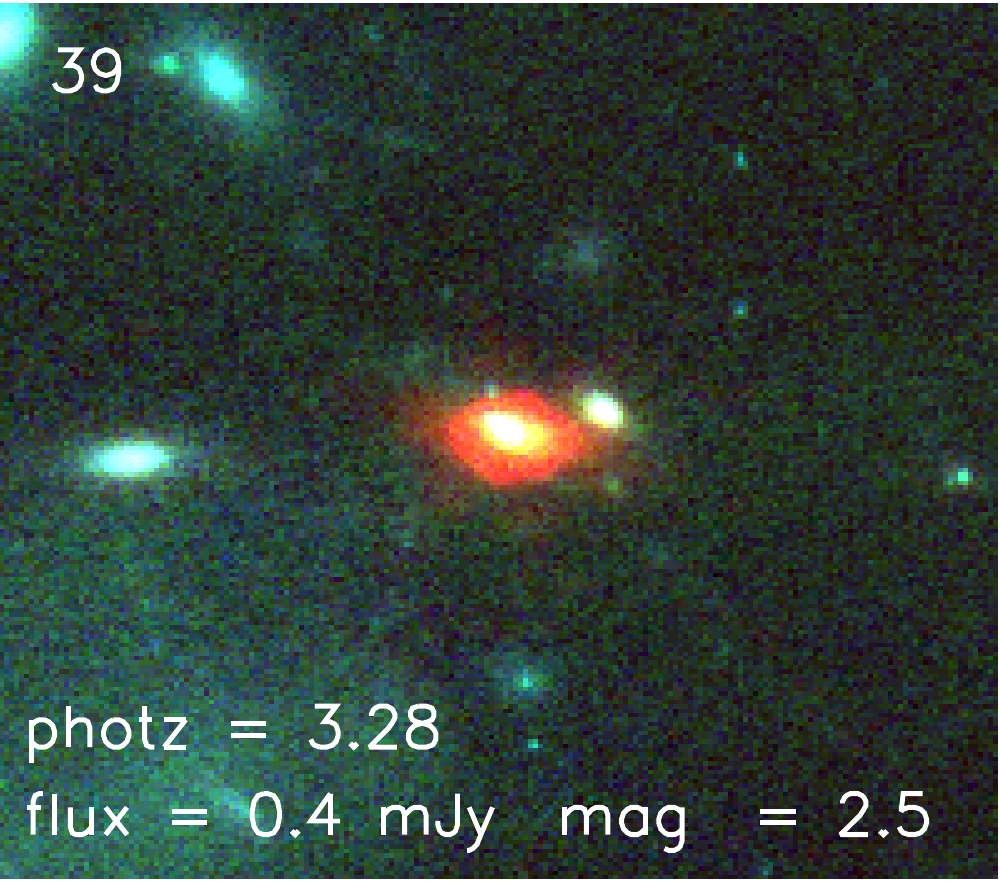}
\hskip -0.2cm
\includegraphics[width=1.8in,angle=0]{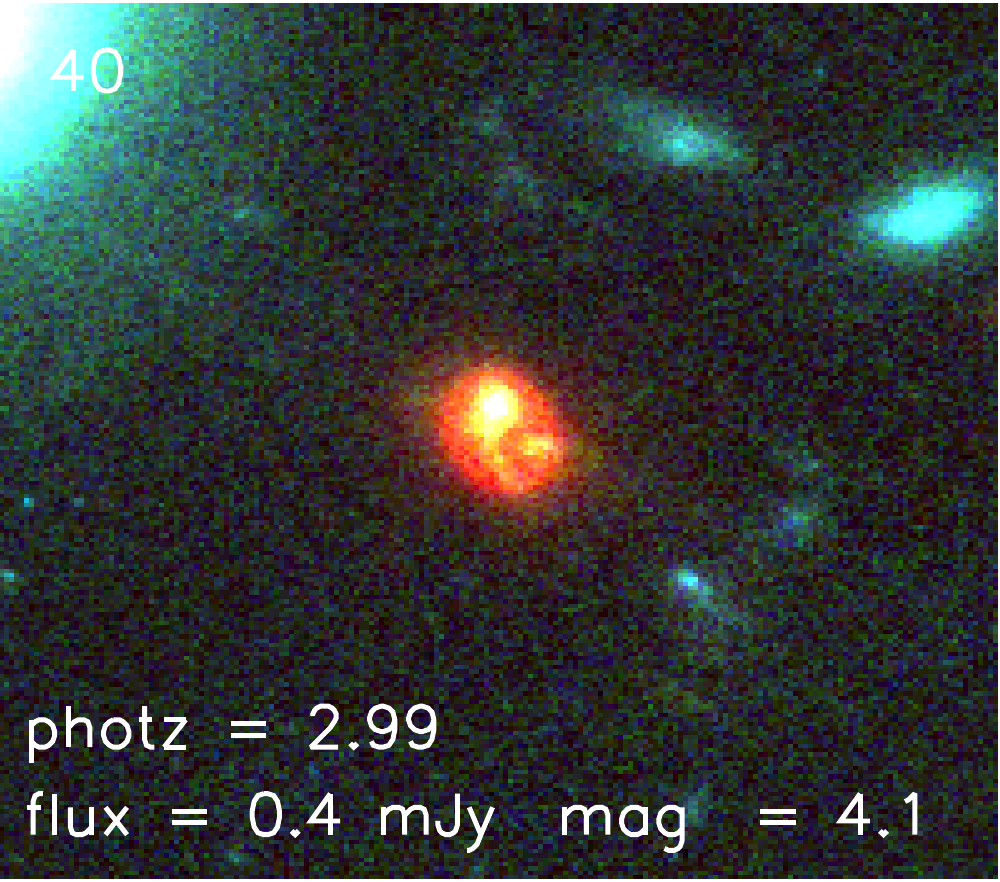}
\caption{ (Cont.)
\label{all_scuba2_images_2}}
\end{figure*}

%---------------------------------------------------------------------
% FIGURE 13 (cont)
%---------------------------------------------------------------------
\begin{figure*}
\setcounter{figure}{12}
\includegraphics[width=1.8in,angle=0]{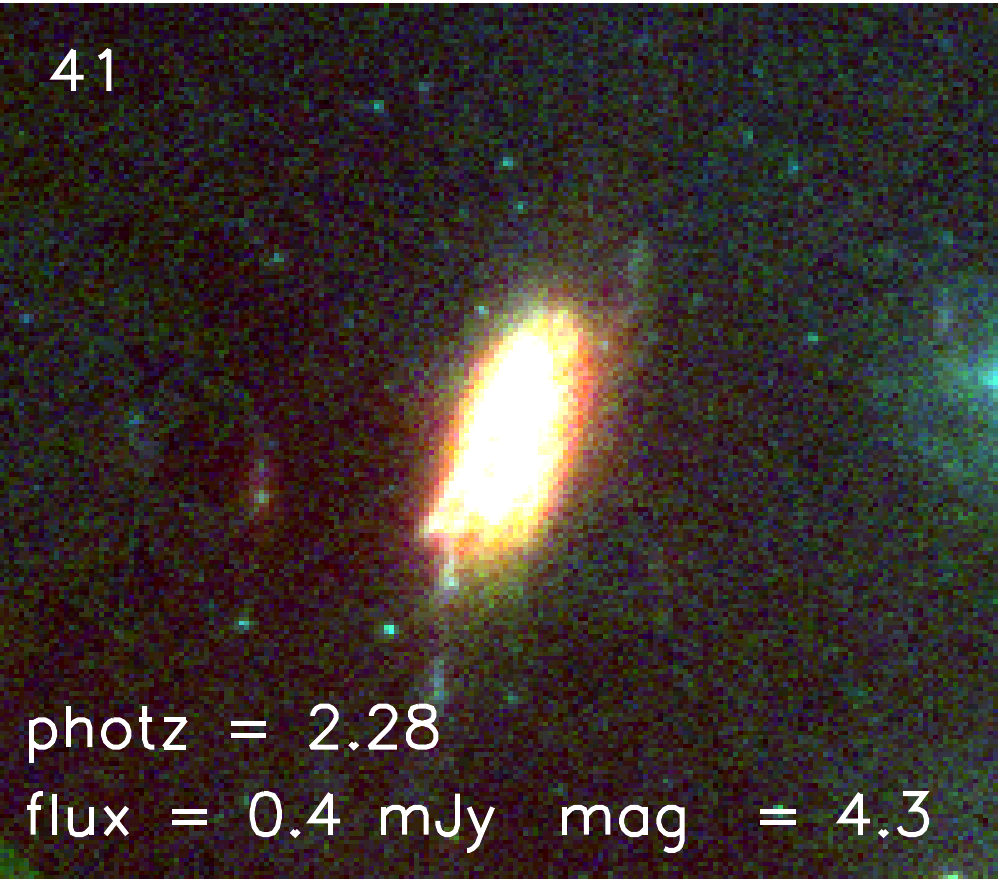}
\hskip -0.2cm
\includegraphics[width=1.8in,angle=0]{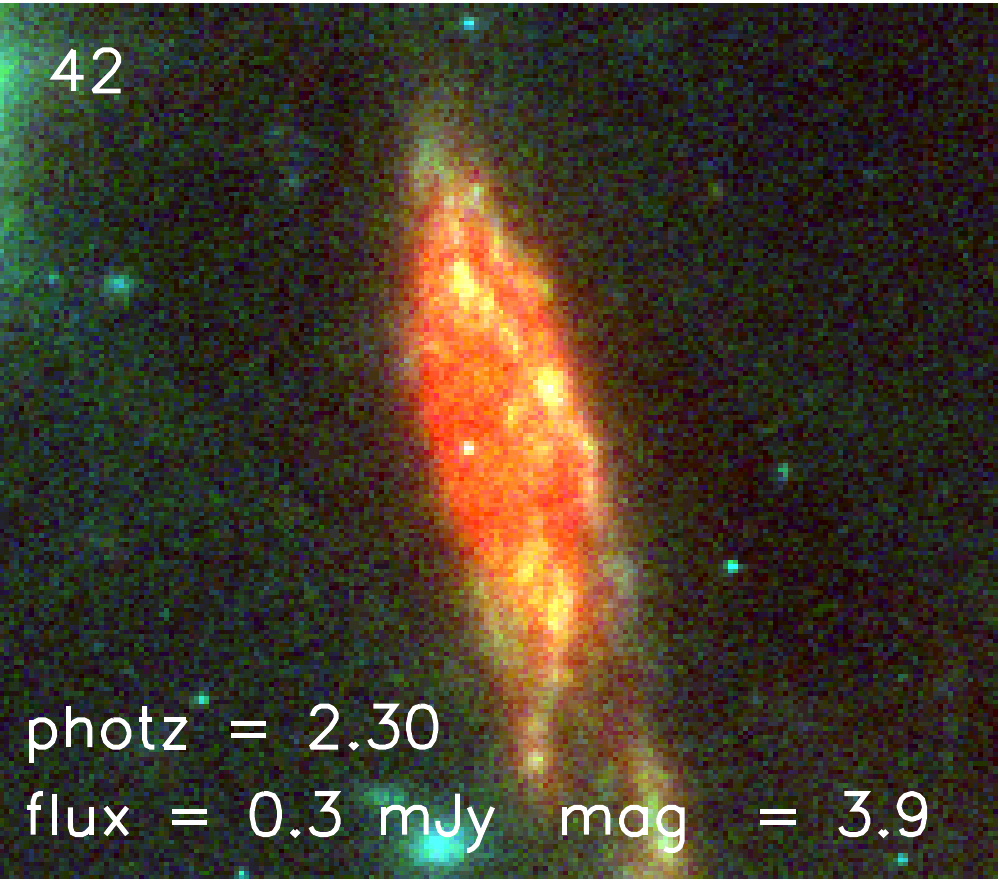}
\hskip -0.2cm
\includegraphics[width=1.8in,angle=0]{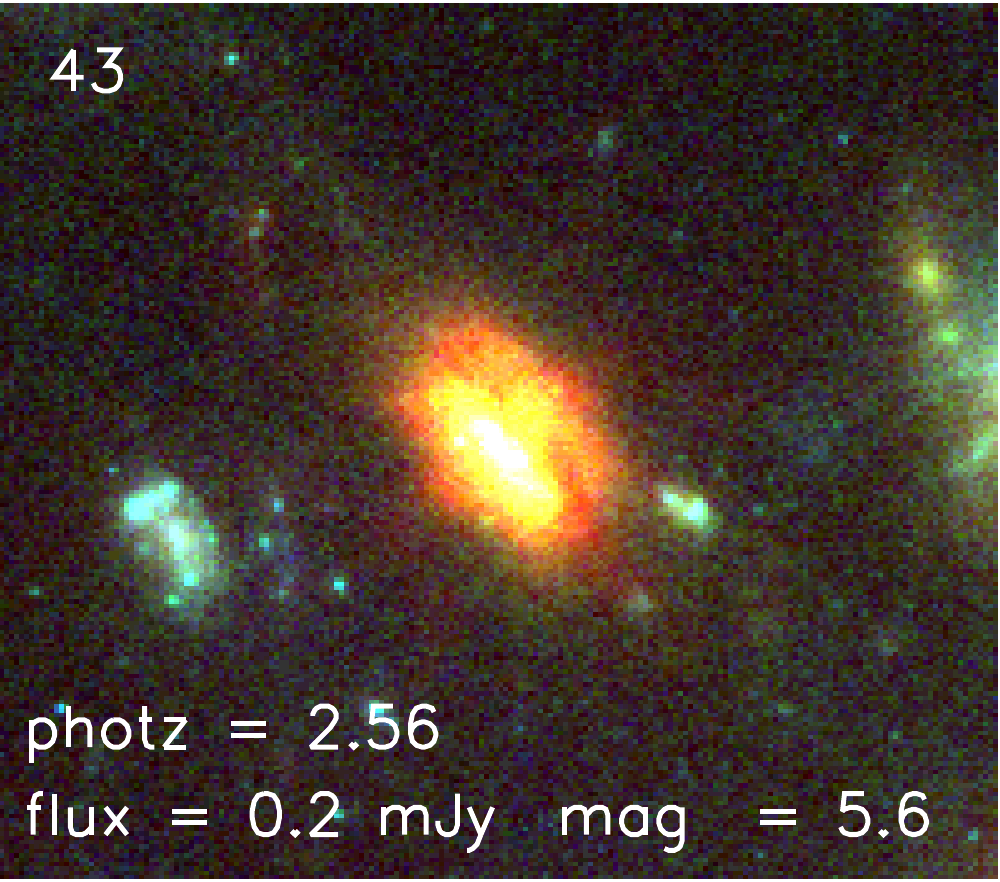}
\hskip -0.2cm
\includegraphics[width=1.8in,angle=0]{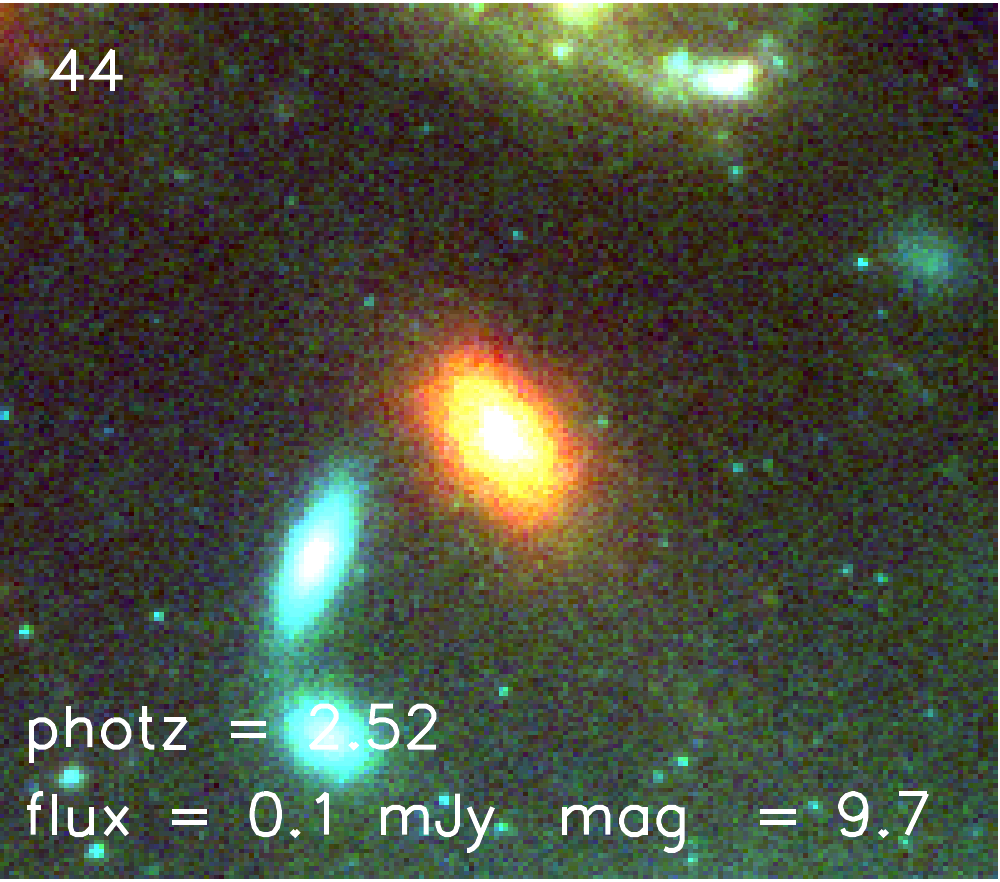}
\caption{ (Cont.)
\label{all_scuba2_images_4}}
\end{figure*}

\end{document}